\documentclass[a4paper,twoside,12pt]{book}

\usepackage[tbtags]{amsmath}
\usepackage{amssymb}
\usepackage{amsthm}
\usepackage{fancyhdr}
\usepackage[dvips]{graphicx}
\usepackage{exscale}
\usepackage{ifthen}
\usepackage{color}

\newcommand{\sh}[1]{
\ifthenelse{\equal{#1}{l}}
  {
   \text{$\not{#1}$}
  }
  {
   \text{$\not{\!#1}$}
  }
}
\newcommand{\lqcd}{\Lambda_{\text{QCD}}}
\newcommand{\Op}{\mathcal{O}}
\newcommand{\intd}{\int\frac{d^dk}{(2\pi)^d}}
\newcommand{\ifp}{\frac{i}{(4\pi)^2}}
\newcommand{\lmu}{\ln\frac{\mu^2}{m_b^2}}
\newcommand{\li}{\text{Li}_2}
\theoremstyle{definition}
\newtheorem{ex}{Example}[section]
\newtheorem{ru}{Rule}

\allowdisplaybreaks[1]

\renewcommand{\sectionmark}[1]%
         {\markright{\thesection\ #1}}
\newcommand{\LMUTitle}[9]{

  \thispagestyle{empty}

  \vspace*{\stretch{1}}
  {\parindent0cm
  \rule{\linewidth}{.7ex}}
  \begin{flushright}
    \vspace*{\stretch{1}}
    \sffamily\bfseries\Huge
    #1\\
    \vspace*{\stretch{1}}
    \sffamily\bfseries\large
    #2
    \vspace*{\stretch{1}}
  \end{flushright}
  \rule{\linewidth}{.7ex}

  \vspace*{\stretch{3}}
  \begin{center}
    \Large Dissertation\\
    \Large an der #4\\
    \Large der Ludwig--Maximilians--Universit\"at\\
    \Large M\"unchen\\
    \vspace*{\stretch{1}}
    \Large vorgelegt von\\
    \Large #2\\
    \Large aus #3\\
    \vspace*{\stretch{2}}
    \Large M\"unchen, den #6
  \end{center}

  \newpage
  \thispagestyle{empty}

  \vspace*{\stretch{1}}

  \begin{flushleft}
    \large Erstgutachter:  #7 \\[1mm]
    \large Zweitgutachter: #8 \\[1mm]
    \large Tag der m\"undlichen Pr\"ufung: #9\\
  \end{flushleft}

  \cleardoublepage
}

\begin{document}

  \frontmatter

  \LMUTitle
      {Hard spectator interactions in $B\to\pi\pi$
       at order $\alpha_s^2$}                        
      {Volker Pilipp}                                
      {M\"unchen}                                    
      {Fakult\"at f\"ur Physik}                      
      {M\"unchen 2007}                               
      {31. Mai 2007}                                 
      {Prof.~Dr.~Gerhard Buchalla}                   
      {Priv.~Doz.~Dr.~Stefan Dittmaier}              
      {30. Juli 2007}                                

  \tableofcontents
  \markboth{Contents}{Contents}


  \cleardoublepage

  \markboth{Zusammenfassung}{Zusammenfassung}
  \addcontentsline{toc}{chapter}{\protect Zusammenfassung}

\chapter*{Zusammenfassung}

In der vorliegenden Arbeit diskutiere ich die \emph{hard spectator
  interaction} Amplitude von $B\to\pi\pi$ zur n\"achstf\"uhrenden 
Ordnung in QCD (d.h.\ $\mathcal{O}(\alpha_s^2)$).
Dieser spezielle Teil der Amplitude, dessen f\"uhrende
Ordnung bei $\mathcal{O}(\alpha_s)$ beginnt, ist im Rahmen der QCD
Faktorisierung definiert. QCD Faktorisierung erm\"oglicht, in
f\"uhrender Ordnung in einer Entwicklung in $\lqcd/m_b$ die kurz- und
die langreichweitige Physik zu trennen, wobei die kurzreichweitige
Physik in einer st\"orungstheoretischen Entwicklung in $\alpha_s$
berechnet werden kann. Gegen\"uber anderen Teilen der Amplitude
erfahren hard spectator interactions formal eine Verst\"arkung durch
die zus\"atzlich zur $m_b$-Skala hinzutretende hartkollineare Skala
$\sqrt{\lqcd m_b}$, die zu einem gr\"o\ss{}eren numerischen Wert von
$\alpha_s$ f\"uhrt. 

Aus rechentechnischer Sicht liegen die haupts\"achlichen
Herausforderungen dieser Arbeit in der Tatsache begr\"undet, dass
die Feynmanintegrale, mit denen wir es zu tun haben, bis zu f\"unf
\"au\ss{}ere Beine haben und drei unabh\"angige Skalenverh\"altnisse
enthalten. Diese Feynmanintegrale m\"ussen in Potenzen in $\lqcd/m_b$
entwickelt werden. Ich werde integration by parts Identit\"aten
vorstellen, mit denen die Anzahl der Masterintegrale reduziert werden
kann. Ebenso werde ich diskutieren, wie man mit
Differenzialgleichungsmethoden die Entwicklung der Masterintegrale in
$\lqcd/m_b$ erh\"alt. Im Anhang ist eine konkrete Implementierung der
integration by parts Identit\"aten f\"ur ein Computeralgebrasystem vorhanden. 

Schlie\ss{}lich diskutiere ich numerische Sachverhalte, wie die
Abh\"angigkeit der Amplituden von der Renormierungsskala und die
Gr\"o\ss{}e der Verzweigungsverh\"altnisse. Es wird sich herausstellen
das die n\"achstf\"uhrende Ordnung der hard spectator interactions
wichtig jedoch klein genug ist, so dass die G\"ultigkeit der
St\"orungstheorie bestehen bleibt. 

  \markboth{Abstract}{Abstract}
  \addcontentsline{toc}{chapter}{\protect Abstract}

\chapter*{Abstract}

In the present thesis I discuss the \emph{hard spectator interaction}
amplitude in $B\to\pi\pi$ at NLO i.e.\ at $\mathcal{O}(\alpha_s^2)$. 
This special part of
the amplitude, whose LO starts at $\mathcal{O}(\alpha_s)$, is defined in
the framework of QCD factorization. QCD factorization allows to separate the 
short- and the long-distance physics in leading power in an expansion
in $\lqcd/m_b$, where the short-distance physics can be
calculated in a perturbative expansion in $\alpha_s$. Compared to
other parts of the amplitude hard
spectator interactions are formally enhanced by the hard collinear
scale $\sqrt{\lqcd m_b}$, which occurs next to the $m_b$-scale and
leads to an enhancement of $\alpha_s$.

From a technical point of view the main challenges of this calculation
are due to the fact that we have to deal with Feynman
integrals that come with up to five external legs and with three 
independent ratios of scales. These Feynman integrals have to be
expanded in powers of $\lqcd/m_b$.
I will discuss integration by parts identities to reduce the 
number of master integrals and differential equations techniques to
get their power expansions. A concrete implementation of integration by
parts identities in a computer algebra system is given in the
appendix.

Finally I discuss numerical issues like scale dependence of the
amplitudes and branching ratios. It will turn out that the NLO
contributions of the hard spectator interactions are important but
small enough for perturbation theory to be valid.

  \mainmatter
  \setcounter{page}{1}
  \chapter{Introduction}
The present situation of particle physics is the following. On the
one hand we have got an extremely successful standard model that
describes physics up to energy scales current accelerators are able to
reach. On the other hand it has limitations and problems, e.g.\ the
arbitrariness of the standard model parameters, the fact that the
Higgs particle has not yet been found, the stabilisation of
the Higgs mass under loop corrections (fine tuning problem) or the
question why the electroweak scale is so much lower than the Planck
scale (hierarchy problem). So most particle physicists expect new
physics to show up at energy scales that are beyond the range of 
present accelerators but will be reached by future colliders. Within
the next year LHC at CERN will start running and in the following
years will collect data from proton proton collisions at a centre of
mass energy of 14 TeV. This will allow us to obtain information about 
new physics by producing not yet observed particles directly. 
On the other hand physics beyond the standard model can be discovered 
by precision measurements of low energy quantities which are
influenced by new physics particles because of quantum effects. The 
currently running experiments BaBar and Belle and after the start of 
LHC also LHCb are dedicated to examine decays of $B$-mesons, where new 
physics is expected to be seen in CP asymmetries. 

However in order to find new
physics by indirect search, some parameters of the standard model have to
be determined more precisely. To this end LHCb will make an important
contribution. Above all the Wolfenstein parameters 
$\bar{\rho}$ and $\bar{\eta}$ \cite{Wolfenstein:1983yz,Buras:1994ec}
that occur in the parametrisation of the Cabibbo-Kobayashi-Maskawa
(CKM) matrix and determine 
its complex phase, which leads to CP asymmetry in the 
standard model, are up to now only very imprecisely known 
\cite{Charles:2004jd}:
\begin{equation}
\bar{\rho}=0.182^{+0.045}_{-0.047}\quad
\bar{\eta}=0.332^{+0.032}_{-0.036}.
\end{equation}
These parameters, which influence weak interactions of quarks, can
be determined with higher accuracy by $B$-meson decays. In order to
reduce their large uncertainties on the experimental side better
statistics is needed, which is expected to be improved in the next few
years, and on the theoretical side we have to get hadronic physics 
under control. This is due to the fact that weak interactions of
quarks, from which $\bar{\rho}$ and $\bar{\eta}$ are measured, are
always spoiled by non-perturbative strong interactions, because quarks
are bound in hadronic states like mesons. Hadronic physics, however, is
governed by the energy scale $\lqcd$, where QCD cannot be handled 
perturbatively.
 
There are several advantages in observing $B$-meson decays. 
One of them is due to the production of $B$-mesons itself:
There exists an extremely clean source to produce $B$-mesons: 
The resonance $\Upsilon(4S)$, a bound state of a $b\bar{b}$ pair, has
a mass that is only slightly larger than twice the mass of the
$B$-meson and decays nearly completely into $B\bar{B}$ pairs. This
resonance is used at the $B$-factories BaBar and Belle.

Another advantage is the possibility to obtain clean information about
the complex phase of the CKM~matrix by measurement of quantum mechanical
oscillations in the $B\!-\!\bar{B}$ system. The lifetime of $B$-mesons,
which is about $1.5\,\text{ps}$, is large enough to observe those
oscillations in the detector \cite{Abe:2004mz,Aubert:2005kf}. By
measuring the time dependent CP violation it is possible to obtain the
CKM angles $\alpha$, $\beta$ and $\gamma$, which determine
$\bar{\rho}$ and $\bar{\eta}$, with  small hadronic uncertainties 
(see e.g.\ chapter 1 of \cite{Harrison:1998yr}). 
The ``golden channel'' $B\to J/\psi K_S$, where the dependence on hadronic
quantities is strongly suppressed by small CKM parameters, allows
a quite precise determination of $\sin(2\beta)=0.687\pm0.032$ 
\cite{Barberio:2006bi}. In the same way the decay $B\to\pi\pi$ could be used
for a precise determination of the angle $\alpha$. However other than in the
``golden channel'' in the case of $B\to\pi\pi$ hadronic physics plays a
subdominant but non-negligible role. 

At this point another convenient property of the $B$-meson comes into
play. The mass of the $b$-quark introduces a hard scale, at which
$\alpha_s$ is small enough to make perturbation theory possible.
However the bound state of the $b$-quark in the $B$-meson is
dominated by physics of the soft scale $\lqcd$, where perturbation theory
breaks down. While inclusive decays can be handled in the framework of
operator product expansion, for exclusive decays, which the present thesis
deals with, the framework of QCD factorization has been
proposed \cite{Beneke:1999br,Beneke:2000ry}. This framework  makes it 
possible to disentangle the soft and hard physics at leading power in an
expansion in $\lqcd/m_b$. Decay amplitudes are then obtained in perturbative
expansions, which come with hadronic parameters that have to be determined in
experiment or by non perturbative methods like QCD sum rules or lattice QCD.

Whereas the $\alpha_s$ corrections for the
transition matrix elements of $B\to\pi\pi$ have been calculated in
\cite{Beneke:2001ev}, the present thesis deals with the
$\mathcal{O}(\alpha_s^2)$ contribution of a specific part of the
amplitude. This part, which consists of the hard spectator interaction
Feynman diagrams, will be defined in the next section. There I will also
argue, that it is reasonable to consider the hard spectator
interactions separately. The calculation of the rest of the
$\mathcal{O}(\alpha_s^2)$ corrections has been partly performed by
Guido Bell in his PhD thesis \cite{Bell:2007tz,Bell:2007tv}. There the complete
imaginary part and a preliminary result of the real part of the
amplitude is given. My calculation of the hard spectator scattering
amplitude is not the first one as it has been calculated recently
by \cite{Beneke:2005vv,Kivel:2006xc}. It is however the first
pure QCD calculation, whereas \cite{Beneke:2005vv,Kivel:2006xc}
used the framework of soft-collinear effective theory (SCET)
\cite{Bauer:2000yr,Bauer:2001yt,Beneke:2002ph} an effective theory, 
where the expansion in $\lqcd/m_b$ is performed at the level of the 
Lagrangian rather than of Feynman integrals. It is the main result 
of this thesis to confirm the results of
\cite{Beneke:2005vv,Kivel:2006xc} and to show by explicit calculation 
that pure QCD and SCET lead to the same result in this special case. 

From a technical point of view the calculation in this thesis consists
of the evaluation of about 60 one-loop Feynman diagrams. The
challenges of this task are due to the fact that these diagrams come
with up to five external legs and three independent ratios of
scales. In order to reduce the number of master integrals and to
perform power expansions of the Feynman integrals, integration by parts
methods and differential equation techniques will prove appropriate
tools. Most parts of the calculation will be performed by a computer
algebra system, whereas the algorithms and the necessary steps to obtain
input results for the programs will be discussed in detail. As I did
not obtain the completed $\mathcal{O}(\alpha_s^2)$ corrections, the
phenomenological part of this thesis is restricted to the reproduction of
the branching ratios numerically obtained in \cite{Beneke:2005vv}. 
Other observables like CP asymmetries are not improved by my partial
result alone.

This thesis is organised as follows:
In chapter \ref{prelim} I start with an introduction to QCD
factorization and define in this framework the hard spectator
scattering amplitude. After defining my notations I demonstrate the
calculation of  the LO of the hard spectator interactions and end the
chapter by explaining the technical details of the integration by parts
methods and differential equation techniques.

Chapter \ref{calcnlo} is the most technical of all. There all of the
Feynman diagrams that contribute are listed and their evaluation is
discussed in detail. Furthermore NLO corrections to the wave functions
and evanescent operators occurring at this order are dealt with.

After presenting the complete analytical results and 
the numerical analysis in chapter \ref{nloresults} I end up with the
conclusions.

\chapter{Preliminaries \label{prelim}}
\section{Hard spectator interactions and QCD factorization}
Though the decay of the $B$-meson is caused by weak interactions,
strong interactions play a dominant role. It is however not possible
to handle the QCD effects completely perturbatively. This is due to
the energy scales that are contained in the $B$-meson: Whereas $\alpha_s$ 
at the mass of the $b$-quark is a small parameter, the bound
state of the quarks leads to an energy scale of $\mathcal{O}(\lqcd)$
which spoils perturbation theory. The idea of QCD factorization
\cite{Beneke:1999br,Beneke:2000ry} is to separate these scales. At
leading power in $\lqcd/m_b$ we obtain the amplitude for $B\to\pi\pi$ in
the following form:
\begin{eqnarray}
  \langle \pi\pi| \mathcal{H} | B \rangle & \sim & 
  F^{B\to\pi}\int_0^1 dx\,T^\text{I}(x)f_\pi\phi_\pi(x)+ \nonumber \\
  &&\int_0^1 dx dy d\xi \, T^\text{II}(x,y,\xi) 
    f_B\phi_{B1}(\xi)f_\pi\phi_\pi(x)f_\pi\phi_\pi(y)
   \label{factform}
\end{eqnarray}
Two different types of quantities enter this formula. On the one hand
the hadronic physics is contained in the form
factor $F^{B\to\pi}$ and the wave functions $\phi_{B1}$ and $\phi_\pi$,
which will be defined in the next section more precisely. These
quantities contain the information about the bound states of the
mesons. They have to be determined by non-perturbative methods like QCD 
sum rules or lattice calculations. Alternatively, because they are at least
partly process independent, they might be extracted in the future from
experiment. On the other hand the hard scattering kernels $T^\text{I}$
and $T^\text{II}$ contain the physics of the hard scale $\mathcal{O}(m_b)$ and 
the hard collinear scale $\mathcal{O}(\sqrt{m_b\lqcd})$ and can be
calculated perturbatively. 

Here I would like to make two remarks to (\ref{factform}): 

First I want to note that (\ref{factform}) is only valid in leading
power in the expansion in $\lqcd/m_b$. Higher orders in this expansion
lead to endpoint singularities i.e.\ the integrals over the
variables $x$, $y$ and $\xi$ diverge at the endpoints. This leads to a
mixture of the physics of the soft scale into $T^\text{I}$ and
$T^\text{II}$ and spoils QCD factorization. However there are 
corrections that are formally of subleading power but numerically
enhanced and cannot be handled within the framework of QCD
factorization. They have to be estimated in the numerical analysis.

The second remark concerns the separation of the hard scattering
kernel into $T^\text{I}$ and $T^\text{II}$. The Feynman diagrams that
contribute to $B\to\pi\pi$ can be distributed into two different
classes. 
\begin{figure}
\begin{center}
\resizebox{\textwidth}{!}{\includegraphics{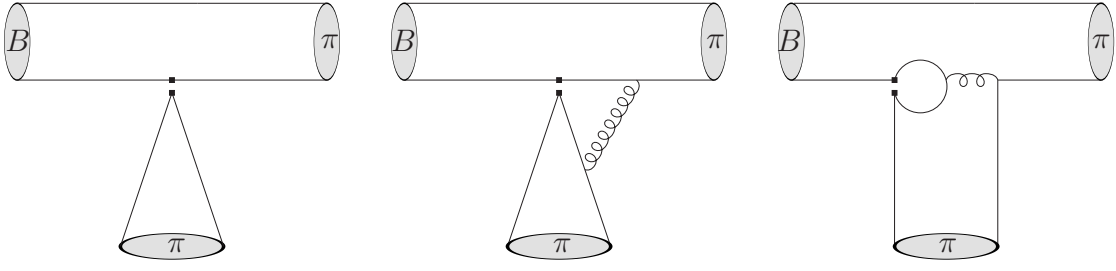}}
\end{center}
\caption{Tree level, vertex correction and penguin contraction. 
         These diagrams contribute to $T^\text{I}$.}
\label{vertex}
\end{figure}
The class of diagrams where there is no hard interaction of the
spectator quark (fig.~\ref{vertex}) contributes to $T^\text{I}$.
\begin{figure}
\begin{center}
\resizebox{\textwidth}{!}{\includegraphics{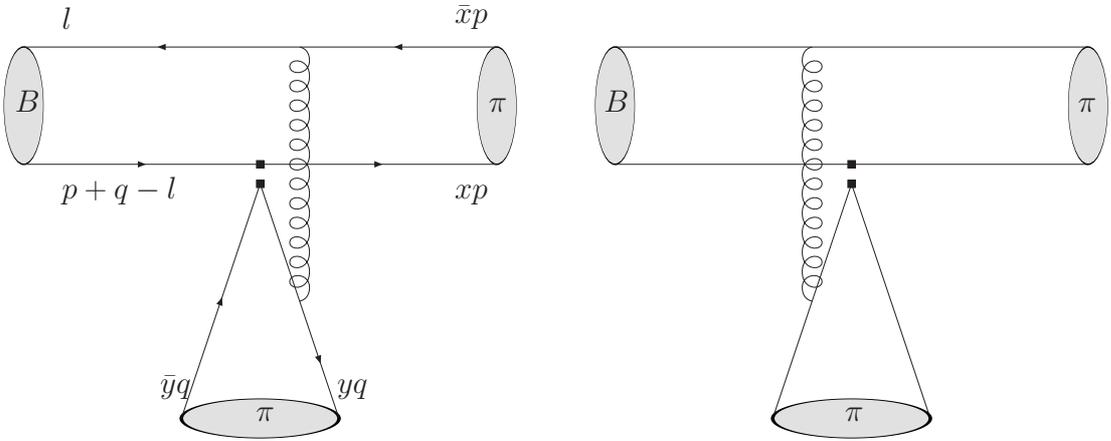}}
\end{center}
\caption{Hard spectator interactions at $\Op(\alpha_s)$. This is the
  LO of $T^\text{II}$}
\label{basic}
\end{figure}
The hard spectator scattering diagrams, which are shown in LO in $\alpha_s$
in fig.~\ref{basic}, contribute to $T^\text{II}$. Through the soft
momentum $l$ of the constituent quark of the $B$-meson the hard collinear
scale $\sqrt{\lqcd m_b}$ comes into play. This leads to the fact 
that in contrast to $T^\text{I}$, which is completely governed by the 
scale $m_b$, $T^\text{II}$ has to be evaluated at the hard-collinear 
scale. This leads to an enhancement of $\alpha_s$ and makes the NLO 
corrections (i.e.\ $\mathcal{O}(\alpha_s^2)$) of the hard spectator 
interaction diagrams more important. These $\alpha_s^2$ corrections of
$T^\text{II}$ are the topic of the present thesis. Hard spectator
scattering corrections to the penguin diagram (third diagram of
fig.~\ref{vertex}) are beyond the scope of this thesis. The
cancellation of the dependence on the renormalisation scale of this
class of diagrams is completely independent of the ``tree amplitude''
i.e.\ the diagrams of fig.~\ref{basic} and higher order $\alpha_s$
corrections. For phenomenological applications, however, they should be
taken into account.
\section{Notation and basic formulas}
\subsection{Kinematics}
For the process $B\to\pi\pi$ we will assign the momenta $p$ and
$q$ to the pions which fulfil
the condition 
\begin{equation}
p^2,q^2=0.
\label{kin1}
\end{equation}
This is the leading power approximation in $\Lambda_\text{QCD}/m_b$ 
as we count the mass of the pion as $\Op(\lqcd)$. 
Let us define two Lorentz vectors $n_+$, $n_-$ by:
\begin{equation}
n_+^\mu\equiv(1,0,0,1),\quad n_-^\mu\equiv(1,0,0,-1).
\label{kin2}
\end{equation}
In the rest frame of the decaying meson $p$ can be defined to be in the
direction of $n_+$ and $q$ to be in the direction of $n_-$. 
Light cone coordinates for the Lorentz vector $z^\mu$ are defined by:
\begin{equation}
z^+\equiv\frac{z^0+z^3}{\sqrt{2}},\quad
z^-\equiv\frac{z^0-z^3}{\sqrt{2}},\quad
z_\perp\equiv (0,z^1,z^2,0)
\label{k5}\end{equation}
So one can decompose $z^\mu$ into:
\begin{equation}
z^\mu=\frac{z\cdot p}{p\cdot q} q^\mu+\frac{z\cdot q}{p\cdot q}
p^\mu+z_\perp^\mu \label{kin3}
\end{equation}
such that 
\begin{equation}
z_\perp\cdot p=z_\perp\cdot q =0.\label{kin4}
\end{equation}

We denote the mass of the $B$-meson with $m_B$ and the mass of the
$b$-quark with $m_b$. The difference $m_B-m_b=\Op(\lqcd)$ such that we
cannot distinguish those masses in leading power. However setting
\begin{equation}
m_b=m_B
\label{kin5}
\end{equation}
in Feynman integrals might lead to additional infrared
divergences. So we have to perform the integral before we can make
the substitution (\ref{kin5}) unless we are sure that we do not 
produce infrared divergences. If we calculate 
Feynman integrals, it is convenient to set 
\begin{equation}
m_B=1
\end{equation}
such that $p\cdot q=\frac{1}{2}$. The dependence on $m_B$ can be
reconstructed by giving the correct mass dimension to the 
physical quantities.

\subsection{Colour factors}
In our calculations we will use the following three colour factors,
which arise from the $\text{SU}(3)$ algebra:
\begin{equation}
  C_N=\frac{1}{2},\quad C_F=\frac{N_c^2-1}{2N_c}\quad\text{and}\quad
  C_G=N_c,
  \label{colorfactor}
\end{equation}
where $N_c=3$ is the number of colours. 

\subsection{Meson wave functions}
The pion light cone distribution amplitude $\phi_\pi$ is defined by
\begin{equation}
  \label{mw1}
  \langle\pi(p)|\bar{q}(z)_\alpha [\ldots] q^\prime(0)_\beta|0\rangle_{z^2=0} =
  \frac{if_\pi}{4}(\sh{p}\gamma_5)_{\beta\alpha}
  \int_0^1 dx\, e^{ix p\cdot z}\phi_\pi(x).
\end{equation}
The ellipsis $[\ldots]$ stands for the Wilson line 
\begin{equation}
[z,0]=\text{P}\exp\left(\int_0^1 dt\,ig_s z\cdot A(z t)\right),
\end{equation}
which makes (\ref{mw1}) gauge invariant.
For the definition of the $B$-meson wave function $\phi_{B1}$ we need
the special kinematics of the process. Following \cite{Beneke:2000ry} let us 
define
\begin{equation}
\Psi_B^{\alpha\beta}(z,p_B)=
\langle 0| \bar{q}_\beta(z)[\ldots]b_\alpha(0)|B(p_B)\rangle
=\int\frac{d^4l}{(2\pi)^4}e^{-il\cdot z}\phi_B^{\alpha\beta}(l,p_B).
\label{mw2}\end{equation}
In the calculation of matrix elements we get terms like:
\begin{equation}
\int\frac{d^4l}{(2\pi)^4}\text{tr}(A(l)\phi_B(l))=
\int\frac{d^4l}{(2\pi)^4}\int d^4z\, e^{il\cdot z}\text{tr}(A(l)\Psi_B(z)).
\label{mw3}
\end{equation}
We will only consider the case that the dependence of the amplitude $A$ on $l$
is like this:
\begin{equation}
A(l)=A(2 l\cdot p)
\end{equation}
In this case we can use the $B$-meson wave function on the light cone
which is given by \cite{Beneke:2000ry}:
\begin{eqnarray}
\lefteqn{\langle 0|
\bar{q}_\alpha(z)[\ldots]b_\beta(0)|B(p_B)\rangle\bigg|_{z^-,z_\perp=0}}
\label{mw4}\\
&&=-\frac{if_B}{4}[(\sh{p}_B+m_b)\gamma_5]_{\beta\gamma}\int_0^1 d\xi\,
e^{-i\xi p_B^-
z^+}[\Phi_{B1}(\xi)+\sh{n}_+\Phi_{B2}(\xi)]_{\gamma\alpha}\nonumber
\end{eqnarray}
where
\begin{equation}
\int_0^1 d\xi\,\Phi_{B1}(\xi)=1\quad\mbox{and}\quad 
\int_0^1 d\xi\,\Phi_{B2}(\xi)=0.
\label{mw5}\end{equation}
It is now straight forward to write down the momentum projector of the
$B$-meson:
\begin{eqnarray}
\lefteqn{
\int\frac{d^4 l}{(2 \pi)^4}\, \text{tr} (A(2 l\cdot p)\hat{\Psi}(l))}
\nonumber\\
&&=\frac{-i f_B}{4}\text{tr}(\sh{p_B}+m_B)\gamma_5
\int_0^1 d\xi\,(\Phi_{B1}(\xi)+\sh{n}_+\Phi_{B2}(\xi))A(\xi m_B^2)
\label{mw6}
\end{eqnarray}
At this point we give the following definitions
\begin{eqnarray}
\frac{m_B}{\lambda_B}&\equiv&\int_0^1\frac{d\xi}{\xi}\phi_{B1}(\xi)
\label{lambdaB}\\
\lambda_n &\equiv& \frac{\lambda_B}{m_B}
\int_0^1\frac{d\xi}{\xi}\ln^n\xi\phi_{B1}(\xi).
\label{lambdan}
\end{eqnarray}

\subsection{Effective weak Hamiltonian}
\label{ewh}
The effective weak Hamiltonian which leads to
$B\to\pi\pi$ is given by \cite{Buchalla:1995vs}:
\begin {equation}
\mathcal{H}_\text{eff}=\frac{G_F}{\sqrt{2}}\sum_{p=u,c}\lambda_p^\prime
\left[C_1 \mathcal{O}_1 + C_2 \mathcal{O}_2
+\sum_{i=3\ldots 6} C_i \mathcal{O}_i+C_{8 g} \Op_{8g}\right]+\text{h.c.}, 
\label{w0}\end{equation}
where $\lambda_p^\prime=V_{pd}^\ast V_{pb}$ and
\begin{eqnarray}
\Op_1 & = & (\bar{d}p)_{V-A}(\bar{p}b)_{V-A},\label{w1}\\
\Op_2 & = & (\bar{d}_ip_j)_{V-A}(\bar{p}_jb_i)_{V-A},
\label{w2}\\
\Op_3   & = & (\bar{d}b)_{V-A}\sum_q(\bar{q}q)_{V-A},
\label{w3}\\
\Op_4   & = &
(\bar{d}_ib_j)_{V-A}\sum_q(\bar{q}_jq_i)_{V-A},
\label{w4}\\
\Op_5   & = & (\bar{d}b)_{V-A}\sum_q(\bar{q}q)_{V+A},
\label{w5}\\
\Op_6   & = &
(\bar{d}_ib_j)_{V-A}\sum_q(\bar{q}_jq_i)_{V+A},\label{w6}\\
\Op_{8g} & = & \frac{g}{8\pi^2}m_b\bar{d}_i\sigma^{\mu\nu}(1+\gamma_5)
T^a_{ij}b_j G_{\mu\nu}^a.\label{w8}
\end{eqnarray}
Explicit expressions for the short-distance coefficients $C_i$ can be
obtained from \cite{Buchalla:1995vs}. 
The decay amplitude of $B\to\pi\pi$ is given by
\begin{equation}
\mathcal{A}(B\to\pi\pi)\equiv\langle\pi\pi|\mathcal{H}_\text{eff}|B\rangle.
\label{w9}
\end{equation}
For later convenience we define 
\begin{equation}
\mathcal{A}(B\to\pi\pi)\equiv
\mathcal{A}(B\to\pi\pi)^\text{I}+\mathcal{A}(B\to\pi\pi)^\text{II}
\label{w10}
\end{equation}
where 
$\mathcal{A}^\text{I}$ ($\mathcal{A}^\text{II}$) belongs to the first (second)
term of (\ref{factform}). Because $\mathcal{A}^\text{I}$ and
$\mathcal{A}^\text{II}$ contain different hadronic quantities, the
renormalisation scale dependence of both of them has to vanish separately. 
So we can set their scales to different values $\mu^\text{I}$ and
$\mu^\text{II}$. As in $\mathcal{A}^\text{I}$ there occurs only the 
mass scale $m_b$
we can set $\mu^\text{I}=m_b$. In $\mathcal{A}^\text{II}$ there occurs also the
hard-collinear scale $\sqrt{\lqcd m_b}$.  As we will see this scale is an
appropriate choice for $\mu^\text{II}$.

In order to separate the QCD effects from the weak physics we write
the matrix elements of the effective weak Hamiltonian in the following
factorised form \cite{Beneke:2001ev}:
\begin{equation}
\langle\pi\pi|\mathcal{H}_\text{eff}|\bar{B}\rangle=
\frac{G_F}{\sqrt{2}}\sum_{p=u,c}\lambda_p^\prime
\langle\pi\pi|\mathcal{T}_p+\mathcal{T}_p^\text{ann}|\bar{B}\rangle
\label{w11}
\end{equation} 
where
\begin{equation}
\begin{split}
\mathcal{T}_p \quad= \quad& 
a_1\delta_{pu}(\bar{u}b)_{V-A}\otimes(\bar{d}u)_{V-A}
\\
+&a_2\delta_{pu}(\bar{d}b)_{V-A}\otimes(\bar{u}u)_{V-A}
\\
+&a_3\sum_q(\bar{d}b)_{V-A}\otimes(\bar{q}q)_{V-A}
\\
+&a_4^p\sum_q(\bar{q}b)_{V-A}\otimes(\bar{d}q)_{V-A}
\\
+&a_5\sum_q(\bar{d}b)_{V-A}\otimes(\bar{q}q)_{V+A}
\\
+&a_6^p\sum_q(-2)(\bar{q}b)_{S-P}\otimes(\bar{d}q)_{S+P}
\end{split}.
\label{w12}
\end{equation}
Note that in contrast to \cite{Beneke:2001ev} the electroweak corrections to
the effective weak Hamiltonian are
not included in the above equations as in the case of $B\to\pi\pi$
they can be safely neglected.
\begin{figure}
\begin{center}
\resizebox{0.5\textwidth}{!}{\includegraphics{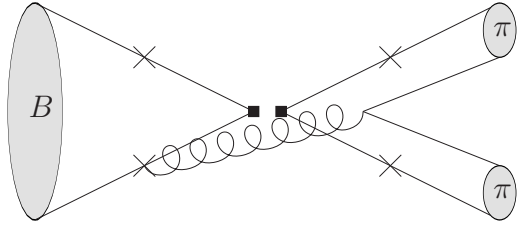}}
\end{center}
\caption{Annihilation topology. The gluon vertex that is marked by the
  cross can alternatively be attached to other crosses.}
\label{ann}
\end{figure}
$\mathcal{T}_p^\text{ann}$ stands for the contributions of the 
annihilation topologies, which are shown in fig.~\ref{ann}. 
These contributions do not occur in leading power and 
cannot be calculated in a model independent way in the 
framework of QCD factorization. For the exact
definition of $\mathcal{T}_p^\text{ann}$ I refer to \cite{Beneke:2001ev}.
The matrix elements of the operators $j_1\otimes j_2$ are defined to
be $\langle\pi\pi|j_1\otimes j_2|\bar{B}\rangle\equiv
\langle\pi|j_1|\bar{B}\rangle\langle\pi|j_2|0\rangle$ or 
$\langle\pi|j_2|\bar{B}\rangle\langle\pi|j_1|0\rangle$ corresponding
to the flavour structure of the $\pi$-mesons. The penguin contractions 
that are shown in the third diagram of fig.~\ref{vertex} and the
contributions of the operators $\Op_3$-$\Op_{8g}$ are by
definition contained in the amplitudes $a_3$-$a_6^p$. As we take in
the present thesis only the ``tree amplitude'' (fig.~\ref{basic}) into
account, we only calculate $\alpha_s^2$ corrections to the amplitudes 
$a_1$ and $a_2$.

The decay amplitudes of
$B\to\pi\pi$ can be written in terms of $a_i$ as follows \cite{Beneke:2001ev}:
\begin{eqnarray}
-\mathcal{A}(\bar{B}^0\to\pi^+\pi^-) &=&
\left[\lambda_u^\prime a_1+\lambda_p^\prime(a_4^p+r_\chi^\pi
  a_6^p)\right]A_{\pi\pi}
\nonumber\\
-\sqrt{2}\mathcal{A}(B^-\to\pi^-\pi^0) &=&
\lambda_u^\prime(a_1+a_2)A_{\pi\pi}
\nonumber\\
\mathcal{A}(\bar{B}^0\to\pi^0\pi^0) &=& 
\left[
-\lambda_u^\prime a_2+\lambda_p^\prime(a_4^p+r_\chi^\pi a_6^p)
\right]A_{\pi\pi}
\label{w13}
\end{eqnarray}
where 
$$A_{\pi\pi}=i\frac{G_F}{\sqrt{2}}(m_B^2-m_\pi^2)f^{B\pi}_+f_\pi$$
and 
\begin{equation}
r_\chi^\pi(\mu)=\frac{2m_\pi^2}{\bar{m}_b(\mu)(\bar{m}_u(\mu)+\bar{m}_d(\mu))}.
\label{w14}
\end{equation}
For the LO and NLO results of the $a_i$ I refer to
\cite{Beneke:2001ev}.

The annihilation contributions are parametrised in the following form
\cite{Beneke:2001ev}:
\begin{eqnarray}
-\mathcal{A}_\text{ann}(\bar{B}^0\to\pi^+\pi^-) &=&
\left[\lambda^\prime_u
  b_1+(\lambda^\prime_u+\lambda^\prime_c)(b_3+2b_4)
\right]B_{\pi\pi}
\nonumber\\
\mathcal{A}_\text{ann}(B^-\to\pi^-\pi^0) &=& 0
\nonumber\\
\mathcal{A}_\text{ann}(\bar{B}^0\to\pi^0\pi^0) &=&
-\mathcal{A}_\text{ann}(\bar{B}^0\to\pi^+\pi^-)
\label{w15}
\end{eqnarray}
where
\begin{equation}
B_{\pi\pi}=i\frac{G_F}{\sqrt{2}}f_Bf_\pi^2.
\label{w16}
\end{equation}
The parameters $b_i$ can be further parametrised by the Wilson
coefficients occurring in (\ref{w0}) and purely hadronic quantities:
\begin{eqnarray}
b_1&=&\frac{C_F}{N_c^2}C_1A_1^i\nonumber\\
b_3&=&\frac{C_F}{N_c^2}
\left[
C_3A_1^i+C_5(A_3^i+A_3^f)+N_cC_6A_3^f
\right]
\nonumber\\
b_4&=&\frac{C_F}{N_c^2}
\left[C_4A_1^i+C_6A_2^i
\right]
\label{w17}
\end{eqnarray}
where the quantities $A_k^{i(f)}$ are given by \cite{Beneke:2001ev}:
\begin{eqnarray}
A_1^i &=& \pi\alpha_s\left[18\left(
X_A-4+\frac{\pi^2}{3}\right)+2{r_\chi^\pi}^2X_A^2\right]
\nonumber\\
A_2^i &=& A_1^i\nonumber\\
A_3^i &=& 0\nonumber\\
A_3^f &=& 12\pi\alpha_sr_\chi^\pi\left(2X_A^2-X_A\right).
\label{w18}
\end{eqnarray}
Here $r_\chi^\pi$ is defined as in (\ref{w14}) and $X_A$ parametrises
an integral that is divergent because of endpoint singularities.
In section \ref{psc} I will give an estimate of $X_A$
for numerical calculations. 
\section{Hard spectator interactions at LO \label{HspLO}}
The leading order of the hard spectator interactions which start at
$\mathcal{O}(\alpha_s)$  is shown in \mbox{fig.~\ref{basic}}. 
The hard spectator scattering kernel
$T^\text{II}$, which does not depend on the wave functions, can be
obtained by calculating the transition matrix element between free
external quarks, to which we assign the momenta shown in
\mbox{fig.~\ref{basic}}. The variables $x, \bar{x}\equiv 1-x, y,
\bar{y}\equiv 1-y$ are the arguments of $T^\text{II}$, which arise from the
projection on the pion wave function (\ref{mw1}). In the sense of
power counting we count all components of $l$ of 
$\mathcal{O}(\lqcd)$, while the components of $p$ and $q$ are 
$\mathcal{O}(m_b)$ or
exactly zero. We define the following quantities
\begin{eqnarray}
\xi     & \equiv & \frac{l\cdot p}{p\cdot q} \nonumber\\
\theta  & \equiv & \frac{l\cdot q}{p\cdot q}. \label{LO1}
\end{eqnarray} 
We will see that in the end the dependence on $\theta$ vanishes 
in leading power such that we can use (\ref{mw6}).

We consider the three cases $\bar{B}^0\to\pi^+\pi^-$,
$\bar{B}^0\to\pi^0\pi^0$ and  $B^-\to\pi^-\pi^0$. 
In the case, that the external quarks come with the flavour content of 
$\bar{B}^0\to\pi^+\pi^-$, the LO hard spectator 
amplitude for the effective operator $\Op_2$ reads:
\begin{eqnarray}
\lefteqn{A_\text{spect.}^{(1)}(\bar{B}^0\to\pi^+\pi^-)\equiv}\nonumber\\
&&\langle \bar{d}(\bar{x}p) u(xp)\, \bar{u}(\bar{y}q)d(yq)|
\Op_2 | \bar{d}(l)b(p+q-l)\rangle_\text{spect.}=\nonumber\\
&&4\pi\alpha_s C_F N_c\frac{1}{\bar{x}\xi m_B^2}
\bar{d}(l)\gamma^\mu d(\bar{x} p)\,
\bar{u}(x p) \gamma^\nu(1-\gamma_5)b(p+q-l)\nonumber\\
&&\bar{d}(y q)\left(
\frac{2\sh{p}g_{\mu\nu}}{\bar{y}}-
\frac{\sh{p}}{y\bar{y}}\gamma_\mu\gamma_\nu\right)(1-\gamma_5)
u(\bar{y}q),
\label{LO2}
\end{eqnarray}
where the quark antiquark states in the input and output channels of
the matrix element form colour singlets. The subscript ``spect.''\ 
means that only diagrams with a hard spectator interaction are taken
into account. The amplitude of $\Op_1$ vanishes to this order in 
$\alpha_s$. In the case of 
$\bar{B}^0\to\pi^0\pi^0$ we get the tree amplitude from the matrix
element of $\Op_1$. The case $B^-\to\pi^-\pi^0$ does not need to
be considered separately, because from isospin symmetry follows
\cite{Gronau:1994rj,Beneke:2001ev}:
\begin{equation}
\sqrt{2}\mathcal{A}(B^-\to\pi^-\pi^0) =
\mathcal{A}(\bar{B}^0\to\pi^+\pi^-)+
\mathcal{A}(\bar{B}^0\to\pi^0\pi^0).
\end{equation}

On the other hand the full amplitude is the convolution of
$T^\text{II}$ with the wave functions, given by (\ref{factform}). 
To extract $T^\text{II}$ from
(\ref{LO2}) we need the wave functions with the same external states we
have used in (\ref{LO2}),
i.e.\ we have to calculate the matrix elements (\ref{mw1}) and
(\ref{mw2}), where the pion or $B$-meson states are replaced by free
external quark states. To the order $\mathcal{O}(\alpha_s^0)$ we get
\begin{eqnarray}
\phi_{\pi^-\alpha\beta}^{(0)}(y^\prime) &\equiv&
\int d(z\cdot q) e^{-i z\cdot q y^\prime}
\langle \bar{u}(\bar{y} q) d(y q)|\bar{d}^i_\beta(z) u^i_{\alpha}(0)|
0\rangle_{z^-,z_\perp=0} \nonumber\\
&=& 2\pi N_c\delta(y^\prime-y)
\bar{d}_\beta(yq)u_\alpha(\bar{y}q)\nonumber\\
\phi_{\pi^+\alpha\beta}^{(0)}(x^\prime) & = &
2\pi N_c\delta(x^\prime-x)\bar{u}_\beta(xp)d_\alpha(\bar{x}p)\label{LO3}\\
\phi_{B\alpha\beta}^{(0)}(l^{\prime -}) &\equiv& 
\int dz^+ e^{il^{\prime -} z^+}
\langle 0|\bar{d}_\beta(z)^i b_\alpha(0)^i|\bar{d}(l)b(p+q-l)
\rangle_{z^-,z_\perp=0}\nonumber\\
&=&2\pi N_c\delta(l^{\prime -}-l^-)\bar{d}_\beta(l)b_\alpha(p+q-l)
\nonumber
\end{eqnarray}
By using 
\begin{equation}
A^{(1)}_\text{spect.}=\int dxdydl^-\, 
\phi^{(0)}_{\pi^+\alpha\alpha^\prime}(x)
\phi^{(0)}_{\pi^-\beta\beta^\prime}(y)
\phi^{(0)}_{B\gamma\gamma^\prime}(l^-)
T^{\text{II}(1)}(x,y,l^-)_{\alpha^\prime\alpha\beta^\prime\beta\gamma^\prime\gamma}
\label{LO3.1}
\end{equation}
we finally obtain:
\begin{eqnarray}
T^{\text{II}(1)}(x,y,l^-)_{\alpha^\prime\alpha\beta^\prime\beta\gamma^\prime\gamma}
&=&
4\pi\alpha_s \frac{C_F}{(2\pi)^3 N_c^2}\frac{1}{\xi\bar{x}m_B^2}
\gamma^\mu_{\gamma^\prime\alpha}
\left[\gamma^\nu(1-\gamma_5)\right]_{\alpha^\prime\gamma}
\nonumber\\
&&\left[\left(\frac{2\sh{p}g_{\mu\nu}}{\bar{y}}-
\frac{\sh{p}}{y\bar{y}}\gamma_\mu\gamma_\nu\right)(1-\gamma_5)\right]_
{\beta^\prime\beta}.
\label{LOkernel}
\end{eqnarray}
It should be noted that only the first summand of the above equation 
contributes after performing the Dirac trace in four dimensions. The
second summand is evanescent. This will be important, when we will
calculate the NLO corrections of the wave functions (see section
\ref{wf}).
 
If we plug the hadronic wave functions defined by (\ref{mw1}) 
and (\ref{mw6}) into (\ref{LO3.1}) i.e.\ we calculate the matrix element 
(\ref{LO2}) between meson states instead of free quark states,
we get for the LO amplitude
\footnote{At this point I want to apologise to the reader because of
  a mismatch in my notation: $A_\text{spect.}$ is used for the
  matrix elements of the effective Operators $\Op_i$ between both free
  external quarks and hadronic meson states. It should become clear
  from the context what is actually meant.}:
\begin{equation}
A^{(1)}_\text{spect.}=
-\frac{if_\pi^2f_BC_F}{4 N_c^2}4\pi\alpha_s
\int_0^1 dxdyd\xi\,\Phi_{B1}(\xi)\phi_\pi(x)\phi_\pi(y) 
\frac{1}{\xi\bar{x}\bar{y}}.
\label{LOresult}
\end{equation}
Following (\ref{factform}) and the conventions of \cite{Beneke:2005vv} we
write our amplitude in the form:
\begin{equation}
A_{\text{spect.} i}=-im_B^2
\int_0^1dx dy d\xi \, T_i^\text{II}(x,y,\xi)
f_B\Phi_{B1}(\xi)f_\pi\phi_\pi(x)
f_\pi\phi_\pi(y).
\label{LO4}
\end{equation}
where in the case of $\bar{B}\to\pi^+\pi^-$ we define
\begin{eqnarray}
A_{\text{spect.}1} &=& \langle \Op_2\rangle_\text{spect.}
\nonumber\\
A_{\text{spect.}2} &=& \langle \Op_1\rangle_\text{spect.}
\label{LO5}
\end{eqnarray}
and in the case $\bar{B}\to\pi^0\pi^0$ we define 
\begin{eqnarray}
A_{\text{spect.}1} &=& \langle \Op_1\rangle_\text{spect.}
\nonumber\\
A_{\text{spect.}2} &=& \langle \Op_2\rangle_\text{spect.}.
\label{LO6}
\end{eqnarray}
Because we use the NDR-scheme which preserves Fierz transformations
for $\Op_1$ and $\Op_2$, 
$T^\text{II}_i$ has the same form for both decay channels. From
(\ref{LOresult}) and (\ref{LO4}) we get:
\begin{eqnarray}
T_1^{\text{II}(1)} &=&
4\pi\alpha_s\frac{C_F}{4N_c^2}\frac{1}{\xi\bar{x}\bar{y}m_B^2}
\nonumber\\
T_2^{\text{II}(1)} &=& 0 \quad .
\label{LO7}
\end{eqnarray}
According to \cite{Beneke:2001ev} the contribution of (\ref{LOresult})
to $a_1$ and $a_2$ (see (\ref{w12})) is given by:
\begin{eqnarray}
a_{1,\text{II}} &=& \frac{C_2C_F\pi\alpha_s}{N_c^2}H_{\pi\pi}
\nonumber\\
a_{2,\text{II}} &=& \frac{C_1C_F\pi\alpha_s}{N_c^2}H_{\pi\pi}
\label{LO8}
\end{eqnarray}
where
\begin{equation}
H_{\pi\pi}=
\frac{f_Bf_\pi}{m_B^2 f_+^{B\pi}}
\int_0^1\frac{d\xi}{\xi}\Phi_{B1}(\xi)
\int_0^1\frac{dx}{\bar{x}}\phi_{\pi}(x)
\int_0^1\frac{dy}{\bar{y}}\phi_{\pi}(y)
\label{LO9}
\end{equation}
and the label $\text{II}$ in (\ref{LO8}) denotes the contribution to
$\mathcal{A}^\text{II}$ as defined in (\ref{w10}).

\section{Calculation techniques for Feynman integrals} 
\subsection{Integration by parts method}
\label{ibpmethod}
Integration by parts (IBP) identities were introduced in 
\cite{Chetyrkin:1981qh,Tkachov:1981wb}.
An algorithm to reduce Feynman integrals by IBP-identities to master
integrals is very well described in
\cite{Laporta:2001dd}. So I will only show the basic principles. Because the
topic of my thesis is a one-loop calculation I will restrict to the
one-loop case, the generalisation to multi loop is straight forward.

The most general form\footnote{Tensor integrals which contain
  expressions like $k^\mu,k^\mu k^\nu,\ldots$ in the numerator can be
  reduced to scalar integrals as described in \cite{Passarino:1978jh}} 
of a one-loop Feynman integral is
\begin{multline}
\intd 
\frac{(k\cdot p_{j_1})^{n_1}\ldots(k\cdot p_{j_l})^{n_l}}
{\left[(k+p_{i_1})^2-M_1^2\right]^{m_1}\ldots
 \left[(k+p_{i_t})^2-M_t^2\right]^{m_t}
}\times\\
\frac{1}{
 \left[k\cdot p_{\tilde{i}_1}+\tilde{M}_1^2\right]^{\tilde{m}_1}
 \ldots
 \left[k\cdot p_{\tilde{i}_u}+\tilde{M}_u^2\right]^{\tilde{m}_u}
}
\equiv \intd 
\frac{s_1^{n_1}\ldots s_l^{n_l}}
{D_1^{m_1}\ldots D_t^{m_t}
 \tilde{D}_1^{\tilde{m}_1}\ldots\tilde{D}_u^{\tilde{m}_u}},
\label{ibp1}
\end{multline} 
where $n_1\ldots n_l, m_1\ldots m_t, 
\tilde{m}_1\ldots \tilde{m}_u \ge 0$,
$j_1\ldots,j_l,i_1\ldots,i_t,\tilde{i}_1,\ldots,\tilde{i}_u
\in \{1,\ldots, n\}$ and $p_1,\ldots,p_n$ are the
momenta which appear in the internal propagator lines. Without 
loss of generality we can assume that  
there is no $k^2$ in the numerator as we can make the
replacement
\begin{equation}
k^2=D_1+M_1^2-p_{i_1}^2-2k\cdot p_{i_1}.
\label{ibp1.2}
\end{equation}
Because of our
special kinematics we have only three linearly independent momenta
$p,q,l$, so all of the momenta $p_1,\ldots,p_n$ are linear
combinations of $ p,q,l$. This will simplify the reduction of the
Feynman integrals. We will define
\begin{equation}
\mathbf{B}\equiv\{\tilde{p}_1,\ldots,\tilde{p}_k\}
\label{ibp1.1}
\end{equation}
to be a basis of $\text{span}\{p_1,\ldots,p_n\}$ where 
$k\le3$ and $\tilde{p}_1,\ldots,\tilde{p}_k\in\{p_1,\ldots,p_n\}$.

Following \cite{Passarino:1978jh} we can reduce
(\ref{ibp1}) by performing algebraic transformations on the
integrands, which are defined in the following three rules:
\begin{ru}
\label{ru1}
Consider the case that there exist $\{c_1,\ldots,c_t\}$ such that  
\begin{equation}
\sum_{j=1}^{t}c_j p_{i_j} = 0\quad \text{and} \quad \sum_{j=1}^{t}c_j=1.
\label{ibp2}
\end{equation}
Now we can make the following simplification ($l\in\{1\ldots,t\}$):
\begin{eqnarray}
\lefteqn{
\frac{k\cdot p_{i_l}}{D_1^{m_1}\ldots D_t^{m_t}}
}\nonumber\\
&&=\frac{1}{2}
\frac{D_l-\sum_{j=1}^tc_j(D_j+M_j^2-p_{i_j}^2)+M_l^2-p_{i_l}^2}
{D_1^{m_1}\ldots D_t^{m_t}}\nonumber\\
&&=\frac{1}{2}\sum_{j=1}^t(\delta_{jl}-c_j)\left[
\frac{1}{D_1^{m_1}\ldots D_j^{m_j-1}\ldots D_t^{m_t}}+
\frac{M_j^2-p_{i_j}^2}{D_1^{m_1}\ldots D_t^{m_t}}
\right]
\label{ibp3}
\end{eqnarray}
and the scalar product $k\cdot p_{i_l}$ has disappeared from the
numerator.
If (\ref{ibp2}) cannot be fulfilled we use the identity
\begin{equation}
k\cdot p_{i_l}=
\frac{1}{2}(D_l-D_1+M_l^2-M_1^2+p_1^2-p_{i_l}^2)+k\cdot p_{i_1}
\label{ibp4}
\end{equation}
to reduce our set of integrals further. This identity does not reduce
the total number of scalar products in the numerator but the number of
different scalar products.
\end{ru}
\begin{ru}
\label{ru1.2}
For scalar products of the form $k\cdot p_{\tilde{i}_j}$ we make the
replacement
\begin{equation}
\frac{k\cdot p_{\tilde{i}_j}}{\tilde{D}_j^{\tilde{m}_j}}=
\frac{1}{\tilde{D}_j^{\tilde{m}_j-1}}-
\frac{\tilde{M}_j^2}{\tilde{D}_j^{\tilde{m}_j}}.
\label{ibp4.1}
\end{equation}
\end{ru}
\begin{ru}
\label{ru2}
Now consider the case that our integrand is of the form
\begin{equation}
\frac{k\cdot p_k}{D_1^{m_1}\ldots D_t^{m_t}
\tilde{D}_1^{\tilde{m}_1}\ldots\tilde{D}_u^{\tilde{m}_u}}
\end{equation}
where $p_k\notin \{p_{i_1},\ldots,p_{i_t}, 
p_{\tilde{i}_1},\ldots,p_{\tilde{i}_u}\}$. In that case we use the
following rule: Choose a set
$\mathbf{b}_1\subset\{p_{\tilde{i}_1},\ldots,p_{\tilde{i}_u}\}$
which is a basis of
$\text{span}\{p_{\tilde{i}_1},\ldots,p_{\tilde{i}_u}\}$. Choose 
$\mathbf{b}_2\subset\{p_{i_1},\ldots,p_{i_t}\}$
such that $\mathbf{b}=\mathbf{b}_1\cup\mathbf{b}_2$ forms a basis of
$\text{span}\{p_{i_1},\ldots,p_{i_t},p_{\tilde{i}_1},\ldots,p_{\tilde{i}_u}\}$.
Complete $\mathbf{b}$ to a basis of $\text{span}\{p_1,\ldots,p_n\}$
by adding elements of $\{p_1,\ldots,p_n\}$ to $\mathbf{b}$.
Then write $p_k$ as a linear combination of this 
new basis and apply (if possible) (\ref{ibp3}), (\ref{ibp4}) or (\ref{ibp4.1}) 
respectively. 
\end{ru}

For the following identities which are called integration by parts
or IBP identities we will use the fact that in dimensional
regularisation an integral over a total derivative with respect to
the loop momentum vanishes. Using the
definitions of (\ref{ibp1}) we get two further rules:
\begin{ru}
\label{ru3}
\begin{eqnarray}
0 & = & \intd \frac{\partial}{\partial k^\mu}
\frac{k^\mu s_1^{n_1}\ldots s_l^{n_l}}
{D_1^{m_1}\ldots D_t^{m_t}
\tilde{D}_1^{\tilde{m}_1}\ldots\tilde{D}_u^{\tilde{m}_u}} 
\nonumber\\
& = & (d+s-2r)
\intd\frac{s_1^{n_1}\ldots s_l^{n_l}}{D_1^{m_1}\ldots D_t^{m_t}
\tilde{D}_1^{\tilde{m}_1}\ldots\tilde{D}_u^{\tilde{m}_u}}
\nonumber\\
&&-\sum_{a=1}^t 2m_a\intd\left[
\frac{(M_a^2-p_{i_a}^2) s_1^{n_1}\ldots s_l^{n_l}}
{D_1^{m_1}\ldots D_a^{m_a+1}\ldots D_t^{m_t}}-
\frac{k\cdot p_{i_a} s_1^{n_1}\ldots s_l^{n_l}}
{D_1^{m_1}\ldots D_a^{m_a+1}\ldots D_t^{m_t}}
\right]\times
\nonumber\\
&&\frac{1}{\tilde{D}_1^{\tilde{m}_1}\ldots\tilde{D}_u^{\tilde{m}_u}}
-\sum_{a=1}^u \tilde{m}_a\intd\frac{k\cdot p_{\tilde{i}_a}}
{\tilde{D}_1^{\tilde{m}_1}\ldots\tilde{D}_a^{\tilde{m}_a+1}
\ldots\tilde{D}_u^{\tilde{m}_u}}
\label{ibp5}
\end{eqnarray} 
where $s\equiv\sum_{i=1}^l n_i$ and $r\equiv\sum_{i=1}^t m_i$. 
\end{ru}
Another identity is:
\begin{ru}
\label{ru4}
\begin{eqnarray}
0 & = & 
\intd \frac{\partial}{\partial k^\mu}
\frac{p_a^\mu s_1^{n_1}\ldots s_l^{n_l}}
{D_1^{m_1}\ldots D_t^{m_t}
\tilde{D}_1^{\tilde{m}_1}\ldots\tilde{D}_u^{\tilde{m}_u}} 
\nonumber\\
& = &
\sum_{b=1}^l n_b p_a\cdot p_{j_b}
\intd 
\frac{s_1^{n_1}\ldots s_b^{n_b-1}\ldots s_l^{n_l}}
{D_1^{m_1}\ldots D_t^{m_t}
\tilde{D}_1^{\tilde{m}_1}\ldots\tilde{D}_u^{\tilde{m}_u}}
\nonumber\\
&&-\sum_{b=1}^t 2m_b\intd
\frac{s_1^{n_1}\ldots s_l^{n_l} (k\cdot p_a+p_{i_b}\cdot p_a)}
{D_1^{m_1}\ldots D_b^{m_b+1}\ldots D_t^{m_t}
\tilde{D}_1^{\tilde{m}_1}\ldots\tilde{D}_u^{\tilde{m}_u}}
\nonumber\\
&&-\sum_{b=1}^u \tilde{m}_b p_a\cdot p_{\tilde{i}_b}
\intd\frac{s_1^{n_1}\ldots s_l^{n_l}}
{D_1^{m_1}\ldots D_t^{m_t}
\tilde{D}_1^{\tilde{m}_1}\ldots\tilde{D}_b^{\tilde{m}_b+1}\ldots
\tilde{D}_u^{\tilde{m}_u}}
\label{ibp6}
\end{eqnarray}
where $p_a\in \mathbf{B}$.
\end{ru}

For the IBP identities (\ref{ibp5}) and (\ref{ibp6}) we have used
the translation invariance of the dimensional regularised 
integral. We get another
class of identities if we use the invariance under Lorentz
transformations. From equation (2.9) of \cite{Gehrmann:1999as} we get
\begin{eqnarray}
0 & = & \bigg(
p_{{i_1}\nu}\frac{\partial}{\partial p_{i_1}^\mu}-
p_{{i_1}\mu}\frac{\partial}{\partial p_{i_1}^\nu}+
\ldots+
p_{{i_t}\nu}\frac{\partial}{\partial p_{i_t}^\mu}-
p_{{i_t}\mu}\frac{\partial}{\partial p_{i_t}^\nu}+
\nonumber\\
&&p_{{\tilde{i}_1}\nu}\frac{\partial}{\partial p_{\tilde{i}_1}^\mu}-
p_{{\tilde{i}_1}\mu}\frac{\partial}{\partial p_{\tilde{i}_1}^\nu}+
\ldots+
p_{{\tilde{i}_u}\nu}\frac{\partial}{\partial p_{\tilde{i}_u}^\mu}-
p_{{\tilde{i}_u}\mu}\frac{\partial}{\partial p_{\tilde{i}_u}^\nu}+
\nonumber\\
&&p_{{j_1}\nu}\frac{\partial}{\partial p_{j_1}^\mu}-
p_{{j_1}\mu}\frac{\partial}{\partial p_{j_1}^\nu}+
\ldots+
p_{{j_l}\nu}\frac{\partial}{\partial p_{j_l}^\mu}-
p_{{j_l}\mu}\frac{\partial}{\partial p_{j_l}^\nu}
\bigg)\times
\nonumber\\
&&\intd
\frac{s_1^{n_1}\ldots s_l^{n_l}}{D_1^{m_1}\ldots D_t^{m_t}
\tilde{D}_1^{\tilde{m}_1}\ldots\tilde{D}_u^{\tilde{m}_u}}.
\label{ibp7}
\end{eqnarray}
We choose $p_i,p_j\in\mathbf{B}$. By multiplying of (\ref{ibp7}) with
$p_i^\mu p_j^\nu$ we get 
\begin{ru}
\label{ru5}
\begin{eqnarray}
0 & = & \intd \bigg[\sum_{a=1}^l 
n_a(k\cdot p_i p_{j_a}\cdot p_j-k\cdot p_j p_{j_a}\cdot p_i)
\frac{s_1^{n_1}\ldots s_a^{n_a-1}\ldots s_l^{n_l}}
{D_1^{m_1}\ldots D_t^{m_t}
\tilde{D}_1^{\tilde{m}_1}\ldots\tilde{D}_u^{\tilde{m}_u}}-
\nonumber\\
&&\sum_{a=1}^t 
2m_a(k\cdot p_i p_{i_a}\cdot p_j-k\cdot p_j p_{i_a}\cdot p_i)
\frac{s_1^{n_1}\ldots s_l^{n_l}}
{D_1^{m_1}\ldots D_a^{m_a+1}\ldots D_t^{m_t}
\tilde{D}_1^{\tilde{m}_1}\ldots\tilde{D}_u^{\tilde{m}_u}}-
\nonumber\\
&&\sum_{a=1}^u 
\tilde{m}_a(k\cdot p_i p_{\tilde{i}_a}\cdot p_j-
k\cdot p_j p_{\tilde{i}_a}\cdot p_i)
\frac{s_1^{n_1}\ldots s_l^{n_l}}
{D_1^{m_1}\ldots D_t^{m_t}
\tilde{D}_1^{\tilde{m}_1}\ldots\tilde{D}_a^{\tilde{m}_a+1}
\ldots\tilde{D}_u^{\tilde{m}_u}}
\bigg].
\nonumber\\
\label{ibp8}
\end{eqnarray}
\end{ru}
An implementation in Mathematica of the IBP identities can be found in
appendix \ref{ibpalgorithm}.


\subsection{Calculation of Feynman diagrams with differential 
equations} \label{diffeq}
In this section I will discuss the extraction of subleading powers of
Feynman integrals with the {\itshape method of differential
  equations} \cite{Remiddi:1997ny,Caffo:1998du,Gehrmann:1999as}. 
This method will prove
to be easy to implement in a computer algebra system. The idea to
obtain the analytic expansion of Feynman integrals by tracing them 
back to differential equations has first been proposed in 
\cite{Remiddi:1997ny}. 
This method, which is demonstrated in \cite{Remiddi:1997ny} by the one-loop
two-point integral and in \cite{Caffo:1998du} by the two-loop sunrise
diagram, uses differential equations with respect to the small or
large parameter, in which the integral has to be expanded. 

In contrast to \cite{Remiddi:1997ny,Caffo:1998du} 
I will discuss the case that setting the small parameter to zero 
gives rise to new divergences. In this case the initial condition 
is not given by the differential equation itself and also cannot be 
obtained by calculation of the simpler integral that is defined by 
setting the expansion parameter to zero.
It is not possible to give a general proof, but it seems to be a rule, 
that one needs the leading power as a ``boundary condition'', which 
can be calculated by the \emph{method of regions} 
\cite{Gorishnii:1989dd,Beneke:1997zp,Smirnov:1990rz,Smirnov:2002pj}. 
The subleading powers can be obtained from the differential equation. 
In the present section I will discuss which conditions the differential 
equation has to fulfil in order for this to work.

\subsubsection{Description of the method}
We start with a (scalar) integral of the form 
\begin{equation}
I(p_1,\ldots,p_n,m_1,\ldots,m_n)=
\intd \frac{1}{D_1\ldots D_n}
\label{l1}
\end{equation}
where the propagators are of the form $D_i=(k+p_i)^2-m_i^2$. We assume
that there is only one mass hierarchy, i.e.\ there are two masses $m\ll
M$ such that all of the momenta and masses $p_i$ and $m_i$ are of
$\mathcal{O}(m)$ or of $\mathcal{O}(M)$. We expand (\ref{l1}) in
$\frac{m}{M}$ by replacing all small momenta and masses by $p_i\to\lambda
p_i$ and expand in $\lambda$. After the expansion the bookkeeping 
parameter $\lambda$ can be set to $1$. 

We obtain a differential equation for $I$ by differentiating the 
integrand in (\ref{l1}) with respect to $\lambda$. This gives rise 
to new Feynman integrals with propagators
of the form $\frac{1}{D_i^2}$ and scalar products $k\cdot p_i$ in the
numerator. Those Feynman integrals, however, can be reduced to the
original integral and to simpler integrals (i.e.\ integrals that contain
less propagators in the denominator) by using integration by parts 
identities.

Finally we obtain for (\ref{l1}) a differential equation of the form
\begin{equation}
\frac{d}{d\lambda}I(\lambda)=h(\lambda)I(\lambda)+g(\lambda)
\label{l2}
\end{equation}
where $h(\lambda)$ contains only rational functions of $\lambda$ and
$g(\lambda)$ can be expressed by Feynman integrals with a reduced
number of propagators. It is easy to see that $h$ and $g$ are unique if 
and only if $I$ and the integrals contained in $g$ are master integrals 
with respect to IBP-identities, i.e.\ they cannot be reduced to simpler
integrals by IBP-identities. If $I(\lambda)$ is divergent in 
$\epsilon=\frac{4-d}{2}$, $I$, $h$ and $g$ have to be expanded in 
$\epsilon$:
\begin{eqnarray}
I &=& \sum_i I_i \epsilon^i
\nonumber\\
h &=& \sum_i h_i \epsilon^i
\nonumber\\
g &=& \sum_i g_i \epsilon^i.
\label{l3}
\end{eqnarray}
Plugging (\ref{l3}) into (\ref{l2}) gives a system of differential
equations for $I_i$, similar to (\ref{l2}). In the next paragraph we
will consider an example for this case.

First let us assume that $h(\lambda)$ and $g(\lambda)$ have the
following asymptotic behaviour in $\lambda$:
\begin{eqnarray}
  h(\lambda) &=& h^{(0)}+\lambda h^{(1)}+\ldots\nonumber\\
  g(\lambda) &=& \sum_j \lambda^j g^{(j)}(\ln\lambda) 
\label{l4}
\end{eqnarray}
i.e.\ $h$ starts at $\lambda^0$, and we allow that $g$ starts at a negative
power of $\lambda$. We count $\ln\lambda$ as $\mathcal{O}(\lambda^0)$
so the $g^{(j)}$ may depend on $\ln\lambda$. This dependence, however, has
to be such that 
\begin{equation}
\lim_{\lambda\to 0}\lambda g^{(j)}(\ln\lambda)=0.
\label{l4.1}
\end{equation}
The condition (\ref{l4.1}) is fulfilled, if the $g^{(j)}$ are of the
form of a \emph{finite} sum
\begin{equation}
\sum_{n=n_0}^m a_n \ln^n\lambda.
\label{l4.2}
\end{equation}
The limit $m\to\infty$ however can spoil the expansion
(\ref{l4}). E.g.\ $e^{-\ln\lambda}=\frac{1}{\lambda}$ so the condition
(\ref{l4.1}) is not fulfilled, which is due to the fact that we must
not change the order of the limits $\lambda\to 0$ and $m\to\infty$.

Further we assume that also
$I(\lambda)$ starts at $\lambda^0$
\begin{equation}
  I(\lambda)= I^{(0)}(\ln\lambda)+\lambda I^{(1)}(\ln\lambda)
+\ldots
\label{l5}
\end{equation}
and plug this into (\ref{l2}) such that  we obtain an equation which gives 
$I^{(i)}$ recursively:
\begin{equation}
  \lambda^i I^{(i)}=\int_0^\lambda d\lambda^\prime {\lambda^\prime}^{i-1}
  \left(\sum_{j=0}^{i-1}
  h^{(j)}I^{(i-1-j)}(\ln\lambda^\prime)+g^{(i-1)}(\ln\lambda^\prime)\right).
\label{l6}
\end{equation}
I want to stress that, because $h$ starts at
$\mathcal{O}(\lambda^0)$, (\ref{l6}) is a recurrence relation, i.e.\
$I^{(j)}$ does not mix into $I^{(i)}$ if $j\ge i$.
As the integral is only well defined if $i\ge 1$,
we need the leading power $I^{(0)}$ as ``boundary condition''
and (\ref{l6}) will give us all the higher powers in $\lambda$. It is
easy to implement (\ref{l6}) in a computer algebra system, because we
just need the integration of polynomials and finite powers of logarithms.

A modification is needed if $h$ starts at $\lambda^{-1}$ i.e.\
\begin{equation}
  h=-\frac{n}{\lambda}h^{(-1)}+\ldots.
\label{l7}
\end{equation}
By replacing $\bar{I}\equiv \lambda^n I$ we obtain the differential
equation
\begin{equation}
  \frac{d}{d\lambda}\bar{I}=\left(\frac{n}{\lambda}+h\right)\bar{I}
  +\lambda^n g
\label{l8}
\end{equation}
which is similar to (\ref{l2}) and leads to
\begin{equation}
  \lambda^{i+n} I^{(i)}=\int_0^\lambda d\lambda^\prime {\lambda^\prime}^{i+n-1}
  \left(\sum_{j=0}^{i+n-1}
  h^{(j)}I^{(i-1-j)}(\ln\lambda^\prime)+g^{(i-1)}(\ln\lambda^\prime)\right),
\label{l9}
\end{equation}
which is valid for $i\ge 1-n$. So, if $I$ starts at
$\mathcal{O}(\lambda^{-n})$, the subleading powers result from the
leading power.  

\subsubsection{Examples}
We start with a pedagogic example:
\begin{ex}
\begin{equation}
I=\intd\frac{1}{k^2(k^2-\lambda)(k^2-1)}
\label{p1}
\end{equation}
\end{ex}
where $\lambda\ll 1$. The exact expression for this integral is given
by:
\begin{equation}
I=\frac{i}{(4\pi)^2}\frac{\ln\lambda}{1-\lambda}=
\frac{i}{(4\pi)^2}\ln\lambda(1+\lambda+\lambda^2+\ldots).
\label{p2}
\end{equation}
We see that $I$ diverges for $\lambda\to 0$. As described e.g.\ in 
\cite{Smirnov:2002pj} we can
obtain the leading power by expanding the integrand in the regions
$k\sim\sqrt{\lambda}$ and $k\sim 1$. This leads in the first region to
\begin{equation}
\intd \frac{-1}{k^2(k^2-\lambda)}=
-\frac{i}{(4\pi)^{2-\epsilon}}\Gamma(1+\epsilon)
\left(\frac{1}{\epsilon}+1-\ln\lambda\right)
\label{p3}
\end{equation}
and in the second region to
\begin{equation}
\intd \frac{1}{k^4(k^2-1)}=
\frac{i}{(4\pi)^{2-\epsilon}}\Gamma(1+\epsilon)
\left(\frac{1}{\epsilon}+1\right)
\label{p4}
\end{equation}
such that we finally obtain 
\begin{equation}
I^{(0)}(\ln\lambda)=\frac{i}{(4\pi)^2}\ln\lambda.
\label{p5}
\end{equation}
This is the result we obtain from the leading power of (\ref{p2}).
We write the derivative of $I$ with respect to $\lambda$ in the
following form:
\begin{equation}
\frac{d}{d\lambda}I=\frac{1}{1-\lambda}
\left[I-\intd\frac{1}{k^2(k^2-\lambda)^2}\right].
\label{p6}
\end{equation}
We obtained the right hand side of (\ref{p6}) by decomposing
$\frac{d}{d\lambda}I$ into partial fractions. Of course this
decomposition is not unique which is due to the fact that $I$ itself
is not a master integral but can be further simplified by partial
fractioning. From (\ref{p6}) and (\ref{l2}) we get:
\begin{eqnarray}
h &=& \frac{1}{1-\lambda}=1+\lambda+\lambda^2+\ldots
\nonumber\\
g &=& \frac{i}{(4\pi)^2}\frac{1}{\lambda(1-\lambda)}
=  \frac{i}{(4\pi)^2}\left(\lambda^{-1}+1+\lambda+\ldots \right)
\label{p7}
\end{eqnarray}
such that the coefficients in the expansion in $\lambda$ according to
(\ref{l4}) do not depend on the power label $(k)$:
\begin{equation}
h^{(k)}=1\quad\text{and}\quad g^{(k)}=\frac{i}{(4\pi)^2}.
\label{p8}
\end{equation}
We obtain for the recurrence relation (\ref{l6}):
\begin{equation}
I^{(k)}=\frac{1}{\lambda^k}\int_0^\lambda d\lambda^\prime\,
{\lambda^\prime}^{k-1}
\left(\sum_{j=0}^{k-1}I^{(k-1-j)}(\ln\lambda^\prime)
+\frac{i}{(4\pi)^2}\right).
\label{p9}
\end{equation}
Using the initial value (\ref{p5}) it is easy to prove by induction
\begin{equation}
I^{(k)}(\ln\lambda)=\frac{i}{(4\pi)^2}\ln\lambda
\quad\forall k\ge 0.
\label{p10}
\end{equation}
This result coincides with (\ref{p2}).

The first nontrivial example, we want to consider, 
is the following three-point integral:
\begin{ex}
\label{example2}
\begin{equation}
I=\intd \frac{1}{k^2(k+un_-+l)^2(k+n_++n_-)^2}.
\label{intex2}
\end{equation}
\end{ex}
Here $n_+$ and $n_-$ are collinear Lorentz vectors, which fulfil
$n_+^2=n_-^2=0$ and $n_+\cdot n_-=\frac{1}{2}$, $u$ is a real number
between $0$ and $1$ and $l$ is a Lorentz vector with $l^2=0$ and
$l^\mu\ll 1$. Furthermore we define 
\begin{equation}
\xi=2l\cdot n_+\quad\text{and}\quad \theta=2l\cdot n_-.
\end{equation}
We expand $I$ in $l$, so we make the replacement
$l\to\lambda l$ and differentiate $I$ with respect to $\lambda$. The integral
is not divergent in $\epsilon$ such that we obtain a
differential equation of the form (\ref{l2}) where the Taylor series
of $h(\lambda)$ starts at $\lambda^0$ as in (\ref{l4}). 
In $g(\lambda)$ only two-point integrals occur, which are easy to
calculate. I do not want to give the explicit expressions for $h$ and
$g$ because they are complicated, their exact form is not needed to
understand this example and they can be handled by a computer algebra
system. Because the leading power of $I$ is of
$\mathcal{O}(\lambda^0)$, (\ref{l6}) gives all of the subleading
powers.

We obtain the leading power as follows: First we have to identify the
regions, which contribute at leading power. If we decompose $k$ into 
\begin{equation}
k^\mu=2 k\cdot n_+n_-^\mu+2 k\cdot n_-n_+^\mu+k_\perp^\mu
\label{l10.1}
\end{equation}
we note that the only regions, which remain at leading power, are
the hard region $k^\mu\sim 1$ and the hard-collinear region 
\begin{eqnarray}
k\cdot n_+ &\sim& 1 \nonumber\\
k\cdot n_- &\sim& \lambda \nonumber\\
k_\perp^\mu &\sim& \sqrt{\lambda}.
\label{l10.2}
\end{eqnarray}
The soft region $k^\mu\sim\lambda$ leads at leading power to a
scaleless integral, which vanishes in dimensional regularisation.
In the hard region we expand the integrand to
\begin{equation}
\frac{1}{k^2(k+un_-)^2(k+n_++n_-)^2}.
\label{l10.3}
\end{equation}
By introducing a convenient Feynman parametrisation we obtain for the
$(4-2\epsilon)$-dimensional integral over (\ref{l10.3}):
\begin{equation}
\frac{i}{(4\pi)^{2-\epsilon}}\Gamma(1+\epsilon)\exp(i\pi\epsilon)
\frac{1}{u}
\left(
\frac{\ln(1-u)}{\epsilon}-\frac{1}{2}\ln^2(1-u)
\right).
\label{l10.4}
\end{equation}
In the hard-collinear region we expand the integrand to
\begin{equation}
\frac{1}{k^2(k+un_-+\theta n_+)^2(2k\cdot n_++1)}.
\label{l10.5}
\end{equation}
The integral over (\ref{l10.5}) gives:
\begin{equation}
\begin{split}
\frac{i}{(4\pi)^{2-\epsilon}}\Gamma(1+\epsilon)\exp(i\pi\epsilon)
\frac{1}{u}
\bigg(&
\frac{-\ln(1-u)}{\epsilon}
+2\li(u)
+\frac{1}{2}\ln^2(1-u)\\
&+\ln u\ln(1-u)+
\ln(1-u)\ln\theta
\bigg).
\end{split}
\label{l10.6}
\end{equation}
So adding (\ref{l10.4}) and (\ref{l10.6}) together we get the leading
power of (\ref{intex2}):
\begin{equation}
I^{(0)}=
\frac{i}{(4\pi)^2}\frac{1}{u}
\left(
2\li(u)+\ln u\ln(1-u)+\ln(1-u)\ln\theta
\right).
\label{l10.7}
\end{equation}
By plugging (\ref{l10.7}) into (\ref{l6}) we obtain
$I$ at $\mathcal{O}(\lambda)$:
\begin{equation}
\begin{split}
&I^{(1)}=\\
&\begin{split}
\frac{i}{(4\pi)^2}\frac{1}{u}\bigg[&
\theta\bigg(
-2+\ln u +\frac{\ln(1-u)\ln\theta}{u}+\frac{\ln(1-u)\ln u}{u}+
\ln\xi+\frac{2\li(u)}{u}
\bigg)-\\
&\begin{split}
\xi\bigg(&\frac{\ln u}{1-u}+2\frac{\ln(1-u)}{u}+
\frac{\ln(1-u)\ln\theta}{u}+
\frac{\ln(1-u)\ln u}{u}+\\
&\frac{\ln\xi}{1-u}+
\frac{2\li(u)}{u}
\bigg)
\bigg].
\end{split}
\end{split}
\end{split}
\label{l10.8}
\end{equation}

Now we want to consider the following four-point integral
\begin{ex}
\label{example1}
\begin{equation}
I=\intd\frac{1}{k^2(k+n_-)^2(k+l-n_+)^2(k+l-un_+)^2},
\label{intex1}
\end{equation}
\end{ex}
where we used the same variables, which were introduced in
(\ref{intex2}). This example is very special, because in this case our
method will allow us to obtain not only the subleading but also 
the leading power in $l$. $I$ is divergent in $\epsilon$ such that we
obtain after the expansion (\ref{l3}) a system of differential
equations of the following form:
\begin{eqnarray}
\frac{d}{d\lambda}I_{-1}&=&h_0I_{-1}+g_{-1}
\nonumber\\
\frac{d}{d\lambda}I_{0}&=&h_0I_{0}+h_1I_{-1}+g_{0}.
\label{intex1.2}
\end{eqnarray}
It turns out that in our example $h$ takes the simple form
\begin{equation}
h=-\frac{2+2\epsilon}{\lambda}
\end{equation}
such that analogously to (\ref{l8}) we can transform (\ref{intex1.2}) into
\begin{eqnarray}
\frac{d}{d\lambda}(\lambda^2 I_{-1})&=&\lambda^2g_{-1}
\nonumber\\
\frac{d}{d\lambda}(\lambda^2I_{0})&=&-2\lambda I_{-1}+\lambda^2g_{0}.
\label{intex1.3}
\end{eqnarray} 
This system of differential equations can easily be integrated to:
\begin{eqnarray}
I^{(i)}_{-1} &=& \frac{1}{\lambda^{i+2}}
\int_0^\lambda d\lambda^\prime\,{\lambda^\prime}^{i+1}g_{-1}^{(i-1)}
\nonumber\\
I^{(i)}_{0} &=& \frac{1}{\lambda^{i+2}}
\int_0^\lambda d\lambda^\prime\,{\lambda^\prime}^{i+1}
\left(-2I_{-1}^{(i)}+g_{0}^{(i-1)}\right)
\label{intex1.4}
\end{eqnarray}
where the superscript $(i)$ denotes the order in $\lambda$ as in
(\ref{l4}) and (\ref{l5}). 
Both $I_{-1}$ and $I_0$ start at $\mathcal{O}(\lambda^{-1})$. Because
(\ref{intex1.4}) is valid for $i\ge -1$, it gives us the leading power
expression, which reads:
\begin{equation}
I^{(-1)}=\frac{i}{(4\pi)^{2-\epsilon}}\Gamma(1+\epsilon)
\frac{2}{u\xi}
\left(
\frac{1}{\epsilon}-1-\frac{\ln u}{1-u}-\ln\xi
\right)
\label{intex1.5}
\end{equation}
where $\xi=2l\cdot n_+$ as in the example above. The exact expression
for (\ref{intex1}) can be obtained from \cite{Duplancic:2000sk}. Thereby
(\ref{intex1.5}) can be tested.

\subsubsection{A simplification for the calculation of the leading power}
In the last paragraph I want to return to Example \ref{example2}.
I will show how we can use
differential equations to prove that the integral (\ref{intex2}) 
depends in leading power only on the soft kinematical variable 
$\theta=2l\cdot n_-$ and not on $\xi=2l\cdot n_+$.
We need derivatives of the integral with respect
to $\xi$ and $\theta$, which we have to express through derivatives with
respect to $l^\mu$. These derivatives can be applied directly to the
integrand, whose dependence on $l^\mu$ is obvious. We start from the
following identities:
\begin{eqnarray}
  n_+^\mu\frac{\partial}{\partial l^\mu} I &=&
  \frac{\partial}{\partial \theta}I+
  \xi \frac{\partial}{\partial l^2}I\nonumber\\
   n_-^\mu\frac{\partial}{\partial l^\mu} I &=&
  \frac{\partial}{\partial \xi}I+
  \theta \frac{\partial}{\partial l^2}I\label{ldiff1}\\
   l^\mu\frac{\partial}{\partial l^\mu} I &=&
  \xi \frac{\partial}{\partial \xi}I+
  \theta \frac{\partial}{\partial \theta}I+
  2l^2 \frac{\partial}{\partial l^2}I\nonumber
 \end{eqnarray}
which lead to
\begin{eqnarray}
 \xi \frac{\partial}{\partial \xi}I &=&
 \frac{1}{2}(-\theta n_+^\mu+\xi n_-^\mu+l^\mu)
 \frac{\partial}{\partial l^\mu} I\nonumber\\
 \theta \frac{\partial}{\partial \theta}I &=&
 \frac{1}{2}(\theta n_+^\mu-\xi n_-^\mu+l^\mu)
 \frac{\partial}{\partial l^\mu}I.\label{ldiff2}
\end{eqnarray}
where we have set $l^2=0$ in (\ref{ldiff2}). 
Using (\ref{ldiff2}) we can show that in leading power
(\ref{intex2}) depends only on $\theta$ and not
on $\xi$. So we can simplify the calculation of the leading power by
making the replacement $l^\mu\to\theta n_+^\mu$. The
proof goes as follows: From (\ref{l5}) we see that the statement
``$I^{(0)}$ does not depend on $\xi$'' is equivalent to 
\begin{equation}
  \xi\frac{\partial}{\partial \xi}I(\xi\lambda,\theta\lambda)=
  \mathcal{O}(\lambda).
\label{diffproof}
\end{equation}
Using the first equation of (\ref{ldiff2}) we get
\begin{equation}
   \xi\frac{\partial}{\partial \xi}I(\xi\lambda,\theta\lambda)=
   \mathcal{O}(\lambda) I(\xi\lambda,\theta\lambda)+\mathcal{O}(\lambda).
\end{equation}
Because we know (e.g.\ from power counting) that
$I(\xi\lambda,\theta\lambda)$ starts at $\lambda^0$, (\ref{diffproof})
is proven.

\chapter{Calculation of the NLO \label{calcnlo}}
\section{Notation}
\subsection{Dirac structure}
In this thesis I used the NDR scheme \cite{Herrlich:1994kh}
such that $\gamma_5$-matrices are anticommuting. We get for
the matrix elements of $\Op_1$ and $\Op_2$ Dirac
structures of the following type:
\begin{equation}
\langle \Op_{1,2}\rangle =
\bar{q}_1(l)\Gamma_1q_1(\bar{x}p)\,
\bar{q}_2(xp)\Gamma_2b(p+q-l)\,
\bar{q}_3(yq)\Gamma_3q_4(\bar{y}q)
\label{dirac1}
\end{equation}
where $q_i$ are $u$- or $d$-quarks. To avoid to specify the flavour,
which depends on the decay mode, I
introduce for (\ref{dirac1}) the following short notation:
\begin{equation}
\Gamma_1\tilde{\otimes}\Gamma_2\otimes\Gamma_3.
\label{dirac2}
\end{equation} 
The equations of motion lead to:
\begin{eqnarray}
\sh{l}\Gamma_1\tilde{\otimes}\Gamma_2\otimes\Gamma_3 &=& 0
\nonumber\\
\Gamma_1\sh{p}\tilde{\otimes}\Gamma_2\otimes\Gamma_3 &=& 0
\nonumber\\
\Gamma_1\tilde{\otimes}\sh{p}\Gamma_2\otimes\Gamma_3 &=& 0
\nonumber\\
\Gamma_1\tilde{\otimes}\Gamma_2(\sh{p}+\sh{q}-\sh{l}+m_b)\otimes\Gamma_3 &=& 0
\nonumber\\
\Gamma_1\tilde{\otimes}\Gamma_2\otimes\sh{q}\Gamma_3 &=& 0
\nonumber\\
\Gamma_1\tilde{\otimes}\Gamma_2\otimes\Gamma_3\sh{q} &=& 0.
\label{dirac3}
\end{eqnarray}

\subsection{Imaginary part of the propagators}
Unless otherwise stated propagators in Feynman integrals always
contain an term $+i\eta$ where $\eta>0$ and we take the limit
$\eta\to0$ after the integration. For example an integral of the form 
\begin{displaymath}
\intd\frac{1}{k^2(k+p)^2}
\end{displaymath}
is just an abbreviation for
\begin{displaymath}
\intd\frac{1}{(k^2+i\eta)((k+p)^2+i\eta)}. 
\end{displaymath}
If the propagator does not contain the integration momentum
quadratically, the $i\eta$ will always be given explicitly.

\section{Evaluation of the Feynman diagrams \label{clodi}}
\begin{figure}[p]
\begin{center}
\resizebox{0.8\textwidth}{!}{\includegraphics{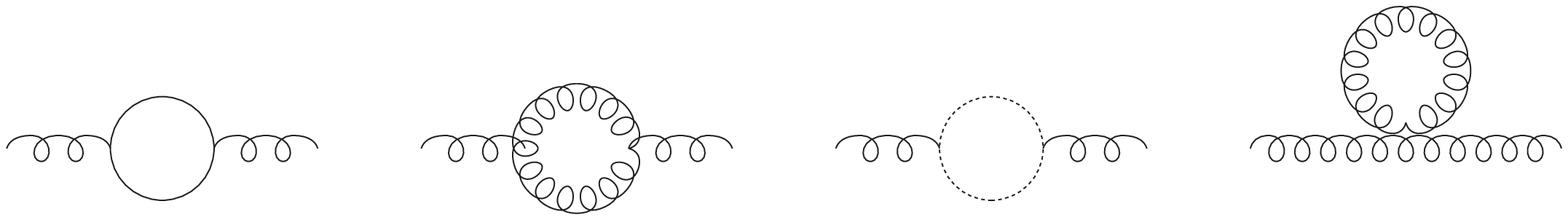}}
\end{center}
\caption{Gluon self energy}
\label{ag}
\end{figure}
\begin{figure}[p]
\begin{center}
\resizebox{.8\textwidth}{!}{\includegraphics{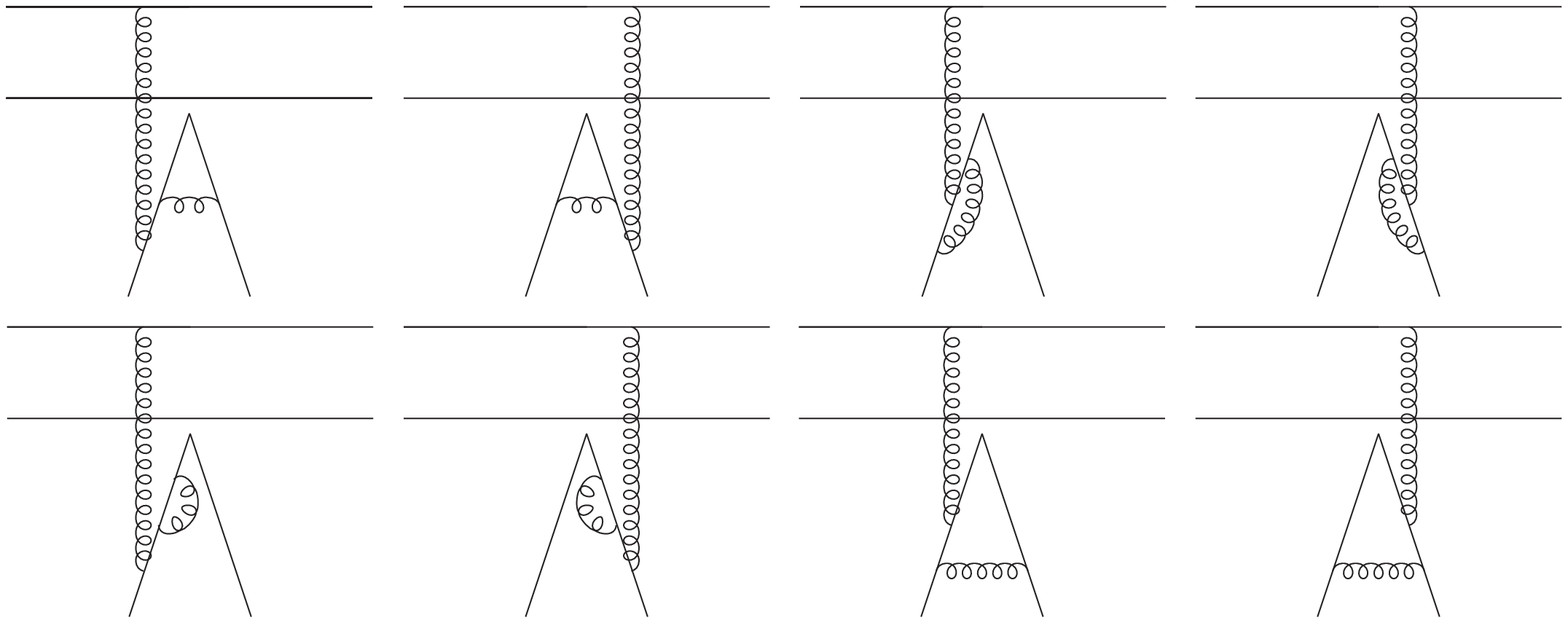}}
\end{center}
\caption{Diagrams aI}
\label{aI}
\end{figure}
\begin{figure}[p]
\begin{center}
\resizebox{.8\textwidth}{!}{\includegraphics{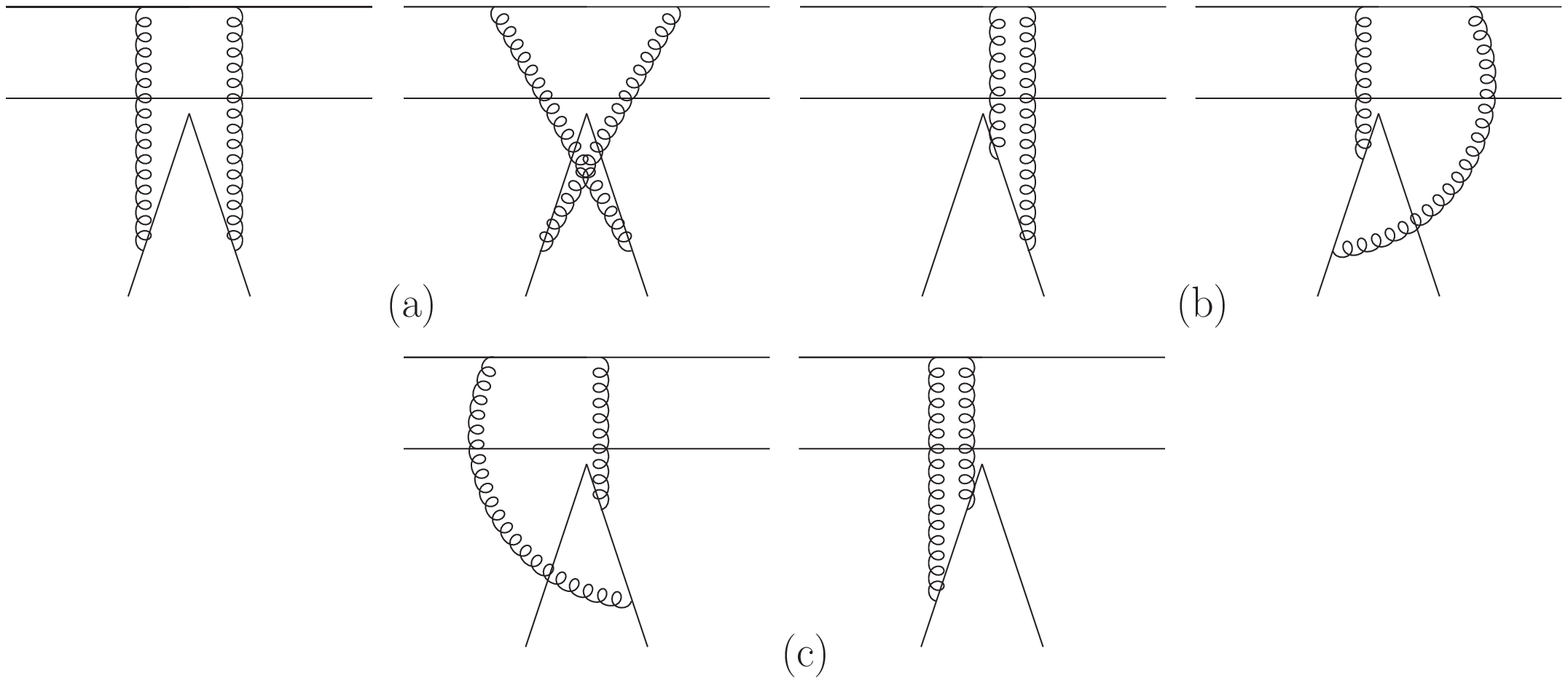}}
\end{center}
\caption{Diagrams aII}
\label{aII}
\end{figure}
\begin{figure}[p]
\begin{center}
\resizebox{.8\textwidth}{!}{\includegraphics{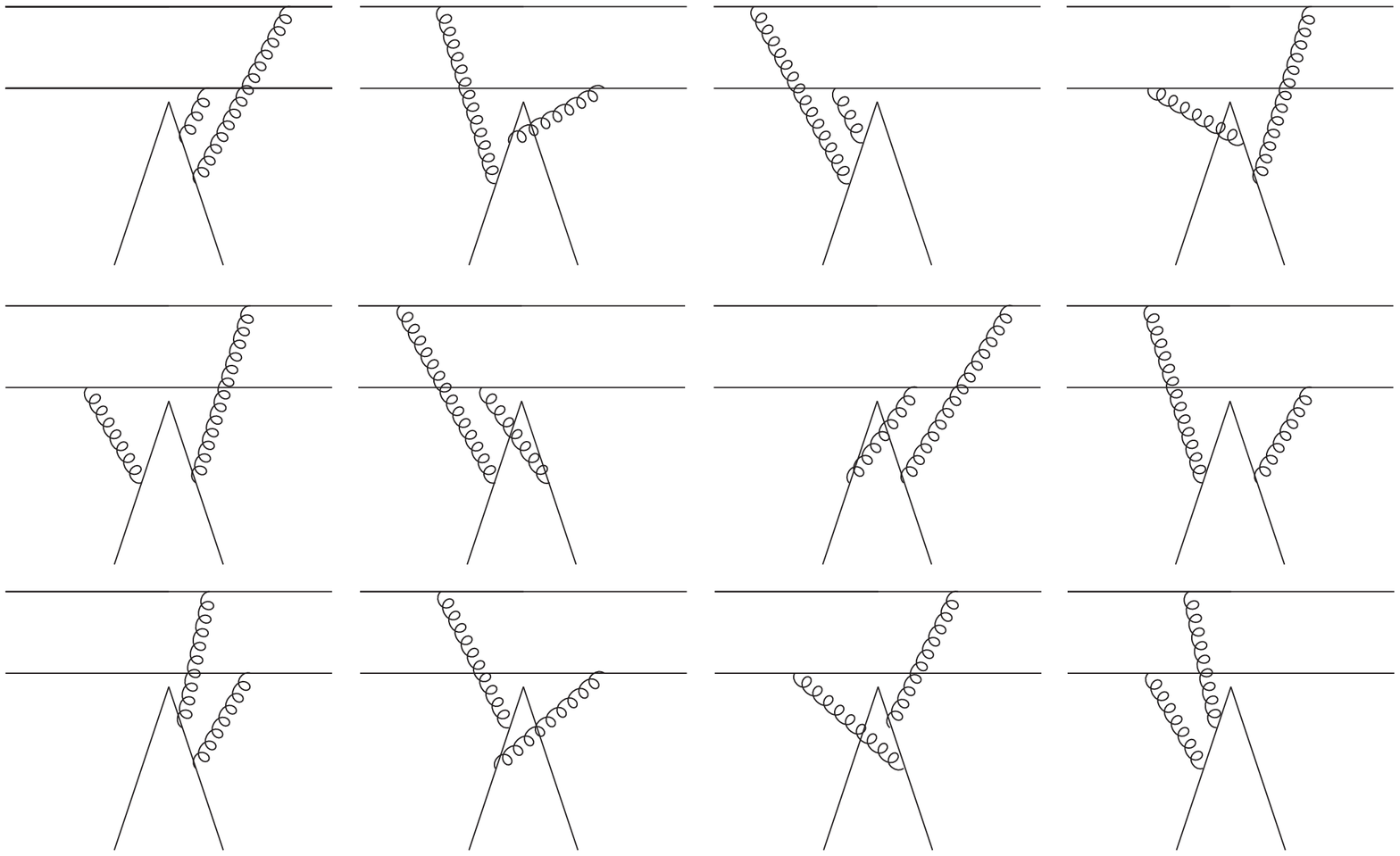}}
\end{center}
\caption{Diagrams aIII}
\label{aIII}
\end{figure}
\begin{figure}[p]
\begin{center}
\resizebox{.8\textwidth}{!}{\includegraphics{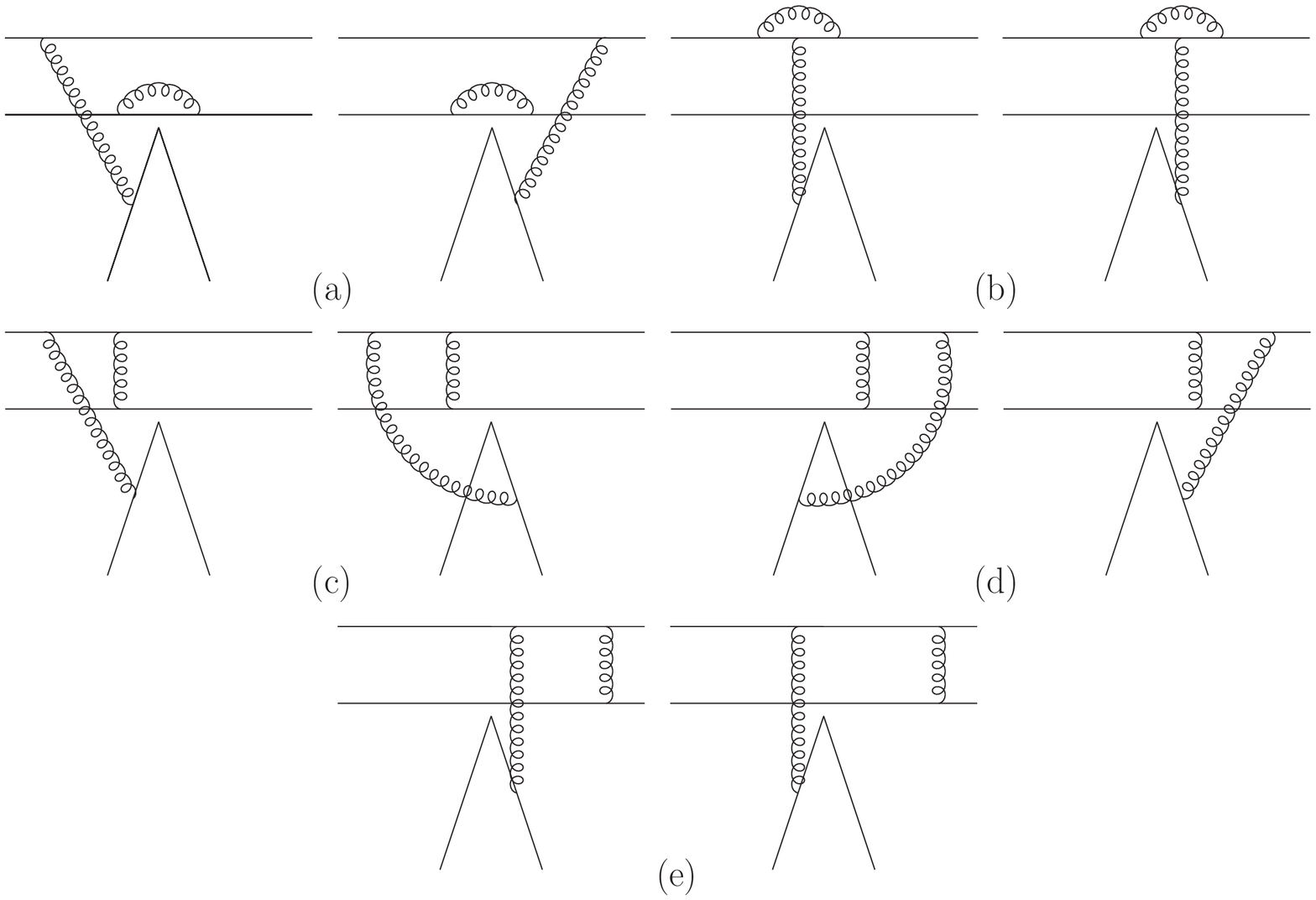}}
\end{center}
\caption{Diagrams aIV}
\label{aIV}
\end{figure}
\begin{figure}[p]
\begin{center}
\resizebox{.8\textwidth}{!}{\includegraphics{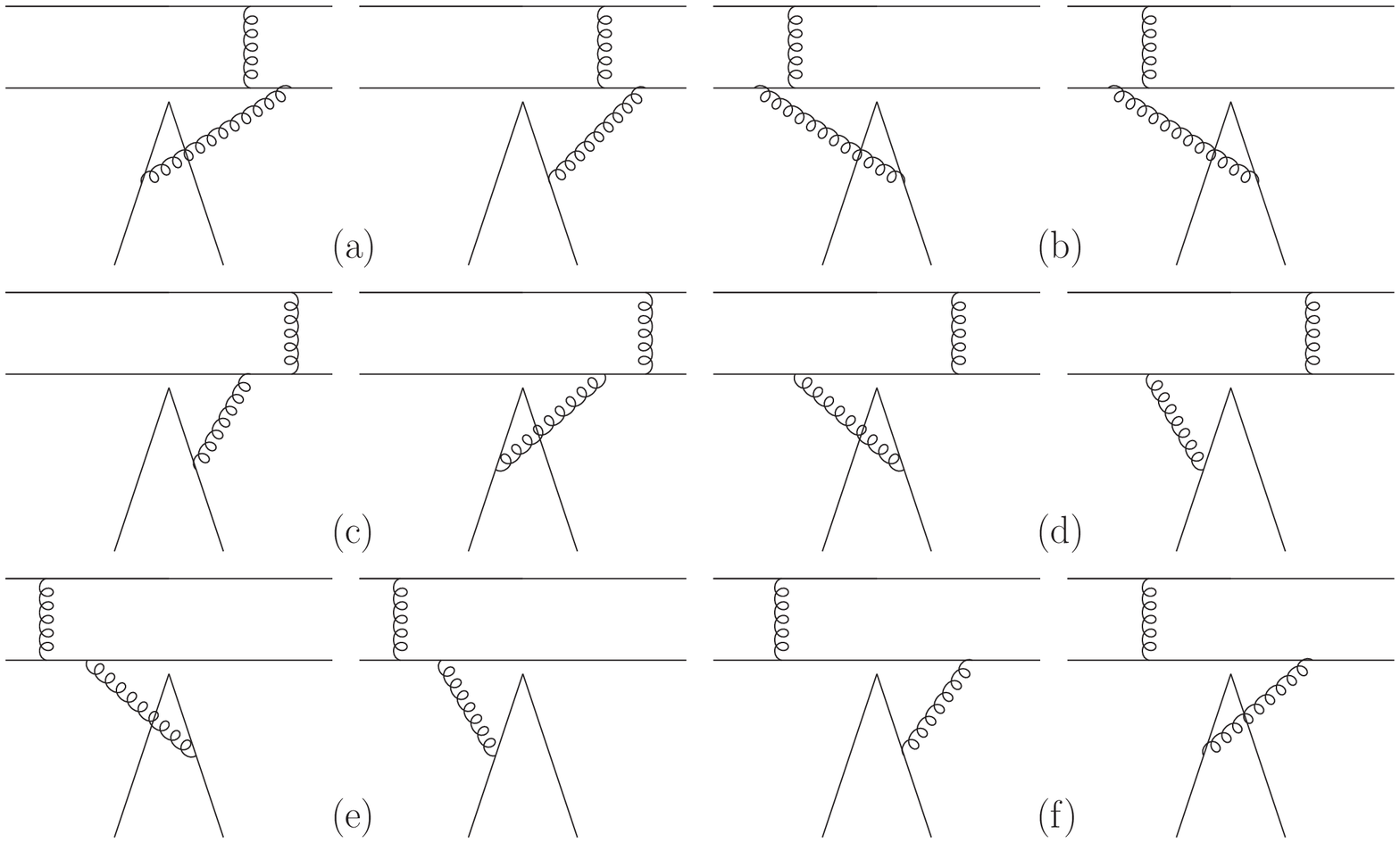}}
\end{center}
\caption{Diagrams aV}
\label{aV}
\end{figure}
\begin{figure}[p]
\begin{center}
\resizebox{.8\textwidth}{!}{\includegraphics{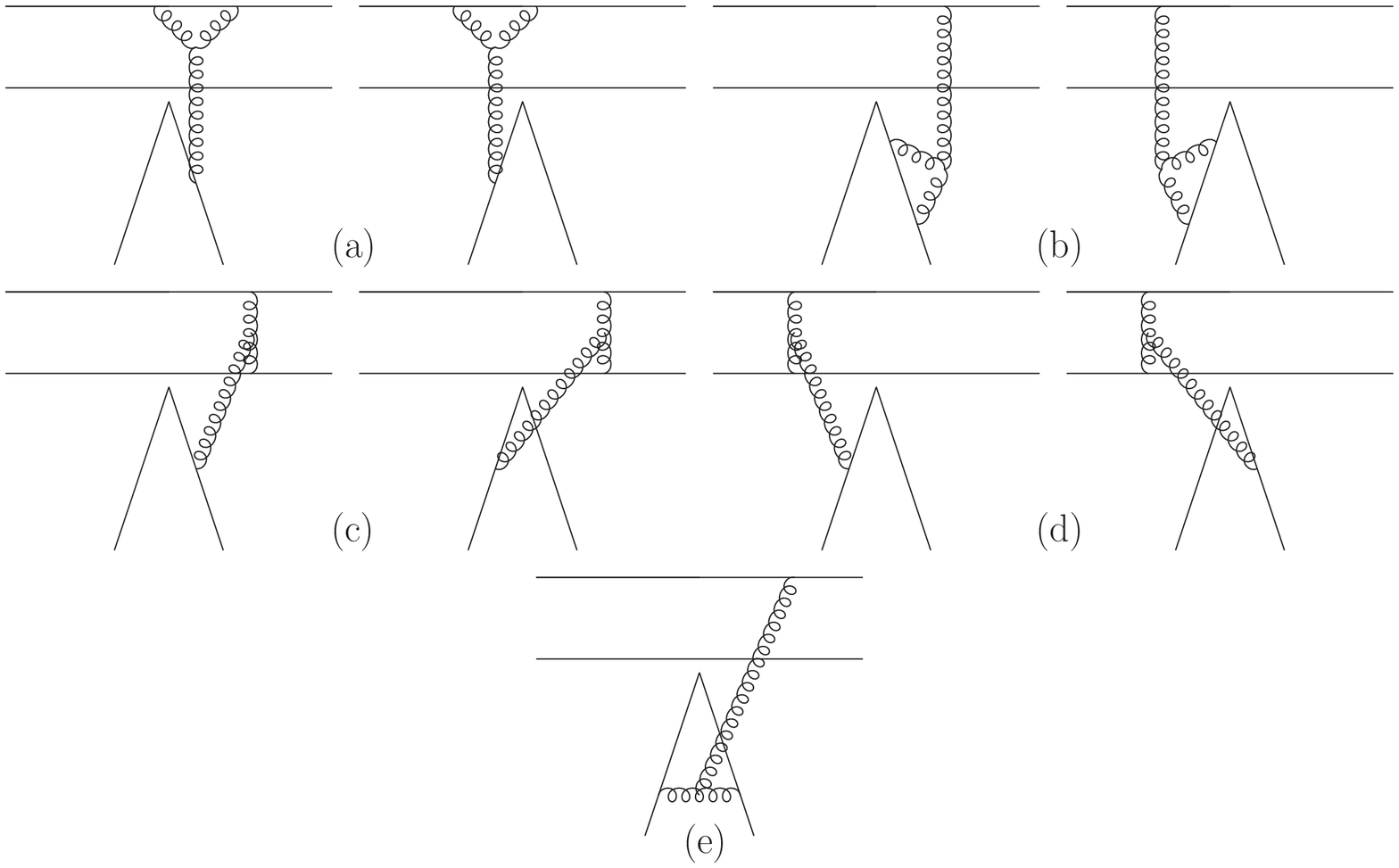}}
\end{center}
\caption{Nonabelian diagrams}
\label{b}
\end{figure}

The diagrams that contribute to $T^\text{II}_1$ at NLO are listed in
fig.~\ref{ag} - \ref{b}. Fig.~\ref{ag} shows the gluon self
energy. This is a subclass of diagrams which factorizes separately. After
adding the counter term for the gluon propagator the 
contribution to $T_1^{\text{II}(2)}$ is:
\begin{eqnarray}
T^{\text{II}(2)}_{\text{gs}} &=&
\alpha_s^2\frac{C_F}{4N_c^2}\frac{1}{\xi\bar{x}\bar{y}}\bigg[
C_N\left(-\frac{20}{3}\lmu+
\frac{16}{3}\ln\xi+\frac{16}{3}\ln\bar{x}-\frac{80}{9}\right)
\nonumber\\
&&+C_FC_G
\left(\frac{5}{3}\lmu-\frac{5}{3}\ln\xi-\frac{5}{3}\ln\bar{x}+\frac{31}{9}
\right)
\bigg].
\label{gs}
\end{eqnarray}
For (\ref{gs}) we have set the number of active quark flavours to
$n_f=5$ and set the mass of the $u$-, $d$-, $s$- and $c$-quark to zero.

For the rest of this section I will consider only those diagrams, for
which  the calculation of Feynman integrals is not straight forward. I
will only show how those Feynman integrals can be calculated in
leading power. For higher powers I refer to the methods shown in
section \ref{diffeq}. It is important to note, that though for the evaluation
of the Feynman integrals arguments depending on power counting have
been used, all of the integrals occuring in this section have been tested
by methods that do not depend on power counting.

The first class of
diagrams with non trivial Feynman integrals are the diagrams in
fig.~\ref{aII}. The two diagrams from fig.~\ref{aII}(a) read
\begin{equation}
\begin{split}
&\text{aII1} = -ig_s^4N_cC_F(C_F-\frac{1}{2}C_G)
\label{aII1}\\
&\quad\times\intd
\frac{\gamma_\mu(\sh{k}-\sh{l})\gamma_\tau
\tilde{\otimes}\gamma_\nu(1-\gamma_5)\otimes
\gamma^\tau(\sh{k}+\bar{x}\sh{p}+y\sh{q}-\sh{l})\gamma^\nu(1-\gamma_5)
(\bar{y}\sh{q}-\sh{k})\gamma^\mu}
{k^2(k-l)^2(k+\bar{x}p-l)^2(k-\bar{y}q)^2(k+\bar{x}p+yq-l)^2}
\end{split}
\end{equation}
\begin{equation}
\begin{split}
&\text{aII2} = -ig_s^4N_cC_F^2
\label{aII2}\\
&\quad\times\intd
\frac{\gamma_\mu(\sh{k}-\sh{l})\gamma_\tau
\tilde{\otimes}\gamma_\nu(1-\gamma_5)\otimes
\gamma^\mu(y\sh{q}-\sh{k})\gamma^\nu(1-\gamma_5)
(\sh{k}+\bar{y}\sh{q}+\bar{x}\sh{p}-\sh{l})\gamma^\tau}
{k^2(k-l)^2(k+\bar{x}p-l)^2(k-yq)^2(k+\bar{x}p+\bar{y}q-l)^2}.
\end{split}
\end{equation}
The denominators of (\ref{aII1}) and (\ref{aII2}) are identical up
to the substitution $y\to\bar{y}$. So the Feynman integrals we have to
calculate are the same. We can reduce our five-point integrals to
four-point integrals by expanding the denominator of the integrand 
into partial fractions
\begin{eqnarray}
\lefteqn{
\frac{1}{k^2(k-l)^2(k+\bar{x}p-l)^2(k-\bar{y}q)^2(k+\bar{x}p+yq-l)^2}
=
}
\nonumber\\
&&\frac{1}{y(\bar{x}-\xi)}\bigg[
\frac{1}{k^2(k-l)^2(k+\bar{x}p-l)^2(k-\bar{y}q)^2}+
\nonumber\\
&&\frac{y}{\bar{y}}\frac{1}
{k^2(k-l)^2(k+\bar{x}p-l)^2(k+\bar{x}p+yq-l)^2}-
\nonumber\\
&&\frac{1}{k^2(k-l)^2(k-\bar{y}q)^2(k+\bar{x}p+yq-l)^2}-
\nonumber\\
&&\frac{y}{\bar{y}}\frac{1}
{(k-l)^2(k+\bar{x}p-l)^2(k-\bar{y}q)^2(k+\bar{x}p+yq-l)^2}
\bigg].
\label{aII1.1}
\end{eqnarray}
Only the first two summands of the right hand side of this equation
give leading power contributions to aII1 and aII2:
\begin{eqnarray}
t_I &\equiv& \frac{1}{k^2(k-l)^2(k+\bar{x}p-l)^2(k-\bar{y}q)^2}
\label{aIItI}\\
t_{II} &\equiv& 
\frac{1}{k^2(k-l)^2(k+\bar{x}p-l)^2(k+\bar{x}p+yq-l)^2}.
\label{aIItII}
\end{eqnarray}
The third one
\begin{equation}
t_{III}\equiv\frac{1}{k^2(k-l)^2(k-\bar{y}q)^2(k+\bar{x}p+yq-l)^2}
\label{aIItIII}
\end{equation}
gives only a leading power contribution in the hard-collinear region
\begin{eqnarray}
k\cdot p &\sim& 1\nonumber\\
k\cdot q &\sim& \lambda\nonumber\\
k_\perp^\mu &\sim& \sqrt{\lambda}
\label{hans}
\end{eqnarray}
and the soft region 
\begin{equation}
k^\mu\sim \lambda,
\end{equation}
where we introduced the counting $l^\mu\sim \lambda$ and set $m_B=1$.
In both regions the leading power of the numerators of aII1 and aII2 
vanishes because of equations of motion. The fourth summand of 
(\ref{aII1.1})
\begin{equation}
t_{IV}\equiv\frac{1}
{(k-l)^2(k+\bar{x}p-l)^2(k-\bar{y}q)^2(k+\bar{x}p+yq-l)^2}
\label{aIItIV}
\end{equation}
does not give a leading power contribution at all. 

The topologies $t_I$ and $t_{II}$ are exactly the denominators of the
Feynman integrals of the last four diagrams of fig.~\ref{aII}:
\begin{eqnarray}
\text{aII3} & = & ig_s^4N_cC_F^2\frac{1}
{\bar{x}y-\bar{x}\xi-y\theta}
\intd\gamma_\mu(\sh{k}-\sh{l})\gamma_\tau
\label{aII3}\\
&&
\tilde{\otimes}\frac{\gamma_\nu(1-\gamma_5)\otimes
\gamma^\tau(\sh{k}+y\sh{q}+\bar{x}\sh{p}-\sh{l})\gamma^\mu
(y\sh{q}+\bar{x}\sh{p}-\sh{l})\gamma^\nu(1-\gamma_5)}
{k^2(k-l)^2(k+\bar{x}p-l)^2(k+\bar{x}p+yq-l)^2}
\nonumber\\
\text{aII4} & = &
ig_s^4N_cC_F(C_F-\frac{1}{2}C_G)\frac{1}
{\bar{x}\bar{y}-\bar{x}\xi-\bar{y}\theta}
\intd\gamma_\mu(\sh{k}-\sh{l})\gamma_\tau
\label{aII4}\\
&&
\tilde{\otimes}\frac{\gamma_\nu(1-\gamma_5)\otimes
\gamma^\nu(1-\gamma_5)(\bar{x}\sh{p}+\bar{y}\sh{q}-\sh{l})\gamma^\mu
(\sh{k}+\bar{x}\sh{p}+\bar{y}\sh{q}-\sh{l})\gamma^\tau}
{k^2(k-l)^2(k+\bar{x}p-l)^2(k+\bar{x}p+\bar{y}q-l)^2}
\nonumber\\
\text{aII5} & = & ig_s^4N_cC_F(C_F-\frac{1}{2}C_G)\frac{1}
{\bar{x}y-\bar{x}\xi-y\theta}
\intd\gamma_\mu(\sh{k}-\sh{l})\gamma_\tau
\label{aII5}\\*
&&
\tilde{\otimes}
\frac{
\gamma_\nu(1-\gamma_5)\otimes
\gamma^\mu(y\sh{q}-\sh{k})\gamma^\tau(\bar{x}\sh{p}+y\sh{q}-\sh{l})
\gamma^\nu(1-\gamma_5)}
{k^2(k-l)^2(k+\bar{x}p-l)^2(k-yq)^2}
\nonumber\\
\text{aII6} &=& ig_s^4N_cC_F^2\frac{1}
{\bar{x}\bar{y}-\bar{x}\xi-\bar{y}\theta}
\times
\label{aII6}\\
&&\intd
\frac{\gamma_\mu(\sh{k}-\sh{l})\gamma_\tau
\tilde{\otimes}\gamma_\nu(1-\gamma_5)\otimes
\gamma^\nu(1-\gamma_5)(\bar{x}\sh{p}+\bar{y}\sh{q}-\sh{l})\gamma^\tau
(\bar{y}\sh{q}-\sh{k})\gamma^\mu}
{k^2(k-l)^2(k+\bar{x}p-l)^2(k-\bar{y}q)^2},
\nonumber
\end{eqnarray}
where the Feynman diagrams are given in the same order as they occur in
fig.~\ref{aII}. So we have
reduced all the Feynman integrals of fig.~\ref{aII} to the topologies
$t_I$ and $t_{II}$. Regarding $t_I$ we need the scalar integral 
\begin{equation}
D_{0I}(l)\equiv\intd t_{I}.
\label{D0I}
\end{equation}
By following the procedure of \cite{Passarino:1978jh} the tensor integrals 
$\intd k^\mu t_{I}$, $\intd k^\mu k^\nu t_{I}$
and $\intd k^\mu k^\nu k^\tau t_{I}$ can be reduced to (\ref{D0I}) and
to the two-point master integrals that are listed in appendix
\ref{mi}. The leading power of (\ref{D0I}) gets only contributions
from the region where
$k$ is soft i.e.\ all components of $k$ are of $\mathcal{O}(\lqcd)$. 
In this region we can expand the integrand of (\ref{D0I}) to
\begin{equation}
\frac{1}{(k^2+i\eta)((k-l)^2+i\eta)(2k\cdot p-\xi+i\eta)(-2k\cdot
  q+i\eta)\bar{x}\bar{y}}.
\label{D0I.1}
\end{equation}
Now we can obtain the leading power of (\ref{D0I}) by integrating 
(\ref{D0I.1}) over all momenta $k$ because there is no other region,
which gives a leading power contribution, besides where $k$ is soft.
The integration of (\ref{D0I.1}) can be easily performed by using the 
following Feynman parametrisation \cite{Falk:1990pz}:
\begin{equation}
\frac{1}{A_0\ldots A_n} = 
\int_0^\infty d^n\lambda\,\frac{n!}
{\left(A_0 +\sum_{i=1}^n \lambda_i A_i \right)^{n+1}}.
\label{diag2}
\end{equation}
Finally we obtain for the leading power of (\ref{D0I})
\begin{equation}
\begin{split}
D_{0I}\doteq\frac{i}{(4\pi)^2}
\frac{\Gamma(1+\epsilon)(4\pi\mu^2)^{\epsilon}}{\bar{x}\bar{y}\xi\theta}
\Bigg(
&
\frac{2}{\epsilon^2}-\frac{2\ln\xi+2\ln\theta+2i\pi}{\epsilon}
\\
&-\pi^2+\ln^2\xi+\ln^2\theta+2\ln\xi\ln\theta+2\pi i(\ln\xi+\ln\theta)
\Bigg)  
\end{split}
\label{diag2.1}
\end{equation}

Regarding the topology $t_{II}$ we need the tensor integrals
$\intd k^\mu t_{II}$, $\intd k^\mu k^\nu t_{II}$ and $\intd k^\mu
k^\nu k^\tau t_{II}$. This topology can be reduced to two-point master
integrals and to the four-point master integral 
\begin{equation}
D_{0II}(l)\equiv\intd t_{II}.
\label{D0II}
\end{equation}
In order to calculate this integral we decompose $t_{II}$ into 
\begin{eqnarray}
t_{II} &=& \frac{1}{y}\bigg[
\frac{1}{k^2(k-l)^2(k+\bar{x}p-l)^2(2k\cdot q +\bar{x}-\theta)}
\nonumber\\
&&-\frac{1}{k^2(k-l)^2(k+\bar{x}p+yq-l)^2(2k\cdot q +\bar{x}-\theta)}
\bigg]
\label{diag1}
\end{eqnarray}
where only the integral over the first summand of (\ref{diag1}) gives
a leading power contribution. This integration is straight forward if
one uses the Feynman parametrisation (\ref{diag2}).

Finally we get the leading power of (\ref{D0II}):
\begin{equation}
\begin{split}
&D_{0II} \doteq\\
&\quad -\ifp
\frac{\Gamma(1+\epsilon)(4\pi\mu^2)^{\epsilon}}{\bar{x}^2y\xi}
\left[
\frac{2}{\epsilon^2}-\frac{2}{\epsilon}(\ln\bar{x}+\ln\xi)
-\frac{\pi^2}{3}+\ln^2\bar{x}+\ln^2\xi
+2\ln\bar{x}\ln\xi
\right].
\label{diag3}
\end{split}
\end{equation}
Actually it turns out that the sum of the diagrams in 
fig.~\ref{aII} vanishes in leading power.

The diagrams of fig.~\ref{aIII} are straight forward to calculate. It
is easy to see that in leading power $l$ does not occur within a loop
integral. So there are only two linearly independent momenta in the
Feynman integrals, which, using similar relations like
(\ref{aII1.1}) and IBP identities, can be reduced to the master integrals
that are listed in appendix \ref{mi}.

The diagrams of fig.~\ref{aIV}(a),(b) are easy to calculate,
because they contain only three-point integrals. 
The next two (fig.~\ref{aIV}(c)) are given by
\begin{eqnarray}
\lefteqn{\text{aIVm1} = ig_s^4N_cC_F(C_F-\frac{1}{2}C_G) 
\intd\gamma^\mu(\sh{k}+\bar{x}\sh{p})\gamma^\tau\tilde{\otimes}}
\label{aIV4}\\
&&\frac{\gamma^\nu(1-\gamma_5)
(\sh{k}+\sh{p}+\sh{q}-\sh{l}+m_b)\gamma_\tau\otimes
\gamma_\nu(1-\gamma_5)
(\sh{k}-\sh{l}+\bar{x}\sh{p}+\bar{y}\sh{q})\gamma_\mu
}
{k^2(k+\bar{x}p)^2(k+\bar{x}p-l)^2(k+\bar{x}p+\bar{y}q-l)^2(k^2+2k\cdot(p+q-l))}
\nonumber\\
\lefteqn{\text{aIVm2} = -ig_s^4N_cC_F(C_F-\frac{1}{2}C_G) 
\intd\gamma^\mu(\sh{k}+\bar{x}\sh{p})\gamma^\tau\tilde{\otimes}}
\label{aIV3}\\
&&\frac{\gamma^\nu(1-\gamma_5)
(\sh{k}+\sh{p}+\sh{q}-\sh{l}+m_b)\gamma_\tau\otimes
\gamma_\mu(\sh{k}-\sh{l}+\bar{x}\sh{p}+y\sh{q})\gamma_\nu(1-\gamma_5)
}
{k^2(k+\bar{x}p)^2(k+\bar{x}p-l)^2(k+\bar{x}p+yq-l)^2(k^2+2k\cdot(p+q-l))}.
\nonumber
\end{eqnarray}
By expanding the denominator of (\ref{aIV4}) or (\ref{aIV3})
into partial fractions we get:
\begin{eqnarray}
\lefteqn{
\frac{1}
{k^2(k+\bar{x}p)^2(k+\bar{x}p-l)^2(k+\bar{x}p+yq-l)^2(k^2+2k\cdot(p+q-l))}
=}
\nonumber\\
&&\frac{1}{\bar{x}-\bar{x}\xi-\theta}\bigg[
-\frac{1}
{k^2(k+\bar{x}p)^2(k+\bar{x}p-l)^2(k+\bar{x}p+yq-l)^2}+
\nonumber\\
&&\frac{1}{y}
\frac{1}
{k^2(k+\bar{x}p)^2(k+\bar{x}p-l)^2(k^2+2k\cdot(p+q-l))}-
\nonumber\\
&&\frac{\bar{y}}{y}
\frac{1}
{k^2(k+\bar{x}p)^2(k+\bar{x}p+yq-l)^2(k^2+2k\cdot(p+q-l))}+
\nonumber\\
&&\frac{x}{\bar{x}}
\frac{1}
{k^2(k+\bar{x}p-l)^2(k+\bar{x}p+yq-l)^2(k^2+2k\cdot(p+q-l))}-
\nonumber\\
&&\frac{x}{\bar{x}}
\frac{1}
{(k+\bar{x}p)^2(k+\bar{x}p-l)^2(k+\bar{x}p+yq-l)^2(k^2+2k\cdot(p+q-l))}
\bigg]
\label{aIV1.1}
\end{eqnarray}
where we get leading power contributions only from:
\begin{eqnarray}
t_I &=& \frac{1}{k^2(k+\bar{x}p)^2(k+\bar{x}p-l)^2(k+\bar{x}p+yq-l)^2}
\label{aIVtI}\\
t_{II} &=& \frac{1}{k^2(k+\bar{x}p)^2(k+\bar{x}p-l)^2(k^2+2k\cdot(p+q-l))}
\label{aIVtII}\\
t_{III} &=& \frac{1}
{(k+\bar{x}p)^2(k+\bar{x}p-l)^2(k+\bar{x}p+yq-l)^2(k^2+2k\cdot(p+q-l))}.
\label{aIVtIII}
\end{eqnarray}
The leading power of (\ref{aIVtI}) can be taken from
(\ref{diag2.1}). We obtain the leading power of (\ref{aIVtII}) by
making the replacement $l\to\xi q$. Alternatively we can use
(\ref{4p7}) in appendix \ref{4pointmass} and take the leading power
afterwards. For (\ref{aIVtIII}) we obtain the leading power by making
the replacement $l\to\theta p$. In contrast to (\ref{aIVtII}) this
replacement is not so obvious and to calculate this integral exactly
is very involved. However we can start with the value, we obtained by
this prescription, and show afterwards by solving a partial
differential equation that it is correct. We define
\begin{equation}
I(x,y,\lambda\xi,\lambda\theta)\equiv\intd t_{III}(x,y,p,q,\lambda l)
\label{aIVtIII.1}
\end{equation}
and derive two differential equations by deriving $I$ with respect to
$x$ and to $y$. As a boundary condition for our differential equations
we calculate $I$ at the point $x=0$ and $y=1$. This can be done by 
decomposing $t_{III}$ into partial fractions and using IBP
identities. We use the fact that
the limits $x\to0$ and $y\to1$ do not lead to extra divergences in
$\epsilon$ and in $\lambda$. So we can solve our differential
equations order by order in $\epsilon$ and $\lambda$. Defining
\begin{equation}
I\equiv\sum_{j,k} I_j^{(k)} \epsilon^j \lambda^k
\label{aIVtIII.2}
\end{equation}
and using the fact that we only need the leading power in $\lambda$ 
we obtain differential equations of the form 
\begin{eqnarray}
\frac{\partial}{\partial x} I_j^{(-1)} &=& 
{h_x}_0 I_j^{(-1)} + {h_x}_1 I_{j-1}^{(-1)} + {g_x}_j^{(-1)}
\nonumber\\
\frac{\partial}{\partial y} I_j^{(-1)} &=& 
{h_y}_0 I_j^{(-1)} +  {g_y}_j^{(-1)}.
\label{aIVtIII.3}
\end{eqnarray}
The coefficients ${h_x}_0$, ${h_x}_1$, ${g_x}_j^{(-1)}$,  ${h_y}_0$ and 
${g_y}_j^{(-1)}$ are straight forward to calculate using IBP
identities and the master integrals are given in appendix \ref{mi}.
It turns out that the leading power integral we obtained by the
prescription $l\to\theta p$ fulfils our boundary condition for
$x=0$ and $y=1$ as well as the set of differential equations
(\ref{aIVtIII.3}).

The sum of the following two diagrams (fig.~\ref{aIV}(d)) is
\begin{eqnarray}
\lefteqn{
\raisebox{-1cm}{\resizebox{2cm}{!}{\includegraphics{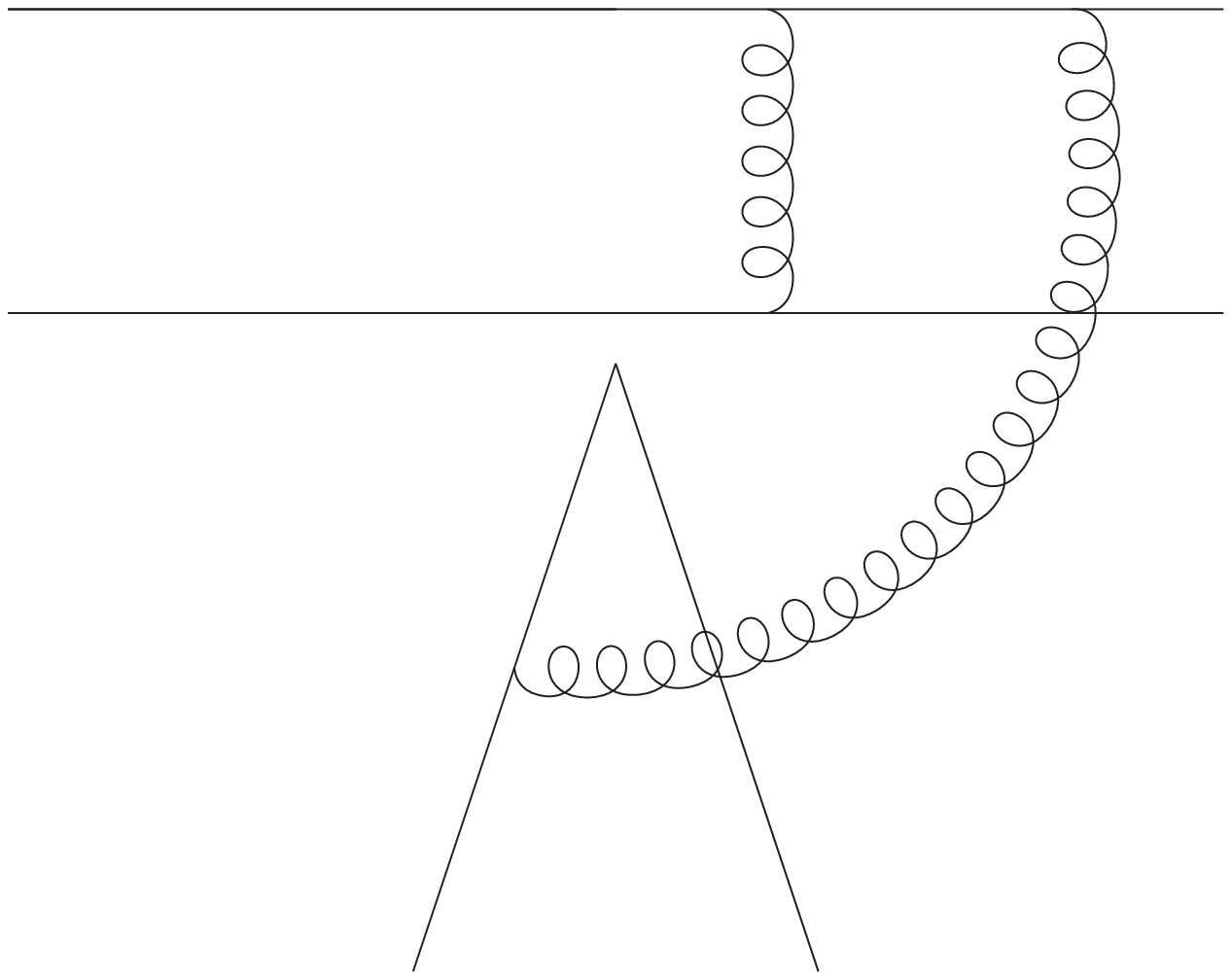}}}
+
\raisebox{-1cm}{\resizebox{2cm}{!}{\includegraphics{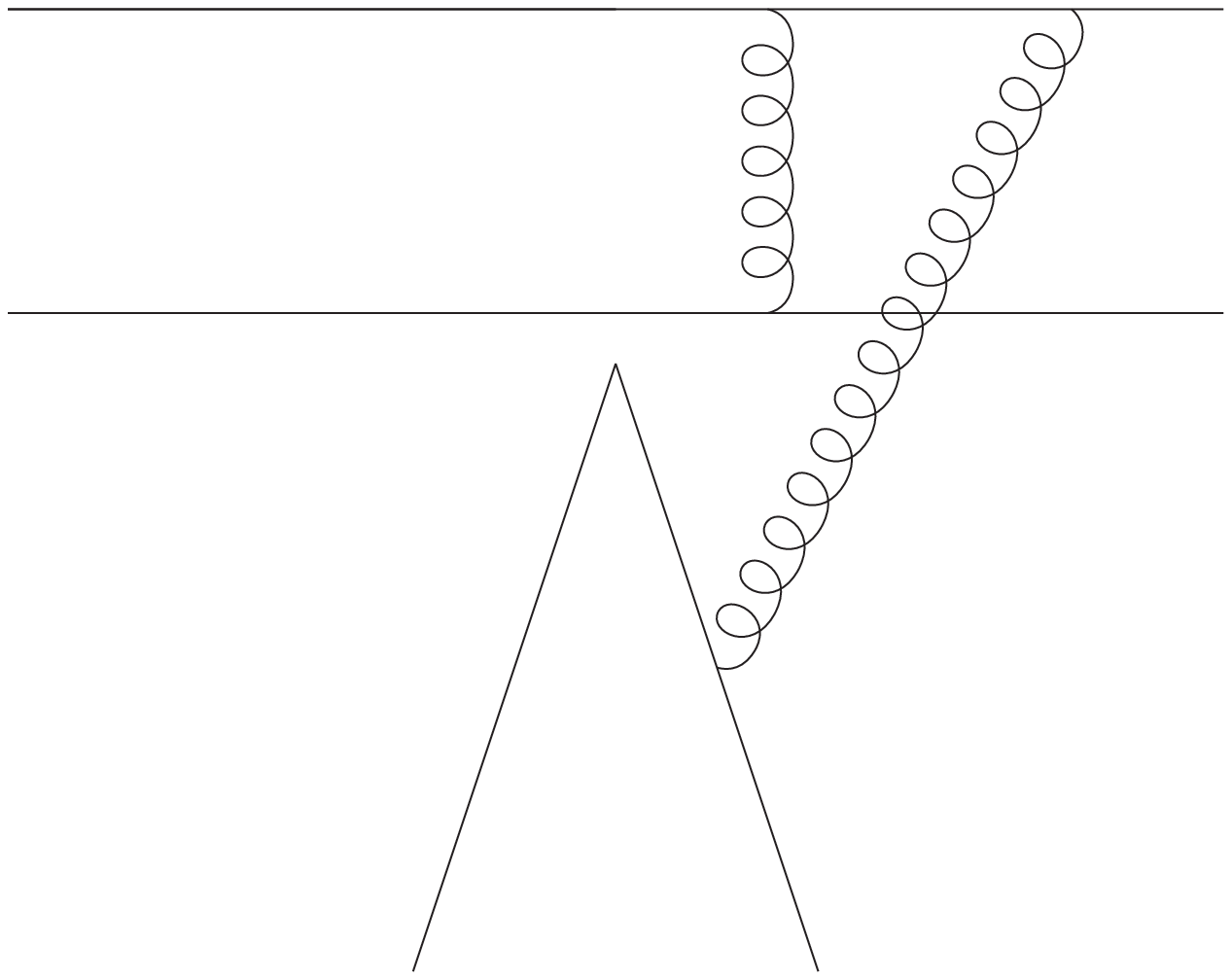}}}
=}
\nonumber\\
&&-ig_s^4N_cC_F(C_F-\frac{1}{2}C_G)
\intd \frac{\gamma^\tau\sh{k}\gamma^\mu\tilde{\otimes}
\gamma_\tau(\sh{k}+x\sh{p}-\sh{l})\gamma^\nu(1-\gamma_5)}
{k^2(k-l)^2(k-\bar{x}p)^2(k+xp-l)^2}\otimes
\nonumber\\
&&\left(
\gamma_\mu\frac{y\sh{q}+\bar{x}\sh{p}-\sh{k}}
{(k-\bar{x}p-yq)^2}\gamma_\nu-
\gamma_\nu\frac{\bar{y}\sh{q}+\bar{x}\sh{p}-\sh{k}}
{(k-\bar{x}p-\bar{y}q)^2}\gamma_\mu
\right)(1-\gamma_5).
\label{aIV56}
\end{eqnarray}
As in the previous case the denominator can be decomposed into partial
fractions and the remaining four-point master integrals can be
calculated in leading power by the replacement $l\to\xi
q$. Alternatively we can use the explicit formulas for four-point
integrals given in \cite{Duplancic:2000sk} and derive the leading power (and
higher powers if necessary) afterwards. This is what I have done in
order to get an independent test of my master integrals.

The last two diagrams of this class (fig.~\ref{aIV}(e)) are given by 
\begin{eqnarray}
\lefteqn{
\raisebox{-1cm}{\resizebox{2cm}{!}{\includegraphics{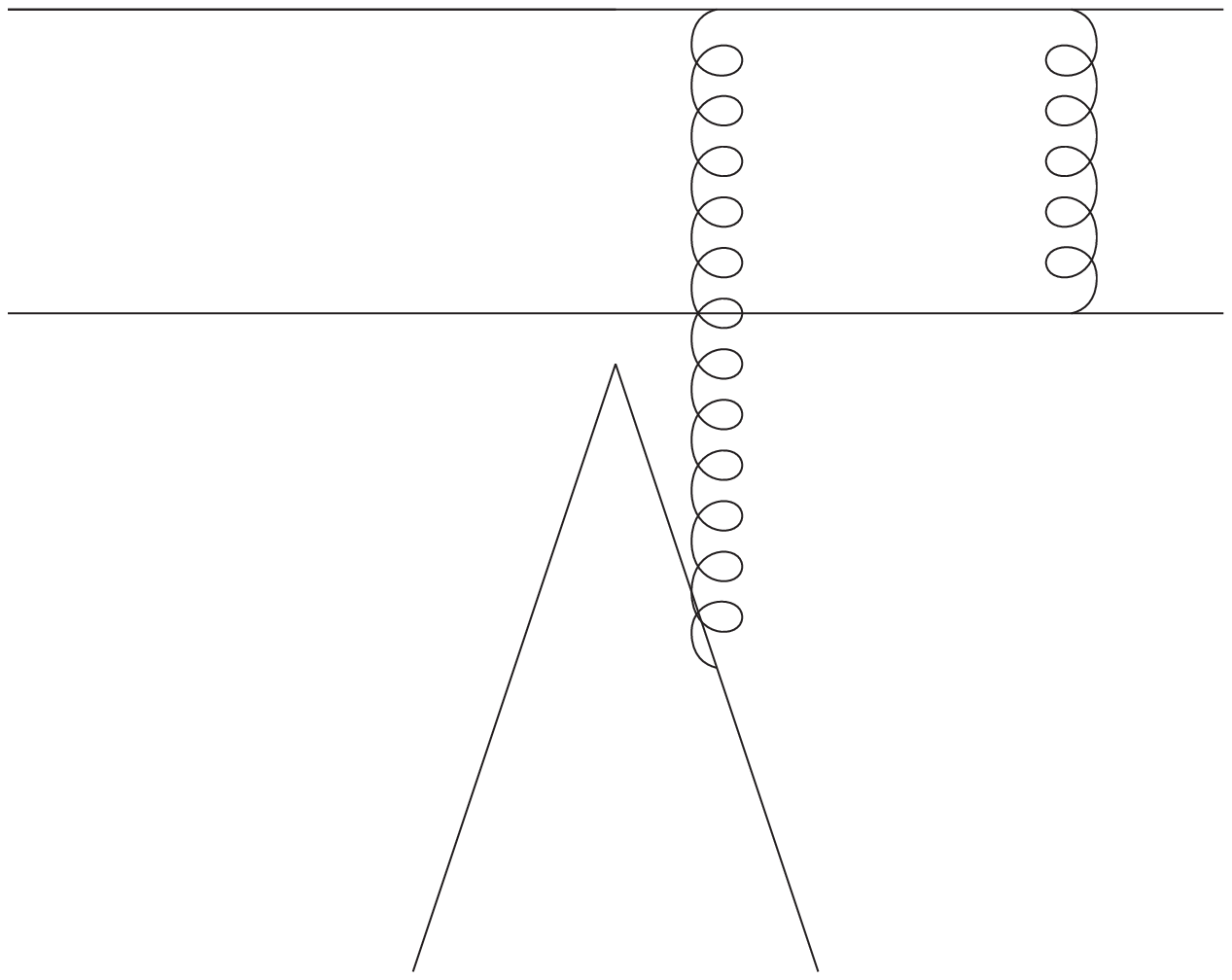}}}
+
\raisebox{-1cm}{\resizebox{2cm}{!}{\includegraphics{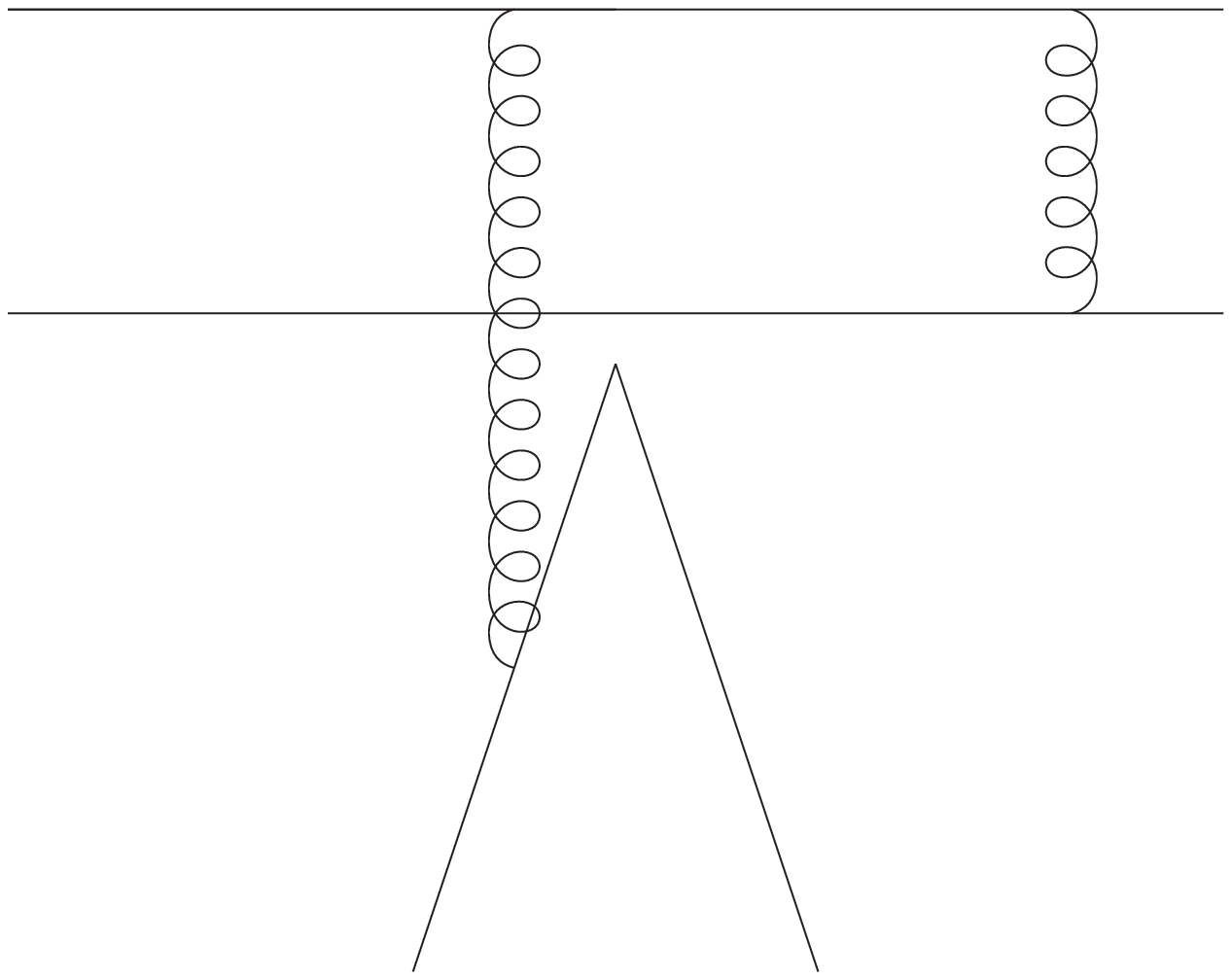}}}
=}
\nonumber\\
&&ig_s^4N_cC_F^2
\intd \frac{\gamma^\mu\sh{k}\gamma^\tau\tilde{\otimes}
\gamma_\tau\sh{k}\gamma^\nu(1-\gamma_5)}
{k^2(k+\bar{x}p)^2(k+p)^2(k+l)^2}\otimes
\nonumber\\
&&\left(
\gamma_\mu\frac{\sh{k}+\sh{l}+y\sh{q}}{(k+l+yq)^2}\gamma_\nu-
\gamma_\nu\frac{\sh{k}+\sh{l}+\bar{y}\sh{q}}{(k+l+\bar{y}q)^2}\gamma_\mu
\right)(1-\gamma_5).
\label{aIV910}
\end{eqnarray}
As in the case before we obtain the leading power of the four-point
master integrals by the replacement $l\to\xi q$ or by using the
formulas in \cite{Duplancic:2000sk}. The denominator of (\ref{aIV910}) however
cannot by decomposed into partial fractions. So we additionally need
the five-point master integral given in appendix \ref{5point}.

Regarding the diagrams in fig.~\ref{aV} there do not occur any
subtleties we have not yet considered in the paragraph above. So we directly
switch to the non-abelian diagrams in fig.~\ref{b}. The first four
(fig.~\ref{b}(a),(b)) do
not lead to any problems because they contain only three-point
integrals. The diagrams in fig.~\ref{b}(c) are given by
\begin{eqnarray}
\lefteqn{
\raisebox{-1cm}{\resizebox{2cm}{!}{\includegraphics{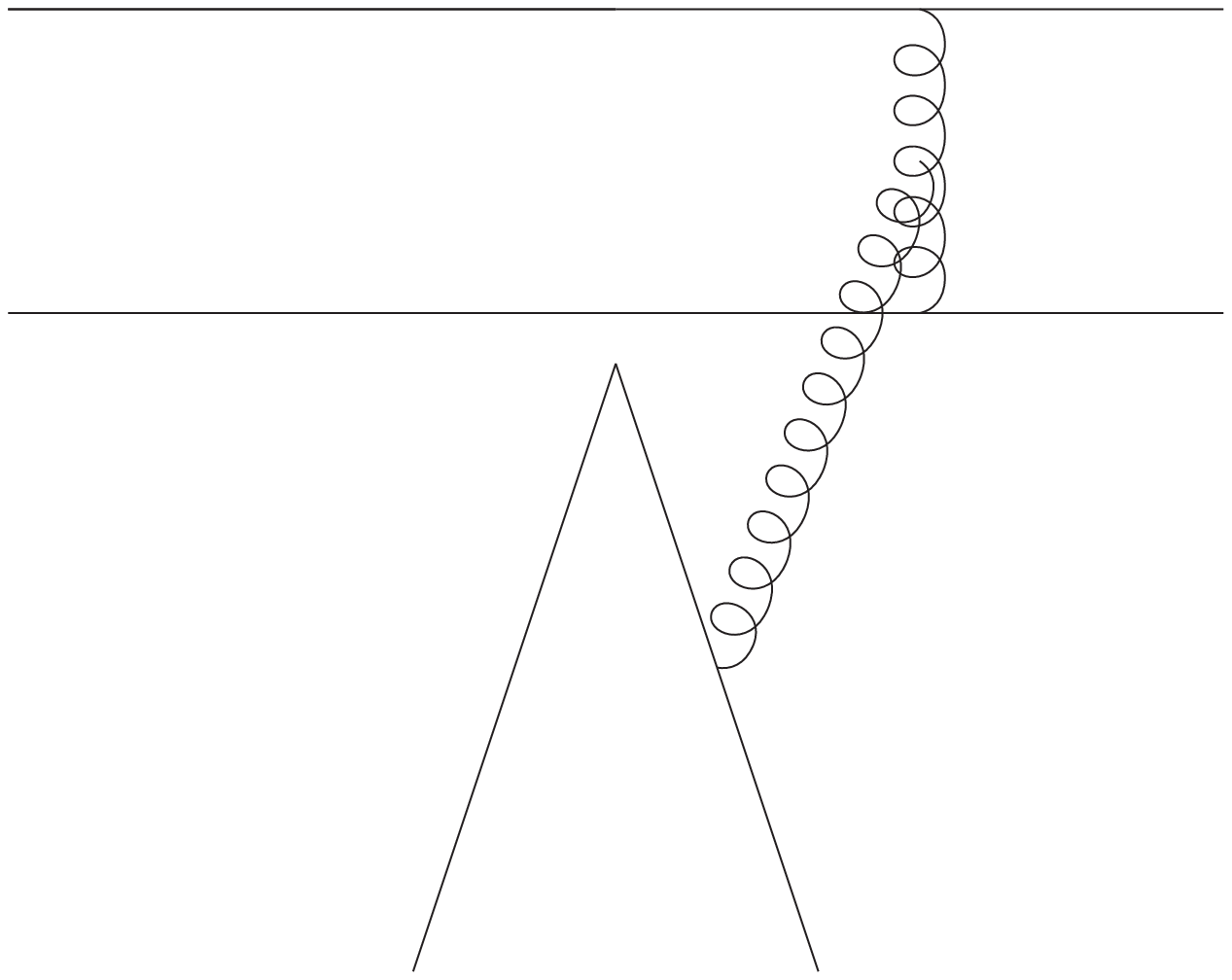}}}
+
\raisebox{-1cm}{\resizebox{2cm}{!}{\includegraphics{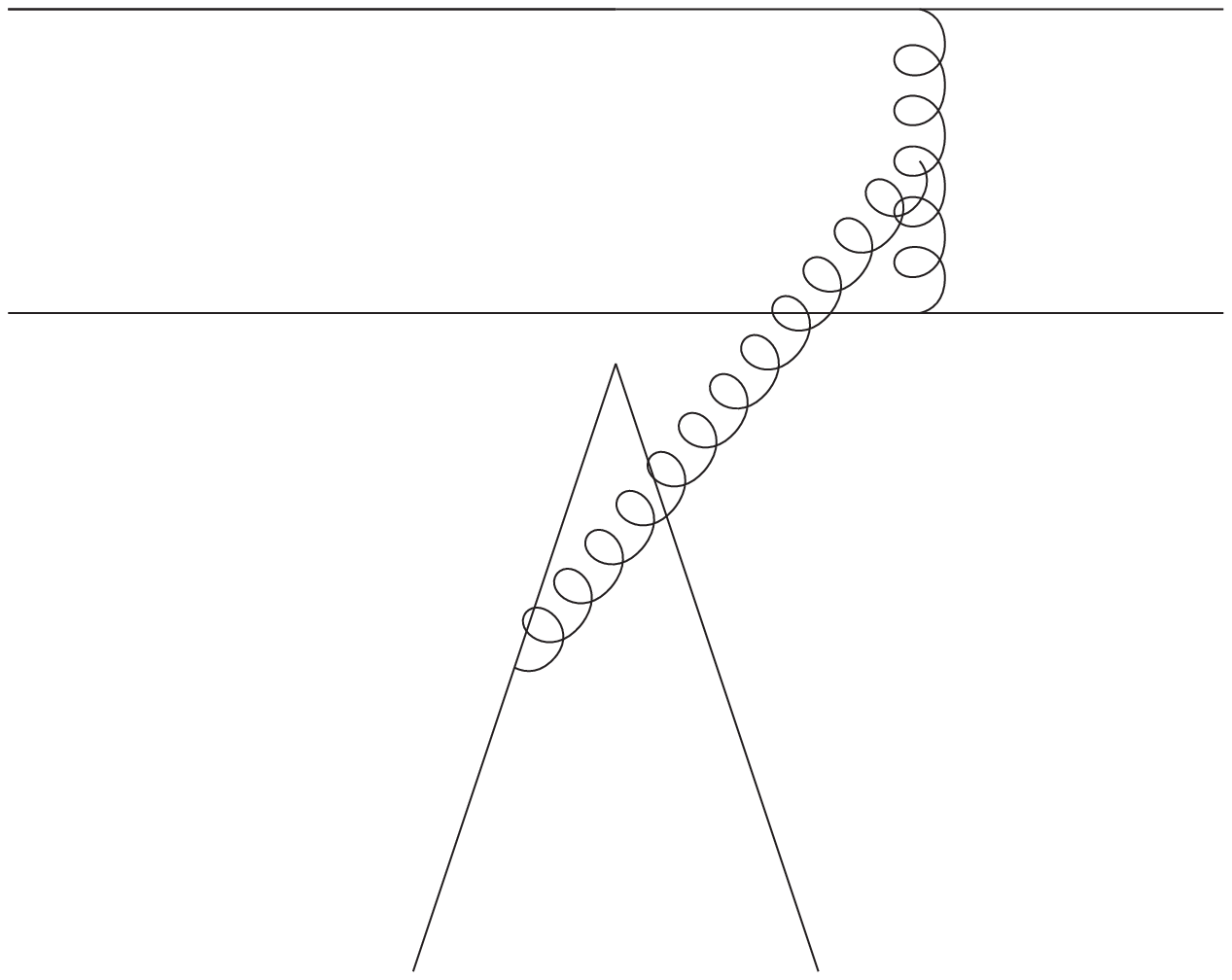}}}
=}
\label{b45}\\
&&-ig_s^4N_cC_FC_G\frac{1}{\bar{x}\xi}
\intd \left(g^{\mu\lambda}(l-k)^\tau+2g^{\lambda\tau}k^\mu
-g^{\tau\mu}(k+2l-2\bar{x}p)^\lambda\right)\times
+\nonumber\\
&&\frac{\gamma_\mu\tilde\otimes
\gamma_\tau(\sh{p}-\sh{l}-\sh{k})\gamma^\nu(1-\gamma_5)}
{k^2(k+l-p)^2(k+l-\bar{x}p)^2}\otimes
\left[
\frac{\gamma_\lambda(\sh{k}+y\sh{q})\gamma_\nu}
{(k+yq)^2}-
\frac{\gamma_\nu(\sh{k}+\bar{y}\sh{q})\gamma_\lambda}
{(k+\bar{y}q)^2}
\right]
(1-\gamma_5).
\nonumber
\end{eqnarray}
The four-point integral we have to solve is nearly the same as
(\ref{intex1}) of example \ref{example1}. We can reduce this integral
to a solution of a differential equation and get in this way 
every power in $\lqcd/m_b$. 

The last diagrams we will consider are those from fig.~\ref{b}(d). In
leading power they read:
\begin{eqnarray}
\lefteqn{
\raisebox{-1cm}{\resizebox{2cm}{!}{\includegraphics{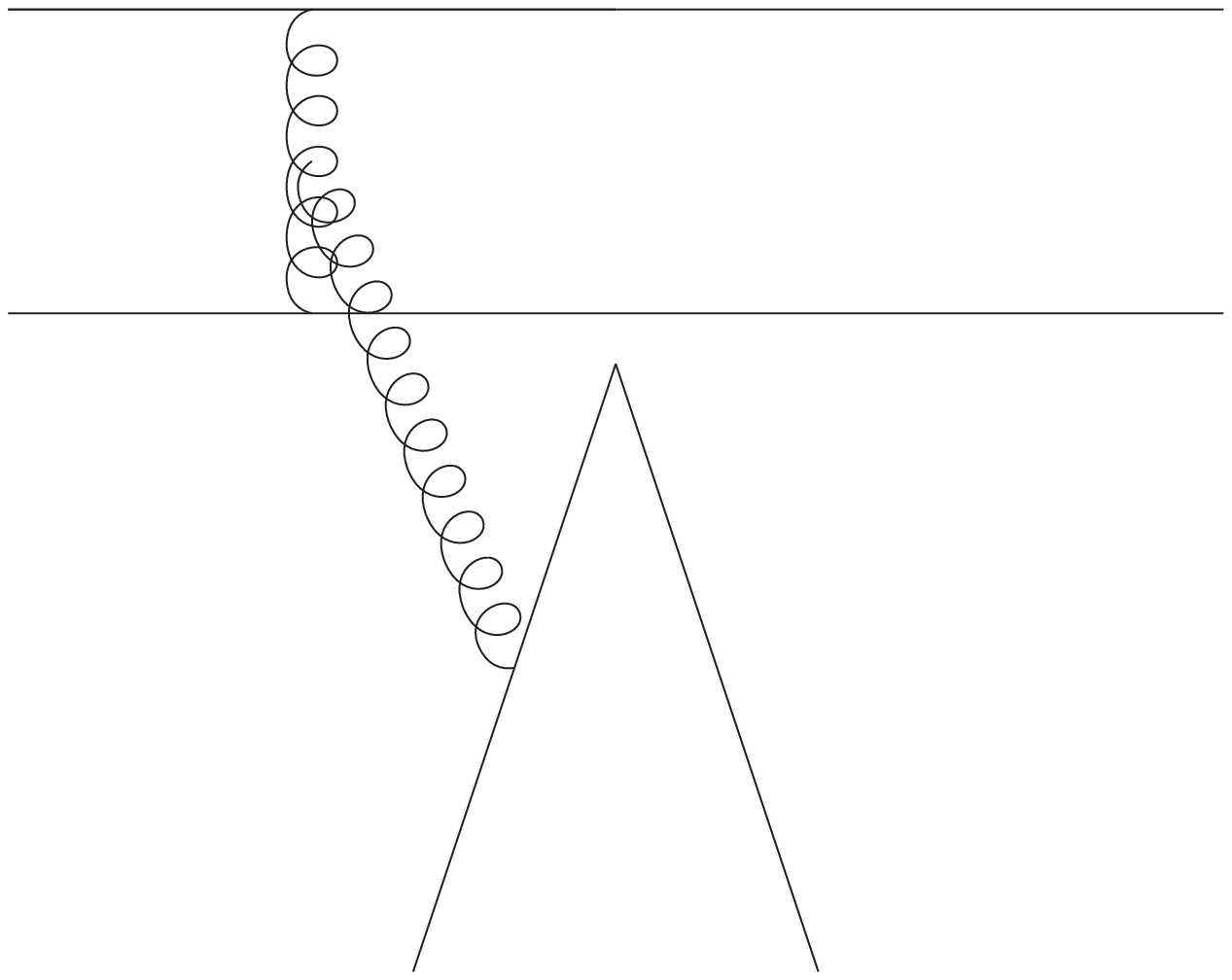}}}
+
\raisebox{-1cm}{\resizebox{2cm}{!}{\includegraphics{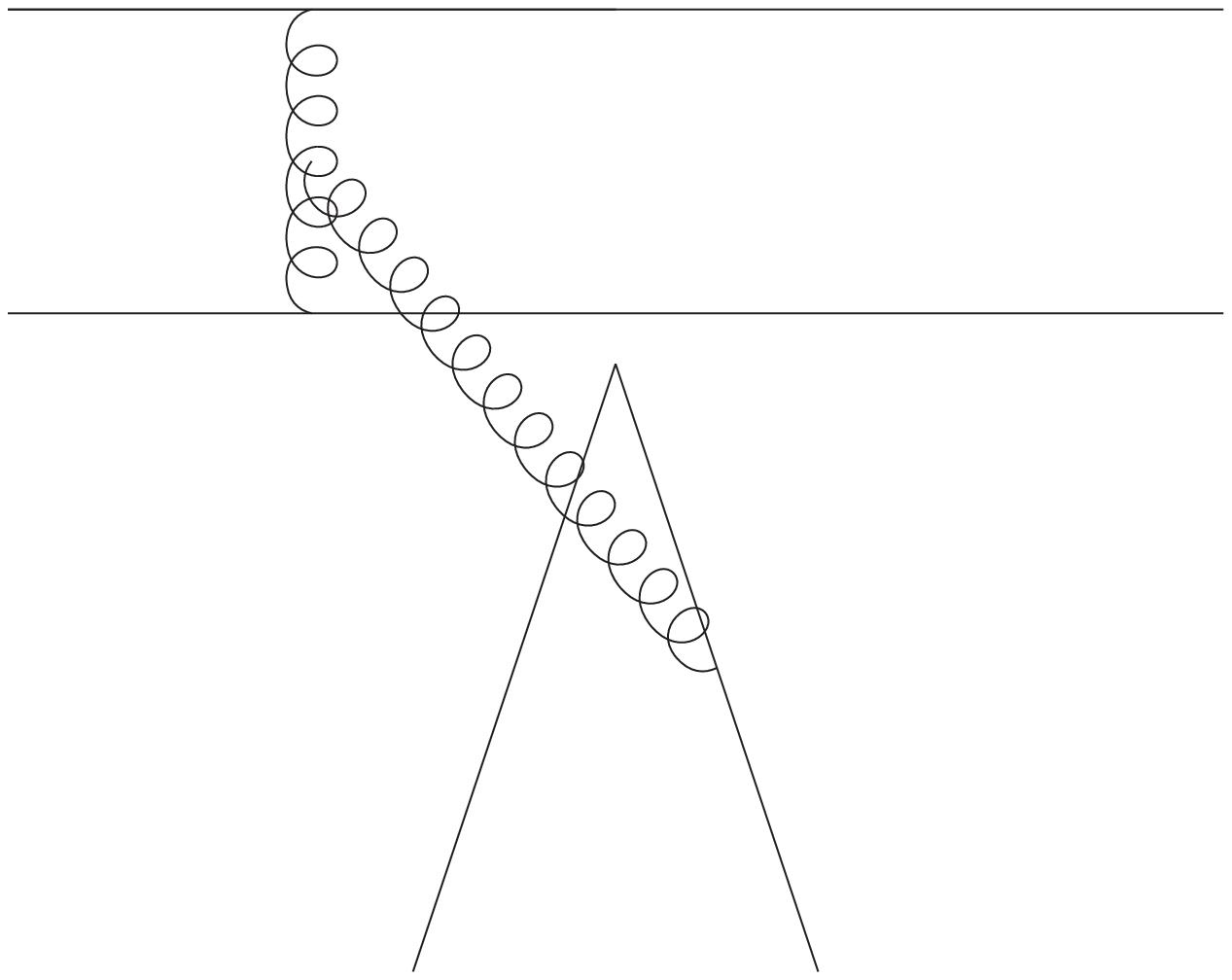}}}
=}
\label{b67}\\
&&ig_s^4N_cC_FC_G\frac{1}{\bar{x}\xi}
\intd \left(g^{\mu\lambda}(k-\bar{x}p)^\tau-2g^{\lambda\tau}k^\mu+
g^{\tau\mu}(k+2\bar{x}p)^\lambda\right)\times
\nonumber\\
&&\frac{\gamma_\mu\tilde{\otimes}\gamma^\nu(1-\gamma_5)
(\sh{k}+\sh{p}+\sh{q}+1)\gamma_\lambda}
{k^2(k+\bar{x}p-l)^2((k+p+q)^2-1)}\otimes
\nonumber\\
&&\left[
\frac{\gamma_\nu(\sh{k}+\bar{x}\sh{p}+\bar{y}\sh{q}-\sh{l})\gamma_\tau}
{(k+\bar{x}p+\bar{y}q-l)^2}
-\frac{\gamma_\tau(\sh{k}+\bar{x}\sh{p}+y\sh{q}-\sh{l})\gamma_\nu}
{(k+\bar{x}p+yq-l)^2}
\right](1-\gamma_5)
\nonumber
\end{eqnarray}
The scalar master integral
\begin{equation}
\intd\frac{1}{k^2(k+\bar{x}p-l)^2(k+\bar{x}p+\bar{y}q-l)^2((k+p+q)^2-1)}
\label{b67.1}
\end{equation}
can be calculated in leading power by setting $\theta=0$ i.e.\ we make
the replacement $l^\mu\to\xi q^\mu$. This can be seen as follows: 
Counting soft momenta as $\mathcal{O}(\lambda)$ and hard momenta as
$\mathcal{O}(1)$ (remember $m_B=1$) 
the regions of space where (\ref{b67.1}) gives a leading power
contribution are 
\begin{equation}
\begin{split}
&k^\mu \sim 1\\
&k^\mu \sim \lambda \\
&k\cdot p \sim \lambda \quad
k^\mu_\perp \sim \sqrt{\lambda} \quad
k\cdot q \sim 1.
\nonumber
\end{split}
\end{equation}
In these regions $l^\mu$ occurs only in the combination $l\cdot p$. 
So we can make the replacement $l^\mu\to\xi q^\mu$. 
Those people who do not believe these arguments are invited
to use the exact expression (\ref{4p6}) for the four-point integral 
with one massive propagator line, which is given in 
appendix~\ref{4pointmass}. After taking the leading power it can
easily be seen that we get the same result as by just making the 
replacement $l^\mu\to\xi q^\mu$. 

The diagrams which contribute to $T_2^\text{II}$ are those of fig.~\ref{aII}
and fig.~\ref{aIII}. The other diagrams drop out because their colour
trace is zero. As in the case of $T_1^\text{II}$ the diagrams of
fig.~\ref{aII} cancel each other in leading power. The remaining
diagrams are easy to calculate because their Feynman integrals are
the same as in the case of $T_1^\text{II}$.
\section{Wave function contributions}
\label{wf}
\subsection{General remarks}
\begin{figure}
\begin{center}
\resizebox{0.5\textwidth}{!}{\includegraphics{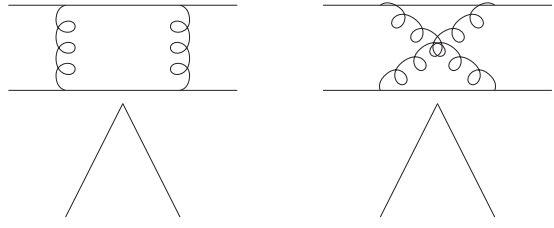}}
\end{center}
\caption{Example for diagrams which obviously belong to the form factor}
\label{ff2}
\end{figure}

It has already been demonstrated in section \ref{HspLO} how in principle we 
can extract the scattering kernel $T^\text{II}$ of (\ref{factform}) 
from the amplitude if we know the wave functions. 
$T^\text{II}$ does not depend on the hadronic
physics and on the form of the wave function $\phi_\pi$ and $\phi_B$ in
particular, so we can get $T^\text{II}$ by calculating the matrix
elements of the effective operators between free quark states carrying
the momenta shown in fig.~\ref{basic} on page \pageref{basic}. Because
we calculate $T^\text{II}$ in NLO we need unlike as in section \ref{HspLO}
the wave functions up to NLO. Let us write the second term of
(\ref{factform}) in the following formal way:
\begin{equation}
\mathcal{A}_\text{spect.}=
\phi_\pi\otimes\phi_\pi\otimes\phi_B\otimes T^\text{II}.
\label{wf1}
\end{equation}
All of the objects arising in (\ref{wf1}) have their perturbative
series in $\alpha_s$, so (\ref{wf1}) becomes
\begin{eqnarray}
\mathcal{A}_\text{spect.}^{(1)} &=&
\phi_\pi^{(0)}\otimes\phi_\pi^{(0)}\otimes\phi_B^{(0)}\otimes
T^{\text{II}(1)}
\label{wf2}\\
\mathcal{A}_\text{spect.}^{(2)} &=&
\phi_\pi^{(1)}\otimes\phi_\pi^{(0)}\otimes\phi_B^{(0)}\otimes
T^{\text{II}(1)}+
\phi_\pi^{(0)}\otimes\phi_\pi^{(1)}\otimes\phi_B^{(0)}\otimes
T^{\text{II}(1)}+
\nonumber\\
&&\phi_\pi^{(0)}\otimes\phi_\pi^{(0)}\otimes\phi_B^{(1)}\otimes
T^{\text{II}(1)}+
\phi_\pi^{(0)}\otimes\phi_\pi^{(0)}\otimes\phi_B^{(0)}\otimes
T^{\text{II}(2)}
\nonumber\\
&\vdots&
\nonumber
\end{eqnarray}
where the superscript $(i)$ denotes the order in $\alpha_s$\footnote{
Please note that the hard spectator scattering kernel starts at
$\mathcal{O}(\alpha_s)$. So we call $T^{\text{II}(1)}$ the LO and 
$T^{\text{II}(2)}$ the NLO.}. 
In order to get $T^{\text{II}(2)}$ we have to calculate
$\mathcal{A}_\text{spect.}^{(2)}$, $\phi_\pi^{(1)}$ and $\phi_B^{(1)}$
for our final states. Then $T^{\text{II}(2)}$ is given by
\begin{eqnarray}
\lefteqn{\phi_\pi^{(0)}\otimes\phi_\pi^{(0)}\otimes\phi_B^{(0)}\otimes
T^{\text{II}(2)}=}
\label{wf3}\\
&&\mathcal{A}_\text{spect.}^{(2)}-
\phi_\pi^{(1)}\otimes\phi_\pi^{(0)}\otimes\phi_B^{(0)}\otimes
T^{\text{II}(1)}-
\phi_\pi^{(0)}\otimes\phi_\pi^{(1)}\otimes\phi_B^{(0)}\otimes
T^{\text{II}(1)}-
\nonumber\\
&&\phi_\pi^{(0)}\otimes\phi_\pi^{(0)}\otimes\phi_B^{(1)}\otimes
T^{\text{II}(1)}
\nonumber
\end{eqnarray}
At this point a subtlety occurs. Let us have a closer look to the
factorization formula (\ref{factform}). By calculating the first 
order in $\alpha_s$ of the partonic form factor $F^{B\to\pi,(1)}$,
which is defined by free quark states instead of hadronic external
states, we see that it can be written in the form
\begin{equation} 
F^{B\to\pi,(1)} = \phi_\pi^{(0)}\otimes\phi_B^{(0)}\otimes
T^{(1)}_\text{formfact.}.
\label{wf4}
\end{equation}
But $T^{(1)}_\text{formfact.}$ is no part of $T^{\text{II}}$. 
So we have to modify (\ref{wf3})
insofar as we have to subtract the right hand side of (\ref{wf4}) from
the right hand side of (\ref{wf3}):
\begin{eqnarray}
\lefteqn{\phi_\pi^{(0)}\otimes\phi_\pi^{(0)}\otimes\phi_B^{(0)}\otimes
T^{\text{II}(2)}=}
\label{wf5}\\
&&\mathcal{A}_\text{spect.}^{(2)}-
\phi_\pi^{(1)}\otimes\phi_\pi^{(0)}\otimes\phi_B^{(0)}\otimes
T^{\text{II}(1)}-
\phi_\pi^{(0)}\otimes\phi_\pi^{(1)}\otimes\phi_B^{(0)}\otimes
T^{\text{II}(1)}-
\nonumber\\
&&\phi_\pi^{(0)}\otimes\phi_\pi^{(0)}\otimes\phi_B^{(1)}\otimes
T^{\text{II}(1)}-
\phi_\pi^{(0)}\otimes\phi_\pi^{(0)}\otimes\phi_B^{(0)}\otimes
T^{(1)}_\text{formfact.}\otimes T^{\text{I}(1)}
\nonumber
\end{eqnarray}
In (\ref{wf5}) we did not include the term 
$F^{B\to\pi,(2)}\otimes T^{\text{I}(0)}$, because it is obviously identical
with the diagrams where the gluons do not interact with the emitted
pion (e.g.\ those of fig.~\ref{ff2}).
Those diagrams where not considered in the last section. So we do not
have to consider them here.

\begin{figure}
\begin{center}
\resizebox{0.5\textwidth}{!}{\includegraphics{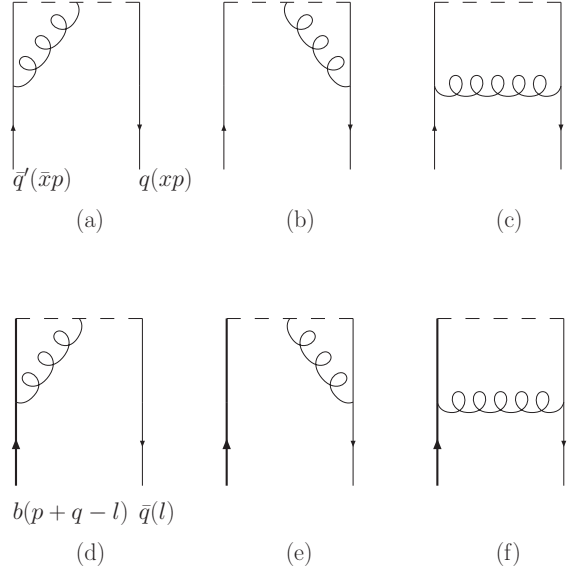}}
\end{center}
\caption{NLO contributions to the meson wave functions. The dashed
  line stands for the eikonal Wilson line which makes the wave
  functions gauge invariant.}
\label{wfpic1}
\end{figure}

The wave functions for free external quark states are given at LO by
(\ref{LO3}). At NLO there exist three possible contractions: The two
external quark states can be connected by a gluon propagator or one
of the external quarks can be connected to the eikonal Wilson line of
the wave function (fig.~\ref{wfpic1}).
The diagrams of fig.~\ref{wfpic1}(a),(b) and (c) give $\mathcal{O}(\alpha_s)$
of the ``pion wave function for free quarks'', i.e.\ we have
replaced the pion final state $\langle \pi(p)|$ in (\ref{mw1}) by the  
free quark state $\langle \bar{q}^\prime(\bar{x}p)q(xp)|$. The Fourier
transformed wave function $\phi_\pi^{(1)}(x^\prime)$ is defined
analogously to (\ref{LO3}). For the diagrams in fig.~\ref{wfpic1} (a), 
(b) and (c) respectively we get:
\begin{eqnarray}
\phi_{\pi\alpha\beta}^{\text{(a)},(1)}(x^\prime) &=&
8\pi^2 i \alpha_s C_F N_c \intd
\frac{\delta(x^\prime-x-\frac{k^+}{p^+})-\delta(x^\prime-x)}{k^2k^+}
\bar{q}_\beta(xp)
\left[\frac{1}{\sh{k}-\bar{x}\sh{p}}\gamma^+q^\prime(\bar{x}p)\right]_\alpha
\nonumber\\
\phi_{\pi\alpha\beta}^{\text{(b)},(1)}(x^\prime) &=&
8\pi^2 i \alpha_s C_F N_c \intd
\frac{\delta(x-x^\prime-\frac{k^+}{p^+})-\delta(x-x^\prime)}{k^2k^+}
\left[\bar{q}(xp)
\gamma^+\frac{1}{\sh{k}-x\sh{p}}
\right]_\beta q^\prime_\alpha(\bar{x}p)
\nonumber\\
\phi_{\pi\alpha\beta}^{\text{(c)},(1)}(x^\prime) &=&
8\pi^2 i \alpha_s C_F N_c \intd
\frac{\delta(x^\prime-x+\frac{k^+}{p^+})}{k^2}
\left[\bar{q}(xp)
\gamma^\tau\frac{1}{x\sh{p}-\sh{k}}
\right]_\beta 
\left[\frac{1}{\sh{k}+\bar{x}\sh{p}}\gamma_\tau
q^\prime(\bar{x}p)\right]_\alpha
\nonumber\\
\label{wf6}
\end{eqnarray}

\subsection{Evanescent operators}
At NLO the convolution of the wave functions with the tree level
kernel $T^{\text{II},(1)}$ gives rise to new Dirac structures, which, however,
can in four dimensions be reduced to the tree level Dirac structures.
So we obtain the tree level Dirac structures plus further evanescent
structures, which vanish for $d=4$ but give finite contributions
if they are multiplied by UV-poles. We define our renormalisation 
scheme such that we subtract the UV-poles and these finite parts of the
evanescent structures. 

The tree level kernel (\ref{LOkernel}) contains two Dirac structures
where the second one is evanescent (after the the projection on the
wave functions). We write $T^{\text{II}(1)}$ in the following form
(that must not be mixed up with the notation (\ref{dirac2})):
\begin{equation}
T^{\text{II}(1)}(x,y,l^-)\equiv\frac{1}{\bar{x}l^-}
\gamma^\mu\tilde{\otimes}\gamma^\nu(1-\gamma_5)
\otimes\left(
\frac{2\sh{p}g_{\mu\nu}}{\bar{y}}-\frac{\sh{p}\gamma_\mu\gamma_\nu}{y\bar{y}}
\right)(1-\gamma_5)
\label{wf6.1}
\end{equation} 
where the symbol $\tilde{\otimes}$ stands for the ``wrong
contraction'' of the Dirac indices
i.e.\ the Dirac indices are given by
\begin{equation}
\left[\Gamma^1\tilde{\otimes}\Gamma^2\otimes\Gamma^3\right]
_{\alpha^\prime\alpha\beta^\prime\beta\gamma^\prime\gamma}
=
\Gamma^1_{\gamma^\prime\alpha}\Gamma^2_{\alpha^\prime\gamma}
\Gamma^3_{\beta^\prime\beta}
\label{wf6.2}
\end{equation}
as in (\ref{LOkernel}). The ``right contraction'' is defined by the
symbol $\otimes$ i.e.\ writing the Dirac indices explicitly
\begin{equation}
\left[\Gamma^1\otimes\Gamma^2\otimes\Gamma^3\right]
_{\alpha^\prime\alpha\beta^\prime\beta\gamma^\prime\gamma}
= \Gamma^1_{\alpha^\prime\alpha}\Gamma^2_{\gamma^\prime\gamma}
\Gamma^3_{\beta^\prime\beta}.
\label{wf6.3}
\end{equation}
In $d=4$ the wrong and the right contraction are related by Fierz
transformations. It is convenient and commonly used to define the
renormalised wave functions in terms of the right contraction,
i.e.\ to define $\phi_\pi^\text{ren.}$ by renormalising the
operator $\bar{q}(z)\gamma^\mu\gamma_5 q^\prime(0)$ instead of 
$\bar{q}(z)_\beta q^\prime(0)_\alpha$. This is why we define our
renormalisation scheme such that only the UV-finite part of the right
contraction operators remains: Using the notation of
(\ref{wf6.1}) -- (\ref{wf6.3}) we define the following operators:
\begin{eqnarray}
\Op_0(x,y,l^-) &\equiv& -\frac{1}{2l^-\bar{x}}
\gamma^\mu(1-\gamma_5)\otimes\gamma_\mu(1-\gamma_5)\otimes
\frac{2\sh{p}}{\bar{y}}(1-\gamma_5)
\label{wf6.4}\\
\Op_1(x,y,l^-) &\equiv& \frac{1}{l^-\bar{x}}
\gamma^\mu\tilde{\otimes}\gamma_\mu(1-\gamma_5)\otimes
\frac{2\sh{p}}{\bar{y}}(1-\gamma_5)
\label{wf6.5}\\
\Op_2(x,y,l^-) &\equiv& \frac{1}{l^-\bar{x}}
\gamma^\mu\tilde{\otimes}\gamma^\nu(1-\gamma_5)\otimes
\frac{-\sh{p}\gamma_\mu\gamma_\nu}{y\bar{y}}(1-\gamma_5).
\label{wf6.6}
\end{eqnarray}
The matrix elements of these operators are defined analogously to 
(\ref{LO3.1}):
\begin{equation}
\langle\Op_i\rangle\equiv
\int dx^\prime dy^\prime dl^{\prime-}\,\phi_{\pi\alpha\alpha^\prime}(x^\prime)
\phi_{\pi\beta\beta^\prime}(y^\prime)
\phi_{B\gamma\gamma^\prime}(l^{\prime-})
\Op_{i\,\alpha^\prime\alpha\beta^\prime\beta\gamma^\prime\gamma}
(x^\prime,y^\prime,l^{\prime-}).
\label{wf6.7}
\end{equation}
Note that $\langle\Op_1+\Op_2\rangle$ is just the convolution of the
tree level kernel (\ref{wf6.1}) with the wave functions. Furthermore
by using Fierz identities it is easy to prove that we have in four dimensions 
\begin{eqnarray}
\langle\Op_0\rangle &=& \langle\Op_1\rangle
\label{wf6.8}\\
\langle\Op_2\rangle &=& 0.
\label{wf6.9}
\end{eqnarray}
So we define the following evanescent operators:
\begin{eqnarray}
E_1&\equiv&\Op_2
\label{wf6.10}\\
E_2&\equiv&\Op_1-\Op_0
\label{wf6.11}\\
E_3&\equiv&\frac{1}{\bar{x}\bar{y}l^-}
\left(
\gamma^\mu\gamma^\nu\gamma^\rho\tilde{\otimes}
\gamma_\rho\gamma_\nu\gamma_\mu(1-\gamma_5)
+\frac{(2-d)^2}{2}
\gamma^\mu(1-\gamma_5)\otimes\gamma_\mu(1-\gamma_5)
\right)
\nonumber\\
&&\otimes
2\sh{p}(1-\gamma_5)
\label{wf6.12}\\
E_4&\equiv&\frac{1}{\bar{x}y\bar{y}l^-}
\gamma^\mu\gamma^\lambda\gamma^\tau\tilde{\otimes}
\gamma_\tau\gamma_\lambda\gamma^\nu(1-\gamma_5)\otimes
\sh{p}\gamma_\mu\gamma_\nu(1-\gamma_5)
\label{wf6.13}
\end{eqnarray}
where we have defined $E_3$ and $E_4$ for later convenience.
Using those operator definitions we define our renormalisation scheme
such that we subtract the UV-pole of $\langle\Op_0\rangle$ and 
the finite parts of
$\langle E_i\rangle$ i.e.\ terms of the form
$\frac{1}{\epsilon_\text{UV}}\langle E\rangle$, where $\langle
E\rangle$ is an arbitrary evanescent structure. 
It is important to note that we do not subtract IR-poles, because they  
depend not only on the operator but also on the external states the
operator is sandwiched in between. They have to vanish in (\ref{wf5}) 
such that the hard scattering kernel is finite. 
Finally we obtain the same result as if we had regularised the 
IR-divergences by small quark and gluon masses because the evanescent
structures vanish in $d=4$.

In the next step we will calculate the convolution integral of
$T^{\text{II},(1)}$ with the NLO wave functions given by (\ref{wf6}), i.e.\ we
have to calculate the renormalised matrix elements of $\Op_1+\Op_2$ at
NLO.

\subsection{Wave function of the emitted pion}
First we consider the renormalisation of the emitted pion wave function:
Because the contribution of the wave functions $\phi_\pi^{(a)}$ 
and $\phi_\pi^{(b)}$ does not change the Dirac structure of the
operators, we do not need to consider evanescent operators when we
calculate the diagrams of fig.~\ref{wfpic1}(a),(b). So for the emitted
pion wave function these diagrams give after renormalisation:
\begin{equation}
\begin{split}
\langle\Op_1^\text{ren.}+\Op_2^\text{ren.}
\rangle^{(1),\text{(a),(b)}}_\text{emitted}
=
\frac{2\alpha_s}{4\pi}C_F\bigg[&
\left(-\frac{1}{\epsilon_\text{IR}}+
2\ln\frac{\mu_\text{UV}}{\mu_\text{IR}}\right)
\frac{\ln\bar{y}+2y}{y}\langle\Op_1\rangle^{(0)}
\\
&-
\left(\frac{1}{\epsilon_\text{IR}}+
2\ln\mu_\text{IR}\right)
\left(2+\ln y+\ln\bar{y}\right)
\langle\Op_2\rangle^{(0)}
\bigg]
\label{wf7}
\end{split}
\end{equation}
where the LO matrix elements $\langle\Op_i\rangle^{(0)}$ can be
obtained from (\ref{LO2}). Note that we kept the IR-pole times
the evanescent matrix element $\langle\Op_2\rangle^{(0)}$. This is
needed for consistency because we also kept similar terms in the
QCD-calculation of $\mathcal{A}_\text{spect.}$. Furthermore it allows
us to show that all IR-divergences vanish.

The diagram in fig.~\ref{wfpic1}(c) mixes different Dirac structures. So
we have to include evanescent operators in the renormalisation. In the
case of the emitted pion wave function the operator $\Op_1$ does not
mix under renormalisation with the evanescent operator $E_1$
(\ref{wf6.10}). We obtain for the renormalised matrix element:
\begin{equation}
\langle\Op_1^\text{ren.}\rangle^{(1),\text{(c)}}_\text{emitted}
=
-\frac{2\alpha_s}{4\pi}C_F\frac{\bar{y}\ln\bar{y}}{y}
\left(-\frac{1}{\epsilon_\text{IR}}+
2\ln\frac{\mu_\text{UV}}{\mu_\text{IR}}\right)
\langle\Op_1\rangle^{(0)}.
\label{wf7.1}
\end{equation}
The matrix element of $E_1$ however has an overlap with $\Op_1$:
\begin{equation}
\begin{split}
\langle E_1\rangle^{(1),\text{(c)}}_\text{emitted}=
\frac{\alpha_s}{4\pi}C_F
\left(
\frac{1}{\epsilon_\text{UV}}-\frac{1}{\epsilon_\text{IR}}+
2\ln\frac{\mu_\text{UV}}{\mu_\text{IR}}
\right)
\bigg[&
\left(-2y\ln y-2\bar{y}\ln\bar{y}\right)
\langle E_1\rangle^{(0)}\\
&-4\epsilon\left(\ln y+\frac{\bar{y}\ln\bar{y}}{y}\right)
\langle\Op_1\rangle^{(0)}
\bigg].
\label{wf7.2}
\end{split}
\end{equation}
The renormalisation prescription tells us to subtract the UV-pole and
the UV-finite part of $\langle E_1\rangle$. So we obtain after
renormalisation:
\begin{equation}
\begin{split}
\langle E_1^\text{ren.}\rangle^{(1),\text{(c)}}_\text{emitted}=
\frac{\alpha_s}{4\pi}C_F
\bigg[&
\left(
-\frac{1}{\epsilon_\text{IR}}+
2\ln\frac{\mu_\text{UV}}{\mu_\text{IR}}
\right)
\left(-2y\ln y-2\bar{y}\ln\bar{y}\right)
\langle E_1\rangle^{(0)}\\
&+4\left(\ln y+\frac{\bar{y}\ln\bar{y}}{y}\right)
\langle\Op_1\rangle^{(0)}
\bigg].
\label{wf7.3}
\end{split}
\end{equation} 
Note that the evanescent operator $E_1$ leads to a finite term $4(\ln
y+\frac{\bar{y}\ln\bar{y}}{y})\langle\Op_1\rangle^{(0)}$, which we
would have missed if we had just dropped the evanescent operators.

\subsection{Wave function of the recoiled pion}
In the next step we consider the NLO contribution of the recoiled pion
wave function. As in the case of the emitted pion the diagrams
fig.~\ref{wfpic1}(a),(b) do not lead to a mixing between the
operators. Therefore we get:
\begin{equation}
\begin{split}
\langle\Op_1^\text{ren.}+\Op_2^\text{ren.}
\rangle^{(1),\text{(a),(b)}}_\text{recoiled}=
\frac{\alpha_s}{4\pi}C_F\frac{2\ln\bar{x}+4x}{x}
\bigg[&
\left(-\frac{1}{\epsilon_\text{IR}}+
2\ln\frac{\mu_\text{UV}}{\mu_\text{IR}}\right)
\langle\Op_1\rangle^{(0)}\\
&-\left(\frac{1}{\epsilon_\text{IR}}+
2\ln\mu_\text{IR}\right)
\langle\Op_2\rangle^{(0)}
\bigg].
\end{split}
\label{wf7.4}
\end{equation}

Other than in the case of the emitted pion the operators $\Op_1$ and
$\Op_2$ mix the spinors of the recoiled pion and the
$B$-meson. Therefore we have to work in the operator basis of $\Op_0$
and the evanescent operators and define our renormalisation scheme
such that the finite parts of the matrix elements of the evanescent
operators vanish. The diagram fig.~\ref{wfpic1}(c) contributes to the
matrix element of the renormalised operator $\Op_0^\text{ren.}$:
\begin{equation}
\langle\Op_0^\text{ren.}\rangle^{(1),\text{(c)}}_\text{recoiled}
=2\frac{\alpha_s}{4\pi}C_F
\left(\frac{1}{\epsilon_\text{IR}}-
2\ln\frac{\mu_\text{UV}}{\mu_\text{IR}}
\right)
\frac{\bar{x}\ln\bar{x}}{x}
\langle\Op_0\rangle^{(0)}.
\label{wf7.5}
\end{equation}
In the case of the evanescent operators we keep the IR-pole:
\begin{eqnarray}
\langle E_1^\text{ren.}\rangle^{(1),\text{(c)}}_\text{recoiled}
&=&
\frac{1}{2}\frac{\alpha_s}{4\pi}C_F
\left(\frac{1}{\epsilon_\text{IR}}+2\ln\mu_\text{IR}\right)
\frac{\bar{x}\ln\bar{x}}{x}
\langle E_4 \rangle^{(0)}
\label{wf7.6}\\
\langle E_2^\text{ren.}\rangle^{(1),\text{(c)}}_\text{recoiled}
&=&
\frac{1}{4}\frac{\alpha_s}{4\pi}C_F
\left(\frac{1}{\epsilon_\text{IR}}+2\ln\mu_\text{IR}\right)
\frac{\bar{x}\ln\bar{x}}{x}
\langle E_3 \rangle^{(0)}
\label{wf7.8}
\end{eqnarray}
At the end of the day we obtain a contribution from diagram
fig.~\ref{wfpic1}(c):
\begin{equation}
\begin{split}
&\langle\Op_0^\text{ren.}+E_1^\text{ren.}+E_2^\text{ren.}
\rangle^{(1),\text{(c)}}_\text{recoiled}=
\\
&\quad
\begin{split}
\frac{1}{2}\frac{\alpha_s}{4\pi}C_F\frac{\bar{x}\ln\bar{x}}{x}
\bigg[&
\left(\frac{1}{\epsilon_\text{IR}}-2\ln\frac{\mu_\text{UV}}{\mu_\text{IR}}
\right)
\langle\frac{1}{\bar{x}l^-}\gamma^\mu\gamma^\lambda\gamma^\tau
\tilde{\otimes}\gamma_\tau\gamma_\lambda\gamma^\nu\otimes
\left(\frac{2\sh{p}g_{\mu\nu}}{\bar{y}}-
\frac{\sh{p}\gamma_\mu\gamma_\nu}{y\bar{y}}
\right)
\rangle^{(0)}
\\
&
+8\langle\Op_1\rangle^{(0)}
\bigg].
\end{split}
\end{split}
\label{wf7.9}
\end{equation}
The very complicated but also very explicit form, in which the above equation
was given, is rather convenient, because on the QCD side the diagrams
in fig.~\ref{aIV}(e) on page \pageref{aIV} come with the same
Dirac structure and cancel the IR-pole of (\ref{wf7.9}).

\subsection{Wave function of the $B$-meson}
The $\alpha_s$ corrections of the wave function of the $B$-meson are
given by the second row of fig.~\ref{wfpic1}. For the diagrams (d),
(e) and (f) respectively they read:
\begin{eqnarray}
\lefteqn{\phi_{B\alpha\beta}^{\text{(d)},(1)}(l^{\prime -}) =}
\label{wf8}\\
&&8\pi^2 i \alpha_s N_c C_F \intd
\frac{\delta(l^{\prime -}-l^--k^-)-\delta(l^{\prime -}-l^-)}{k^2k^-}
\left[\bar{q}(l)\gamma^-\frac{1}{\sh{k}+\sh{l}}\right]_\beta
b_\alpha(p+q-l)
\nonumber\\
\lefteqn{\phi_{B\alpha\beta}^{\text{(e)},(1)}(l^{\prime -}) =}
\nonumber\\*
&&8\pi^2 i \alpha_s N_c C_F \intd
\frac{\delta(l^{\prime -}-l^--k^-)-\delta(l^{\prime -}-l^-)}{k^2k^-}
\times
\nonumber\\*
&&\bar{q}_\beta(l)\left[
\frac{1}{\sh{k}-\sh{p}-\sh{q}+\sh{l}+m_b}\gamma^-b(p+q-l)
\right]_\alpha
\nonumber\\
\lefteqn{\phi_{B\alpha\beta}^{\text{(f)},(1)}(l^{\prime -}) =}
\nonumber\\
&&8\pi^2 i \alpha_s N_c C_F \intd
\frac{\delta(l^{\prime -}-l^--k^-)}{k^2}
\times
\nonumber\\
&&\left[\bar{q}(l)\gamma^\mu\frac{1}{\sh{k}+\sh{l}}\right]_\beta
\left[
\frac{1}{\sh{p}+\sh{q}-\sh{l}-\sh{k}-m_b}\gamma_\mu b(p+q-l)
\right]_\alpha
\nonumber
\end{eqnarray}

In the case of the $B$-meson only the diagrams in
fig.~\ref{wfpic1}(d),(e) give rise to UV-poles. Those diagrams however
do not lead to a mixing of $\Op_0$ and the evanescent operators and we
do not have to deal with evanescent operators.

First let's have a look at the convolution integral which belongs to
the diagram in fig.~\ref{wfpic1}(f)
\footnote{In order not to confuse the reader I stress that the symbols
  $\tilde{\otimes}$, $\otimes$ in this and the following equations are 
  meant in terms of
  (\ref{dirac2}) and not of (\ref{wf6.2}).}:
\begin{eqnarray}
\lefteqn{\langle\Op_1+\Op_2\rangle^{(1),\text{(f)}}_B=}
\label{wf9}\\
&&(4\pi)^2i\alpha_s^2N_cC_F^2\frac{1}{\bar{x}}
\intd\frac{1}{2(k+l)\cdot p\,k^2}
\times
\nonumber\\
&&\gamma^\tau\frac{\sh{k}+\sh{l}}{(k+l)^2}\gamma^\mu\tilde{\otimes}
\gamma^\nu(1-\gamma_5)
\frac{\sh{p}+\sh{q}-\sh{l}-\sh{k}+m_b}{k^2-2k\cdot(p+q-l)}
\gamma_\tau\otimes\left(
\frac{2\sh{p}}{\bar{y}}g_{\mu\nu}-
\frac{\sh{p}}{y\bar{y}}\gamma_\mu\gamma_\nu\right)(1-\gamma_5).
\nonumber
\end{eqnarray}
\begin{figure}
\begin{center}
\resizebox{0.5\textwidth}{!}{\includegraphics{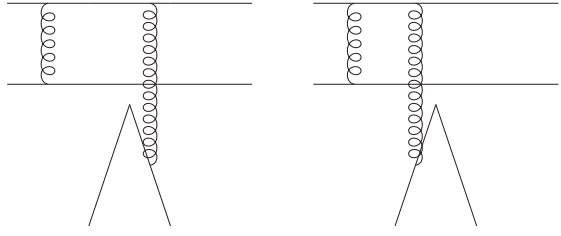}}
\end{center}
\caption{Two diagrams which correspond in leading power to the
  contribution of the $B$-meson wavefunction (\ref{wf9}).}
\label{wfpic2}
\end{figure}
In leading power (\ref{wf9}) is identical to the contribution of the
two diagrams shown in fig.~\ref{wfpic2}, which is given by:
\begin{equation}
\begin{split}
&-(4\pi)^2i\alpha_s^2N_cC_F^2 
\intd
\frac{1}{k^2(k+l-\bar{x}p)^2}
\\
&\quad\times\gamma^\tau\frac{\sh{k}+\sh{l}}{(k+l)^2}\gamma^\mu
\tilde{\otimes}
\gamma^\nu(1-\gamma_5)
\frac{\sh{p}+\sh{q}-\sh{l}-\sh{k}+m_b}{k^2-2k\cdot(p+q-l)}
\gamma_\tau
\\
&\quad\otimes
\left(
\gamma_\mu
\frac{y\sh{q}+\bar{x}\sh{p}-\sh{k}-\sh{l}}{(yq+\bar{x}p-k-l)^2}
\gamma_\nu-
\gamma_\nu
\frac{\bar{y}\sh{q}+\bar{x}\sh{p}-\sh{k}-\sh{l}}{(\bar{y}q+\bar{x}p-k-l)^2}
\gamma_\mu\right)(1-\gamma_5).
\end{split}
\label{wf10}
\end{equation}
In (\ref{wf10}) the leading power comes from the region where $k$ is
soft. In this region of space the integrand gets the form of the
integrand in (\ref{wf9}), so both contributions cancel. As we did not
include the diagrams of fig.~\ref{wfpic2} in the last section we can skip
the contribution of (\ref{wf9}) here. 

The remaining contributions are the diagrams in fig.~\ref{wfpic1}(d)
and (e). Together they read:
\begin{eqnarray}
\lefteqn{\langle\Op_1+\Op_2\rangle^{(1),\text{(d),(e)}}_B
=}
\label{wf11}\\
&&\alpha_s^2N_cC_F^2\frac{1}{\xi\bar{x}}
\gamma^\mu\tilde{\otimes}\gamma^\nu(1-\gamma_5)\otimes
\left(
\frac{2\sh{p}}{\bar{y}}g_{\mu\nu}-
\frac{\sh{p}}{y\bar{y}}\gamma_\mu\gamma_\nu
\right)\times
\nonumber\\
&&\left(
\left(\frac{1}{\epsilon_\text{UV}}+2\ln\frac{\mu_\text{UV}}{m_b}\right)
(4+2\ln\xi)-
2\left(\frac{1}{\epsilon_\text{IR}}+2\ln\frac{\mu_\text{IR}}{m_b}\right)
+4-\frac{2\pi^2}{3}-2\ln^2\xi\right).
\nonumber
\end{eqnarray}

\subsection{Form factor contribution}
\begin{figure}[t]
\begin{center}
\resizebox{0.5\textwidth}{!}{\includegraphics{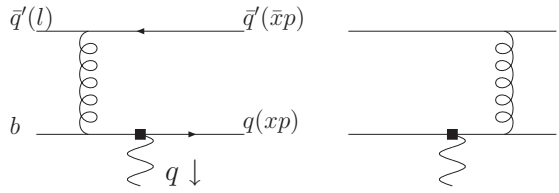}}
\end{center}
\caption{$\alpha_s$ contributions to the form factor}
\label{wfpic3}
\end{figure}
Finally we have to calculate the contribution of (\ref{wf4}). It is
given by
\begin{eqnarray}
\mathcal{A}_\text{formfact.}
&\equiv&
\phi_\pi^{(0)}\otimes\phi_\pi^{(0)}\otimes\phi_B^{(0)}\otimes
T^{(1)}_\text{formfact.}\otimes T^{\text{I}(1)}
\equiv
\label{wf12}\\
&&\frac{C_F\alpha_s}{4\pi}
f^{(1),\nu}\,\bar{q}_e(yq)\gamma_\nu(1-\gamma_5)q^\prime_e(\bar{y}q)
\,T^{(1)}(y).
\nonumber
\end{eqnarray}
The form factor $f^{(1),\nu}$ is the $\alpha_s$ correction of the
matrix element 
\begin{equation}
\label{wf13}
\langle\bar{q}^\prime(\bar{x}p)q(xp)|
\bar{q}\gamma^\nu(1-\gamma_5)b
|b(p+q-l)\bar{q}^\prime(l)\rangle,
\end{equation}
where $\bar{q}_e^\prime$ and $q_e$ are the spinors with the flavour
quantum numbers of the emitted pion and $T^{(1)}(y)$ is given by
\cite{Beneke:2001ev}:
\begin{eqnarray}
T^{(1)}(y) &=&
-6\left(\frac{1}{\epsilon}+\lmu\right)-18+3\left(
\frac{1-2y}{\bar{y}}\ln y-i\pi\right)+
\label{wf14}\\
&&\left[
2\li(y)-\ln^2y+\frac{2\ln y}{\bar{y}}-(3+2i\pi)\ln y 
-(y\leftrightarrow\bar{y})
\right].
\nonumber
\end{eqnarray}
We get $f^{(1),\nu}$ by evaluating the diagrams in
fig.~\ref{wfpic3}. Using the notation of (\ref{dirac2}) we finally
obtain:
\begin{eqnarray}
\mathcal{A}_\text{formfact.} &=& \alpha_s^2N_cC_F^2
\frac{1}{\bar{x}\xi}
\gamma^\mu\tilde{\otimes}
\left(\gamma_\mu\frac{\sh{l}}{\xi}\gamma^\nu(1-\gamma_5)
-\gamma^\nu(1-\gamma_5)\frac{x\sh{p}+\sh{q}+1}{\bar{x}}\gamma_\mu\right)
\otimes
\nonumber\\
&&\gamma_\nu(1-\gamma_5)T^{(1)}(y).
\label{wf15}
\end{eqnarray}

\chapter{NLO results \label{nloresults}}
\section{Analytical results for $T^\text{II}_1$ and  $T^\text{II}_2$}
After the analysis of the last chapter we finally obtain the
$\mathcal{O}(\alpha_s^2)$ 
results for the hard spectator scattering kernels $T^\text{II}_{1,2}$ 
which are defined by (\ref{LO4}). Those expressions appear in convolution
integrals with wave functions, where $x$, $y$ and $\xi$ are the
integration variables as defined in (\ref{LO4}). The ultraviolet
divergences are renormalised in the $\overline{\text{MS}}$-scheme. The
infrared divergences drop out after subtracting the wave function
contributions from the amplitude.
The infrared finiteness together with the finiteness of the 
convolution integrals ensures that the framework of QCD-factorization 
works at this order in $\alpha_s$.  

The explicit $\mathcal{O}(\alpha_s^2)$ contributions for
$T^\text{II}_{1,2}$ read (see next page):
\vfill
\pagebreak[4]
\begin{eqnarray}
\text{Re}T^{\text{II}(2)}_1&=&-\frac{\alpha_s^2C_F}{4N_c^2m_B^2\xi}\times
\label{T1re}\\
&&\Bigg[C_N\bigg(-\frac{16 \ln  \xi }{3 \bar{x} \bar{y}}-\frac{16 \ln  \bar{x}}{3 \bar{x} \bar{y}}+\frac{40 \ln  \frac{\mu }{m_b}}{3 \bar{x} \bar{y}}+\frac{80}{9 \bar{x} \bar{y}}\bigg)\nonumber\\
&&\begin{split}
+C_F\bigg(
&\left(\frac{4 \ln  \xi }{\bar{x} \bar{y}}+\frac{4 \ln
    \bar{x}}{\bar{x} \bar{y}}+\frac{4 \ln  \bar{y}}{\bar{x}
    \bar{y}}+\frac{30}{\bar{x} \bar{y}}\right) 
\ln\frac{\mu }{m_b}
\\
&-\frac{\ln ^2\xi }{\bar{x} \bar{y}}
+\ln\xi \left(-\frac{2 \ln  x}{\bar{x}^2 \bar{y}}-
\frac{2 \ln  \bar{x}}{\bar{x} \bar{y}}-\frac{5}{\bar{x} \bar{y}}\right)
\\
&+\left(-\frac{2 \bar{x}^2}{(y-\bar{x})^3}-\frac{4
    \bar{x}}{(y-\bar{x})^2}-
    \frac{2}{y-\bar{x}}-\frac{2 x}{(y-x) \bar{x}}-\frac{2}{y \bar{x}^2}+
    \frac{2 (5 x-2)}{\bar{y} \bar{x}^2}\right) \li x
\\
&+\left(-\frac{2 \bar{x}^2}{(y-\bar{x})^3}-\frac{4 \bar{x}}{(y-\bar{x})^2}-\frac{2}{y-\bar{x}}+\frac{2 x}{(y-x) \bar{x}}-\frac{4}{\bar{x}}+\frac{2}{y \bar{x}^2}+\frac{4}{\bar{y} \bar{x}^2}\right) \li y
\\
&+\left(\frac{2 (x-2)}{\bar{x}^2 \bar{y}}+\frac{2}{\bar{x}}\right) \li(x y)
\\
&+\left(-\frac{2 \bar{x}^2}{(y-\bar{x})^3}-\frac{4 \bar{x}}{(y-\bar{x})^2}-\frac{2}{y-\bar{x}}\right) \li\left(-\frac{x y}{\bar{x}}\right)
\\
&+\left(\frac{2 x}{(y-x) \bar{x}}+\frac{2}{\bar{x} \bar{y}}\right) 
  \li\left(-\frac{y \bar{x}}{\bar{y}}\right)
+\left(-\frac{2}{\bar{x}}+\frac{2}{\bar{x}^2 \bar{y}}+
 \frac{2}{\bar{x}^2 y}\right) \li(x \bar{y})
\\
&+\left(-\frac{2 x}{(y-x) \bar{x}}-\frac{2}{\bar{x} \bar{y}}\right) 
\li\left(-\frac{x \bar{y}}{\bar{x}}\right)
\\
&+\left(\frac{2 \bar{x}^2}{(y-\bar{x})^3}+\frac{4\bar{x}}{(y-\bar{x})^2}+
\frac{2}{y-\bar{x}}\right) \li\left(-\frac{\bar{x} \bar{y}}{y}\right)
\\
&+\left(-\frac{2}{\bar{y} \bar{x}}-\frac{2}{\bar{x}}\right) \ln  x \ln  y
+\frac{2 (3 x-2) \ln  x \ln  \bar{x}}{\bar{x}^2 \bar{y}}
+\left(\frac{2}{\bar{x}}+\frac{2}{\bar{y} \bar{x}^2}\right)\ln x \ln\bar{y}
\\
&+\left(-\frac{2 \bar{x}^2}{(y-\bar{x})^3}-\frac{4\bar{x}}{(y-\bar{x})^2}-
\frac{2}{y-\bar{x}}-\frac{2}{\bar{x}}+\frac{2}{y \bar{x}^2}+
\frac{2 x}{\bar{y} \bar{x}^2}\right) \ln y \ln\bar{y}
\\
&-\frac{2 \ln  \bar{x} \ln  \bar{y}}{\bar{x} \bar{y}}
+\frac{\ln ^2 x}{\bar{x} \bar{y}}
+\frac{\ln ^2 y}{\bar{x} \bar{y}}
-\frac{\ln ^2 \bar{x}}{\bar{x} \bar{y}}
-\frac{2 \ln ^2 \bar{y}}{\bar{x} \bar{y}}
\\
&+\left(-\frac{4-3 x}{\bar{x}^2 \bar{y}}-\frac{3}{\bar{x}}\right) \ln  x
+\left(\frac{2 (3 x-2)}{\bar{x}^2 \bar{y}}+
 \frac{2 \bar{x}}{(y-\bar{x})^2}+
 \frac{3}{y-\bar{x}}+\frac{3}{\bar{x}}\right) \ln  y
\\
&+\left(\frac{-9 x-1}{x \bar{x} \bar{y}}+\frac{2\bar{x}}{(y-\bar{x})^2}+
  \frac{3}{y-\bar{x}}-\frac{1-3 x}{x^2 y \bar{x}}\right) \ln  \bar{x}
\\
&+\left(-\frac{1}{\bar{x} \bar{y}}-\frac{4}{\bar{x}^2 y}\right) \ln\bar{y}
 +\left(\frac{4}{x \bar{x} \bar{y}}+\frac{4}{x \bar{x}^2 y}\right) \ln (1-x y)
\\
&+\left(-\frac{3 x-1}{x^2 y \bar{x}}+
  \frac{-3 x^2-2 x-1}{x^2 (y-\bar{x})}-
  \frac{3}{\bar{x}}-\frac{2}{\bar{x}^2 \bar{y}}+
  \frac{2 \left(x^2-1\right)}{x (y-\bar{x})^2}\right) \ln (1-x \bar{y})
\\
&+\frac{\pi ^2 \bar{x}^2}{3 (y-\bar{x})^3}+
  \frac{2 \pi ^2 \bar{x}}{3 (y-\bar{x})^2}+
  \frac{\pi ^2}{3 (y-\bar{x})}+\frac{\pi ^2}{3 \bar{x}}-
  \frac{2 \left(2 \pi ^2 x+63 x-63\right)}{3 \bar{y} \bar{x}^2}
\bigg)
\end{split}
\nonumber\\
&&\begin{split}
-\frac{1}{2}C_G\bigg(
&\frac{80 \ln  \frac{\mu }{m_b}}{3 \bar{x} \bar{y}}
+\left(-\frac{2 \ln  x}{\bar{x}^2 \bar{y}}-\frac{22}{3 \bar{x}\bar{y}}\right) 
 \ln  \xi 
\\
&+\left(-\frac{2 x}{(y-x) \bar{x}}+\frac{2 \bar{x}}{(y-\bar{x})^2}+\frac{4}{y-\bar{x}}+\frac{2 (5 x-2)}{\bar{x}^2 \bar{y}}-\frac{2}{y \bar{x}^2}\right) \li x
\\
&+\left(-\frac{2 (x-3)}{\bar{x}^2 \bar{y}}+\frac{2 \bar{x}}{(y-\bar{x})^2}+\frac{4}{y-\bar{x}}+\frac{2 x}{(y-x) \bar{x}}-\frac{4}{\bar{x}}+\frac{2}{y \bar{x}^2}\right) \li y
\\
&+\left(\frac{2 (x-2)}{\bar{x}^2 \bar{y}}+\frac{2}{\bar{x}}\right) \li(x y)
+\left(\frac{2 \bar{x}}{(y-\bar{x})^2}+\frac{4}{y-\bar{x}}+
 \frac{2}{\bar{x}}\right) \li\left(-\frac{x y}{\bar{x}}\right)
\\
&+\left(\frac{2 x}{(y-x) \bar{x}}+\frac{2}{\bar{x}}\right) 
  \li\left(-\frac{y \bar{x}}{\bar{y}}\right)
+\left(-\frac{2}{\bar{x}}+\frac{2}{\bar{x}^2 \bar{y}}+
 \frac{2}{\bar{x}^2 y}\right) \li(x \bar{y})
\\
&+\left(-\frac{2 x}{(y-x) \bar{x}}-\frac{2}{\bar{x}}\right) 
  \li\left(-\frac{x \bar{y}}{\bar{x}}\right)
\\
&+\left(-\frac{2 \bar{x}}{(y-\bar{x})^2}-\frac{4}{y-\bar{x}}-
  \frac{2}{\bar{x}}\right) \li\left(-\frac{\bar{x} \bar{y}}{y}\right)
\\
&-\frac{2 \ln  x \ln  y}{\bar{x} \bar{y}}
+\frac{2 (3 x-2) \ln  x \ln  \bar{x}}{\bar{x}^2 \bar{y}}
+\left(\frac{2}{\bar{x} \bar{y}}-\frac{2}{\bar{x}}\right)\ln y \ln\bar{x}
\\
&+\frac{2 \ln  x \ln  \bar{y}}{\bar{x}^2 \bar{y}}
+\left(\frac{2 \bar{x}}{(y-\bar{x})^2}+\frac{4}{y-\bar{x}}-
  \frac{2}{\bar{x}}+\frac{2}{y \bar{x}^2}+
  \frac{2}{\bar{y} \bar{x}^2}\right) \ln  y \ln  \bar{y}
\\
&+\frac{2 \ln  \bar{x} \ln  \bar{y}}{\bar{x}}
+\frac{\ln ^2 x}{\bar{x} \bar{y}}
+\frac{\ln ^2 y}{\bar{x}}
-\frac{\ln ^2 \bar{x}}{\bar{x} \bar{y}}
-\frac{\ln ^2 \bar{y}}{\bar{x}}
\\
&-\frac{(4-3 x) \ln  x}{\bar{x}^2 \bar{y}}
+\left(-\frac{3-5 x}{\bar{x}^2 \bar{y}}-\frac{2}{y-\bar{x}}\right) \ln  y
\\
&+\left(\frac{2}{x \bar{x} \bar{y}}+\frac{2}{x \bar{x}^2 y}\right) \ln (1-x y)
+\left(\frac{2}{x y \bar{x}}-\frac{31}{3 \bar{x} \bar{y}}-
 \frac{2}{y-\bar{x}}\right) \ln\bar{x}
\\
&-\frac{2 \ln\bar{y}}{y \bar{x}^2}
+\left(\frac{2 (x+1)}{x (y-\bar{x})}-\frac{2}{x y \bar{x}}-
 \frac{2}{\bar{x}^2 \bar{y}}\right) \ln (1-x \bar{y})
\\
&-\frac{2 \left(3 \pi ^2 x+166 x+3 \pi ^2-166\right)}
 {9 \bar{x}^2 \bar{y}}-
 \frac{\pi ^2 \bar{x}}{3 (y-\bar{x})^2}-\frac{2 \pi ^2}{3(y-\bar{x})}+
 \frac{\pi ^2}{3 \bar{x}}
\bigg)\Bigg]
\nonumber\\
\end{split}

\end{eqnarray}
\begin{eqnarray}
\text{Im}T^{\text{II}(2)}_1&=&-\frac{2\pi\alpha_s^2C_F}{4N_c^2m_B^2\xi}\times
\label{T1im}\\
&&\begin{split}
\Bigg[C_F\bigg(
&-\frac{x \ln  x}{\bar{x}^2 \bar{y}}
+\left(\frac{\bar{x}^2}{(y-\bar{x})^3}+
 \frac{2 \bar{x}}{(y-\bar{x})^2}+\frac{1}{y-\bar{x}}+
 \frac{1}{\bar{y} \bar{x}}\right) \ln  y
\\
&+\left(-\frac{\bar{x}^2}{(y-\bar{x})^3}-
  \frac{2 \bar{x}}{(y-\bar{x})^2}-
  \frac{1}{y-\bar{x}}-\frac{x}{(y-x) \bar{x}}-
  \frac{1}{\bar{y} \bar{x}}\right) \ln  \bar{x}
\\
&+\left(\frac{x}{(y-x) \bar{x}}+\frac{1}{\bar{x} \bar{y}}\right) \ln  \bar{y}
-\frac{\bar{x}}{(y-\bar{x})^2}-
 \frac{3}{2 (y-\bar{x})}+\frac{2}{\bar{y} \bar{x}}
\bigg)
\end{split}
\nonumber\\
&&\begin{split}
-\frac{1}{2}C_G\bigg(
&-\frac{x \ln  x}{\bar{x}^2 \bar{y}}
+\left(-\frac{\bar{x}}{(y-\bar{x})^2}-\frac{2}{y-\bar{x}}\right) \ln  y
\\
&+\left(-\frac{x}{(y-x) \bar{x}}+
  \frac{\bar{x}}{(y-\bar{x})^2}+
  \frac{2}{y-\bar{x}}-\frac{1}{\bar{x} \bar{y}}\right) \ln  \bar{x}
\\
&+\frac{x \ln  \bar{y}}{(y-x) \bar{x}}
+\frac{3}{2 \bar{x} \bar{y}}+\frac{1}{y-\bar{x}}
\bigg)\Bigg]
\\
\end{split}\nonumber

\end{eqnarray}
\begin{eqnarray}
\text{Re}T^\text{II}_2&=&-\frac{\alpha_s^2C_F C_N}{4N_c^2m_B^2\xi}\times\label{T2re}\\
&&\begin{split}
\Bigg[
&\frac{12 \ln  \frac{\mu }{m_b}}{\bar{x} \bar{y}}
+\left(-\frac{2 \bar{x}^2}{(y-\bar{x})^3}-\frac{4 \bar{x}}{(y-\bar{x})^2}-\frac{2}{y-\bar{x}}-\frac{2 x}{(y-x) \bar{x}}-\frac{2}{\bar{y} \bar{x}}\right) \li x
\\
&+\frac{2 x \li y}{(y-x) \bar{x}}
+\left(-\frac{2 \bar{x}^2}{(y-\bar{x})^3}-\frac{4 \bar{x}}{(y-\bar{x})^2}-\frac{2}{y-\bar{x}}\right) \li\left(-\frac{x y}{\bar{x}}\right)
\\
&+\left(\frac{2 x}{(y-x) \bar{x}}+\frac{2}{\bar{x} \bar{y}}\right) \li\left(-\frac{y \bar{x}}{\bar{y}}\right)
\\
&+\left(\frac{2 \bar{x}^2}{(y-\bar{x})^3}+\frac{4 \bar{x}}{(y-\bar{x})^2}+\frac{2}{y-\bar{x}}+\frac{2}{\bar{y} \bar{x}}\right) \li \bar{y}
\\
&+\left(-\frac{2 x}{(y-x) \bar{x}}-\frac{2}{\bar{x} \bar{y}}\right) \li\left(-\frac{x \bar{y}}{\bar{x}}\right)
\\
&+\left(\frac{2 \bar{x}^2}{(y-\bar{x})^3}+\frac{4 \bar{x}}{(y-\bar{x})^2}+\frac{2}{y-\bar{x}}\right) \li\left(-\frac{\bar{x} \bar{y}}{y}\right)
\\
&-\frac{2 \ln  x \ln  y}{\bar{x} \bar{y}}
+\frac{2 \ln  x \ln  \bar{y}}{\bar{x} \bar{y}}
+\frac{\ln ^2 y}{\bar{x} \bar{y}}
-\frac{\ln ^2 \bar{y}}{\bar{x} \bar{y}}
-\frac{(2-3 x) \ln  x}{\bar{x}^2 \bar{y}}
\\
&+\left(\frac{x-2}{\bar{x}^2 \bar{y}^2}+\frac{2 \bar{x}}{(y-\bar{x})^2}+\frac{3}{y-\bar{x}}+\frac{x}{\bar{x}^2 \bar{y}}\right) \ln  y
\\
&+\left(\frac{2}{x \bar{x} \bar{y}}+\frac{2}{x \bar{x}^2 y}\right) \ln (1-x y)
+\left(\frac{2 \bar{x}}{(y-\bar{x})^2}+\frac{3}{y-\bar{x}}\right) \ln  \bar{x}
\\
&+\left(\frac{-3 x^2-2 x-1}{x^2 (y-\bar{x})}-\frac{1}{x^2 \bar{x}^2 \bar{y}}+\frac{2 \left(x^2-1\right)}{x (y-\bar{x})^2}+\frac{1}{x \bar{x}^2 \bar{y}^2}\right) \ln (1-x \bar{y})
\\
&+\left(-\frac{3}{\bar{x} \bar{y}}-\frac{2}{\bar{x}^2 y}\right) \ln  \bar{y}
+\frac{16}{\bar{x} \bar{y}}
\Bigg]
\end{split}\nonumber

\end{eqnarray}
\pagebreak[4]
\begin{eqnarray}
\text{Im}T^\text{II}_2&=&-\frac{2\pi\alpha_s^2C_F C_N}{4N_c^2m_B^2\xi}
\times\label{T2im}\\
&&\begin{split}
\Bigg[
&\left(\frac{\bar{x}^2}{(y-\bar{x})^3}+\frac{2 \bar{x}}{(y-\bar{x})^2}+\frac{1}{y-\bar{x}}+\frac{1}{\bar{y} \bar{x}}\right) \ln  y
\\
&+\left(-\frac{\bar{x}^2}{(y-\bar{x})^3}-\frac{2 \bar{x}}{(y-\bar{x})^2}-\frac{1}{y-\bar{x}}-\frac{x}{(y-x) \bar{x}}-\frac{1}{\bar{y} \bar{x}}\right) \ln  \bar{x}
\\
&+\frac{x \ln  \bar{y}}{(y-x) \bar{x}}
-\frac{\bar{x}}{(y-\bar{x})^2}-\frac{3}{2 (y-\bar{x})}+\frac{3}{2 \bar{y} \bar{x}}
\nonumber
\Bigg]
\end{split}

\end{eqnarray}

The $\alpha_s^2$ corrections of the hard spectator interactions have
already been calculated in \cite{Beneke:2005vv,Kivel:2006xc}. However
both of these calculations have been performed in the framework of
SCET, while my result is a pure QCD calculation. In order to compare
(\ref{T1re})-(\ref{T2im}) to \cite{Beneke:2005vv,Kivel:2006xc} we have to
take into account the definition of $\lambda_B$. The SCET calculation
naturally uses the $\lambda_B$, defined by the HQET field for the
$b$-meson, while I define $\lambda_B$ by QCD-fields. Those two
definitions differ at $\mathcal{O}(\alpha_s)$, which has been discussed
in appendix \ref{matchlb}. The difference in the logarithmic moments
of the $B$-meson wave function does not play a role, because these
moments occur first at NLO. Using (\ref{match10}) I figured out with
the help of a computer algebra system, that
(\ref{T1re})-(\ref{T2im}) reproduce the results of  
\cite{Beneke:2005vv,Kivel:2006xc}. 

The main difference between the present QCD calculation and the
framework of SCET
is the way how to make the expansion in $\lqcd/m_b$. While
in the QCD calculation this expansion takes place at the level of the
amplitude and of Feynman integrals, in SCET the Lagrangian is expanded
in powers of $\lqcd/m_b$. This leads to the fact that the structure of
the calculation of \cite{Beneke:2005vv,Kivel:2006xc} is completely
different from the present calculation such that comparing
intermediate results like single Feynman diagrams is not possible. 
So it is allowed to state that the analytical coincidence of the
present result with the results published before gives
more independent test of \cite{Beneke:2005vv,Kivel:2006xc} than 
than a SCET calculation could provide.

\section{Scale dependence}
It is instructive to have a closer look to the scale dependence of
(\ref{T1re})-(\ref{T2im}). As stated in section \ref{ewh} the scale
dependence of $\mathcal{A}^\text{II}$ vanishes. In our case we can prove
this up to $\mathcal{O}(\alpha_s^2)$
i.e.\ 
\begin{equation}
\frac{d}{d\ln\mu}\mathcal{A}^\text{II}=\mathcal{O}(\alpha_s^3).
\label{sd0}
\end{equation}
We need the scale dependence of the following quantities, which I took
from \cite{Buchalla:1995vs}:
\begin{eqnarray}
\frac{d}{d\ln\mu}C_1 &=&
\frac{\alpha_s}{4\pi}\left(12C_NC_2+(12C_F-6C_G)C_1\right)
\label{sd1}\\
\frac{d}{d\ln\mu}C_2 &=&
\frac{\alpha_s}{4\pi}\left(12C_NC_1+(12C_F-6C_G)C_2\right)
\label{sd2}\\
\frac{d}{d\ln\mu}\alpha_s &=&-\frac{\alpha_s^2}{4\pi}2\beta_0
\label{sd3}
\end{eqnarray} 
where $\beta_0=\frac{11C_G-4n_fC_N}{3}$ and
$n_f=5$ the number of active flavours.
Regarding the wave functions we need the scale dependence of their
convolution integrals with the LO kernels:
\begin{eqnarray}
\frac{d}{d\ln\mu}\int_0^1\frac{dx}{\bar{x}}\phi_\pi(x,\mu)
&=& \frac{\alpha_s}{\pi}C_F\int_0^1dx\frac{3+2\ln\bar{x}}{2\bar{x}}
\phi_\pi(x,\mu)
\label{sd4}\\
\frac{d}{d\ln\mu}\int_0^1\frac{d\xi}{\xi}\phi_{B1}(\xi,\mu)
&=&\frac{\alpha_s}{4\pi}C_F\int_0^1\frac{d\xi}{\xi}(4\ln\xi+6)
\phi_{B1}(\xi,\mu).
\label{sd5}
\end{eqnarray} 
(\ref{sd4}) can be obtained using the renormalisation group equation (RGE)
for light-cone distribution amplitudes, which can be found in
\cite{Beneke:2000ry,Lepage:1980fj}. 
We get (\ref{sd5}) from \cite{Lange:2003ff}, where the RGE
for the $B$-meson light-cone wave function, defined in the framework
of HQET, is given. We get the RGE for the pure QCD defined wave function
by matching the nonlocal heavy to light current with QCD. The matching
coefficient is given in Appendix \ref{matchlb}.  

In the case of $\bar{B}^0\to\pi^+\pi^-$ we obtain for the $\mu$-dependent part
of $\mathcal{A}^\text{II}$:
\begin{equation}
\begin{split}
&\
\mathcal{A}^\text{II}(\bar{B}^0\to\pi^+\pi^-)=
-i\frac{G_F}{\sqrt{2}}\lambda_u^\prime f_\pi^2 f_B
\int_0^1 dxdyd\xi\,\phi_\pi(x)\phi_\pi(y)\phi_{B1}(\xi)
\frac{1}{\bar{x}\bar{y}\xi}\times
\\
&\begin{split}
\quad
\Bigg[&
\frac{\pi\alpha_sC_F}{N_c^2}C_2
-\frac{\alpha_s^2C_F}{4N_c^2}
\Bigg(
\ln\mu\bigg(
12C_NC_1
\\
&\quad+C_2\bigg(
\frac{40}{3}C_N+
C_F\left(30+4\ln\bar{x}+4\ln\bar{y}+4\ln\xi\right)
-\frac{40}{3}G_G\bigg)\bigg)
\\
&+(\ldots)\Bigg)\Bigg]
\end{split}
\end{split}
\label{scaledep}
\end{equation}
where the ellipsis $(\ldots)$ stands for $\mu$-independent terms.
Using (\ref{sd1})-(\ref{sd5}) it is easily seen that (\ref{sd0}) is
fulfilled. In the case of $\bar{B}^0\to\pi^0\pi^0$ one just has to
interchange $C_1$ and $C_2$.
\section{Convolution integrals and factorizability}
By looking at the hard scattering kernels of (\ref{T1re})-(\ref{T2im}) it
is not obvious that there remain no singularities in the convolution
integrals  over
wave functions (\ref{LO4}). It is however possible to perform the
integration analytically, which proves the factorizabilty.

Regarding the $B$-meson wave function we will obtain the result in terms
of the quantities $\lambda_B$ and $\lambda_n$, which are defined in
(\ref{lambdaB}) and (\ref{lambdan}). 
The $\pi$-meson wave function is given in terms of Gegenbauer
polynomials:
\begin{equation}
\phi_\pi(x)=6x\bar{x}\left[1+\sum_{n=1}^\infty
  a_n^\pi C_n^{(3/2)}(2x-1)\right].
\label{gegenbauer}
\end{equation}
Because of the symmetry properties of the pion we set
$a_{2n-1}^\pi=0$, furthermore we neglect $a_n^\pi$ for $n>2$. So we
need only the second Gegenbauer polynomial which is given by:
\begin{equation}
 C_2^{(3/2)}(x)=\frac{15}{2}x^2-\frac{3}{2}.
\label{gegenbauer2}
\end{equation}
Using (\ref{LO4}) we get for the NLO of $A_\text{spect}$:
\begin{equation}
\begin{split}
&A_\text{spect. 1}^{(2)}=\alpha_s^2\frac{if_\pi^2f_B}{4N_c^2}C_F
\frac{m_B}{\lambda_B}\times
\\
&\begin{split}
\quad\Bigg[&
C_N\left(120\ln\frac{\mu}{m_b}-48\lambda_1+152\right)
\\
&\begin{split}
+C_F\bigg(&(162+36\lambda_1)\ln\frac{\mu}{m_b}-9\lambda_2+
(-54+6\pi^2)\lambda_1+\frac{1566}{5}-\frac{1008}{5}\zeta(3)+27\pi^2
\\
&+i\left(-9\pi+\frac{18}{5}\pi^3\right)
\bigg)
\end{split}
\\
&\begin{split}
-\frac{1}{2}C_G\bigg(&240\ln\frac{\mu}{m_b}+(-102+6\pi^2)\lambda_1+
 \frac{2101}{5}-
 \frac{1008}{5}\zeta(3)+18\pi^2
\\
&+i\left(9\pi+\frac{18}{5}\pi^3\right)
\bigg)
\end{split}
\\
&\begin{split}
+a_2^\pi\bigg\{&
C_N\left(240\ln\frac{\mu}{m_b}-96\lambda_1+404\right)
\\
&\begin{split}
+C_F\bigg(&(174+72\lambda_1)\ln\frac{\mu}{m_b}-18\lambda_2+
\left(-\frac{741}{2}+42\pi^2\right)\lambda_1-\frac{14809}{35}
\\
&-\frac{45072}{35}\zeta(3)+204\pi^2
+i\left(-338\pi+\frac{1362}{35}\pi^3\right)
\bigg)
\end{split}
\\
&\begin{split}
-\frac{1}{2}C_G\bigg(&480\ln\frac{\mu}{m_b}+(-504+42\pi^2)\lambda_1+
\frac{22299}{35}-\frac{43992}{35}\zeta(3)+161\pi^2
\\
&+i\left(-292\pi+\frac{1482}{35}\pi^3\right)\bigg)
\bigg\}
\Bigg]
\end{split}
\end{split}
\end{split}
\end{split}
\label{aspect12}
\end{equation}
and
\begin{equation}
\begin{split}
&A_\text{spect. 2}^{(2)}=\alpha_s^2\frac{if_\pi^2f_B}{4N_c^2}C_FC_N
\frac{m_B}{\lambda_B}\times
\\
&\begin{split}
\quad\Bigg[&
108\ln\frac{\mu}{m_b}+\frac{1467}{10}+\frac{252}{5}\zeta(3)-6\pi^2
+i\left(54\pi-\frac{12}{5}\pi^3\right)
\\
&+a_2^\pi\bigg(216\ln\frac{\mu}{m_b}+\frac{40281}{140}+
\frac{29268}{35}\zeta(3)-112\pi^2+i\left(118\pi-\frac{108}{35}\pi^3\right)
\bigg)\Bigg].
\end{split}
\end{split}
\label{aspect22}
\end{equation}
The finiteness of the above equations proves factorization of the hard
spectator interactions at NLO.

Including the contributions of 
$A_\text{spect. 1}^{(2)}$ and $A_\text{spect. 2}^{(2)}$ 
the quantities
$a_{1,\text{II}}$ and $a_{2,\text{II}}$ defined in (\ref{w12}) and
(\ref{LO8}) are
\begin{eqnarray}
a_{1,\text{II}} &=& 
\frac{i}{f_\pi f^{B\pi}_+m_B^2}(C_2A_\text{spect. 1}^{(1)}+
C_2A_\text{spect. 1}^{(2)}
+C_1A_\text{spect. 2}^{(2)})
\nonumber\\
a_{2,\text{II}} &=& 
\frac{i}{f_\pi f^{B\pi}_+ m_B^2}(
C_1A_\text{spect. 1}^{(1)}+
C_1A_\text{spect. 1}^{(2)}
+C_2A_\text{spect. 2}^{(2)}),
\label{aspect3}
\end{eqnarray}
where
\begin{equation}
A_\text{spect. 1}^{(1)}=\frac{-i C_F\pi\alpha_s}{N_c^2}
\frac{f_Bf_\pi^2m_B}{\lambda_B}9(1+a_2^\pi)^2.
\end{equation}
\section{Numerical analysis}
\subsection{Input parameters}
%
%
\begin{table}[t]
\begin{tabular*}{\textwidth}{c}
\hline
CKM-parameters\\
\begin{tabular*}{\textwidth}{c@{\extracolsep\fill}cccc}
  $V_\text{ud}$ \cite{Charles:2004jd}& $V_\text{cd}$ & 
  $V_\text{cb}$ \cite{Charles:2004jd}&
  $|V_\text{ub}/V_\text{cb}|$ \cite{Charles:2004jd}& $\gamma$\\
  0.974 & $-0.23$ & 0.041 & 0.09$\pm$0.025 & (70$\pm$20)deg\\
\end{tabular*}\\ \\
Parameters of the $B$-meson\\
\begin{tabular*}{\textwidth}{c@{\extracolsep\fill}cccccc}
  $m_B$ & $f_B$ \cite{Jamin:2001fw}  & $\frac{f_B}{f_+^{B\pi}\lambda_B}$
  \cite{Khodjamirian:2006st} 
  & $\lambda_1$ \cite{Beneke:2005vv} & $\lambda_2$ \cite{Beneke:2005vv}
  & $\tau_{B^\pm}$ & $\tau_{B^0}$ \\
  5.28GeV & \hspace{1ex}(210$\pm$19)MeV & $1.56\pm0.17$ 
  & $-3.2\pm1$ & $11\pm4$ 
  & 1.67 ps & 1.54 ps 
\end{tabular*}\\ \\
Parameters of the $\pi$-meson\\
\begin{tabular*}{\textwidth}{c@{\extracolsep\fill}cccc}
  $f_+^{B\pi}$ \cite{Abada:2000ty,Dalgic:2006dt,Khodjamirian:2000ds}& 
  $f_\pi$ & $m_\pi$ & 
  $a_1^\pi$ & $a_2^\pi$ \cite{Gockeler:2005jz,Braun:2006dg}\\
  $0.28\pm0.05$ & 131MeV & 130MeV &  0 & $0.3\pm0.15$
\end{tabular*}\\ \\ 
Quark and W-boson masses\\
\begin{tabular*}{\textwidth}{c@{\extracolsep\fill}ccc}
  $m_b(m_b)$ & $m_c(m_b)$ & $m_t(m_t)$ \cite{Beneke:2001ev} & $M_W$\\
  4.2 GeV & (1.3$\pm$0.2) GeV & 167 GeV & 80.4 GeV
\end{tabular*}\\ \\
Coupling constants\\
\begin{tabular*}{\textwidth}{c@{\extracolsep\fill}ccc}
  &$\Lambda^{(5)}_{\overline{\text{MS}}}$ & $G_F$ &\\
  &225 MeV & $1.16639\times 10^{-5}\,\text{GeV}^{-2}$&
\end{tabular*}\\
\hline
\end{tabular*}
\caption{Input parameters, which were used in the numerical
  analysis. All parameters given without explicit citation
  can be found in \cite{Yao:2006px}. Unless otherwise stated scale
  dependent quantities are given at $\mu=1\text{GeV}$.} 
\label{params}
\end{table}
For my numerical analysis I use the parameters given in table
\ref{params}. The decay constant $f_B$ and the ratio
$\frac{f_B}{f^{B\pi}_+\lambda_B}$ have been obtained by QCD sum rules
in \cite{Jamin:2001fw} and \cite{Khodjamirian:2006st} respectively. The
logarithmic moments $\lambda_1$ and $\lambda_2$ where calculated in
\cite{Beneke:2005vv} using model light-cone wave functions for the
$B$-meson
\cite{Khodjamirian:2005ea,Braun:2003wx,Grozin:1996pq,Lee:2005gz}.
For the form factor $f_+^{B\pi}$ I use the value from
\cite{Khodjamirian:2000ds}, which has been obtained by QCD sum
rules. This value is consistent with quenched and recent unquenched
lattice calculations \cite{Abada:2000ty,Dalgic:2006dt}. The first
Gegenbauer moment of the pion wave function is zero due to G-parity
while the second moment has been obtained by lattice simulations
\cite{Gockeler:2005jz,Braun:2006dg}.

\subsection{Power suppressed contributions \label{psc}}
In our numerical analysis we include two contributions which are
suppressed in leading power but numerically enhanced: The twist-3
contributions and the annihilation topologies.

The twist-3 contributions come from the twist-3 contributions of the
pion wave functions and are suppressed by the factor
$$
r_\chi^\pi(\mu)=\frac{2m_\pi^2}{\bar{m}_b(\mu)(\bar{m}_u(\mu)+\bar{m}_d(\mu))}
$$ 
(see (\ref{w14})), which is formally of $\mathcal{O}(\lqcd/m_b)$ but
numerically about $0.9$. These contributions modify the hard spectator
scattering function $H_{\pi\pi}$ (\ref{LO9}) to \cite{Beneke:2001ev}
\begin{equation}
H_{\pi\pi}=
\frac{f_Bf_\pi}{m_B^2 f_+^{B\pi}}
\int_0^1\frac{d\xi}{\xi}\Phi_{B1}(\xi)
\int_0^1\frac{dx}{\bar{x}}\phi_{\pi}(x)
\left(
\int_0^1\frac{dy}{\bar{y}}\phi_{\pi}(y)
+\frac{\bar{x}}{x}r^\pi_\chi X_H
\right)
\label{num1}
\end{equation}
where $X_H$ is parametrised by:
\begin{equation}
X_H=\left(1+\rho_He^{i\phi_H}\right)\ln\frac{m_b}{\Lambda_h}
\label{num2}
\end{equation}
with 
\begin{equation}
0 \le \rho_H \le 1.
\label{num3}
\end{equation}
For numerical calculations we set
\begin{equation}
\ln\frac{m_b}{\Lambda_h}\approx 2.4.
\label{num4}
\end{equation}

The annihilation contributions are parametrised in
(\ref{w15})-(\ref{w18}). Analogously to (\ref{num2}) $X_A$ is
parametrised by
\begin{equation}
X_A=\left(1+\rho_Ae^{i\phi_A}\right)\ln\frac{m_b}{\Lambda_h}
\label{num5}
\end{equation}
with 
\begin{equation}
0 \le \rho_A \le 1.
\label{num6}
\end{equation}

\subsection{Amplitudes $a_1$ and $a_2$ \label{qcdampl}}
%
\begin{figure}
\begin{center}
\input{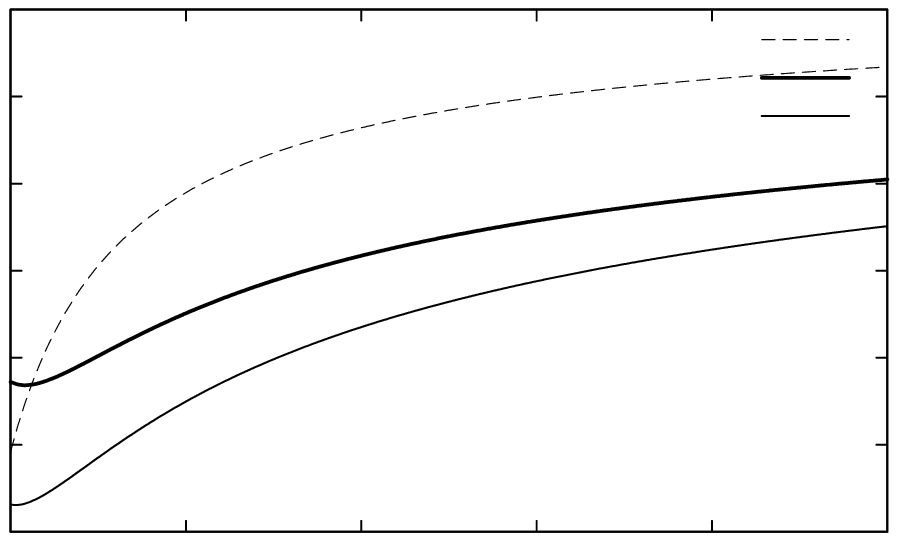}
\input{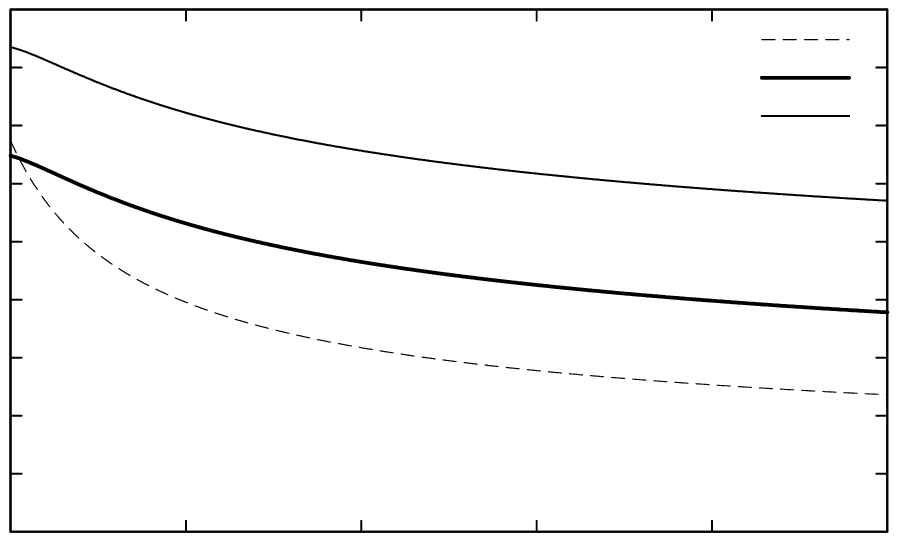}
\input{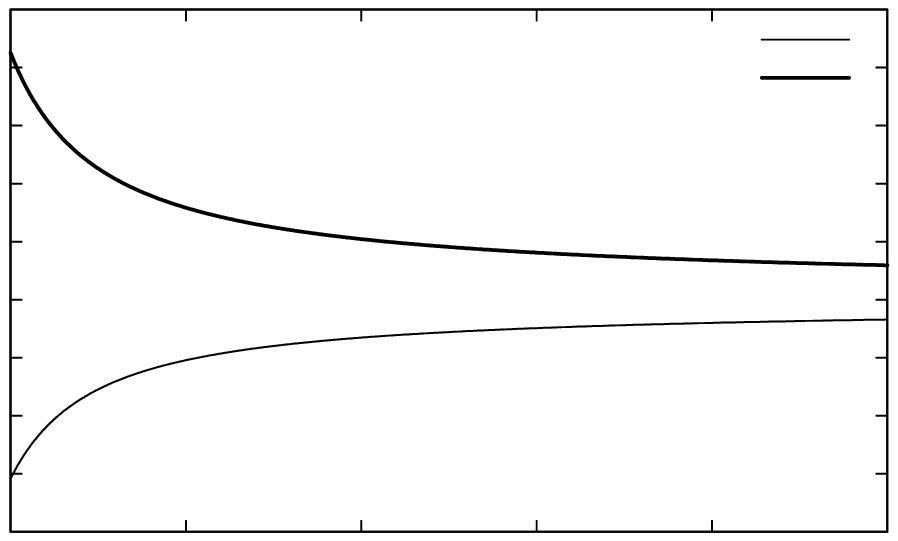}
\end{center}
\caption{Contribution of the hard spectator corrections to $a_1$ and
  $a_2$ as a function of the renormalisation scale $\mu$. The upper
  two figures show the real part, where the LO is given by the dashed 
  line, while the sum of LO and NLO is shown by the thick solid line. 
  The twist-3 corrections are included in the graph given by the thin 
  solid line. The third figure shows the imaginary part, which occurs
  first at $\mathcal{O}(\alpha_s^2)$. So no distinction between LO and
  NLO is made.
}
\label{numpic1}
\end{figure}
The amplitudes $a_1$ and $a_2$ are defined in (\ref{w12}). Their hard
scattering parts $a_{1,\text{II}}$ and $a_{2,\text{II}}$, i.e.\ the
parts of $a_1$ and $a_2$, which contribute to $\mathcal{A}^\text{II}$ 
(see (\ref{w10})), are plotted in fig.~\ref{numpic1} as functions of
the renormalisation scale $\mu$. The strong dependence on $\mu$ of the
real part of LO is reduced at NLO. Taking the twist-3 contributions 
into account does not increase the $\mu$-dependence too much. The
imaginary part, which occurs first at NLO, is strongly dependent on
the renormalisation scale.
An appropriate choice for the scale of the hard scattering amplitude
is the hard collinear scale
\begin{equation}
\mu_\text{hc}=1.5\text{GeV}
\label{num7}
\end{equation} 
In the following numerical calculations we will evaluate
$a_{1,\text{II}}$ and $a_{2,\text{II}}$ at $\mu_\text{hc}$. The vertex 
corrections $\mathcal{A}^\text{I}$ will be evaluated at 
\begin{equation}
\mu_b=4.8\text{GeV}.
\label{num8}
\end{equation}
Using the parameters of table \ref{params} we obtain
\begin{equation}
\begin{split}
a_1 =& 1.015+[0.039+0.018i]_V+[-0.012]_\text{tw3}+
[-0.029]_\text{LO}
\\
&+[-0.010-0.031i]_\text{NLO} 
\\
a_2 =& 0.184+[-0.171-0.080i]_V+[0.038]_\text{tw3}+[0.096]_\text{LO}
\\
&+[0.021+0.045i]_\text{NLO}.
\end{split}
\label{num9}
\end{equation}
These equations are given in a form similar to (61) and
(62) in \cite{Beneke:2005vv}. The first number gives the tree
contribution, the vertex corrections are indicated by the label $V$,
the twist-3 contributions, which come from the last part of
(\ref{num1}), are labelled by $\text{tw3}$. The hard scattering part is
separated into LO and NLO. The hadronic input parameters I used are
slightly different from \cite{Beneke:2005vv} and in contrast to
\cite{Beneke:2005vv} I evaluated all quantities, which belong to the hard
scattering amplitude, at the hard collinear scale
$\mu_\text{hc}$. This is why the values I get for $a_1$ and $a_2$ are
different from \cite{Beneke:2005vv}.

The hard scattering amplitudes $a_{1,\text{II}}$ and $a_{2,\text{II}}$
together with their numerical errors read:
\begin{equation}
\begin{split}
a_{1,\text{II}} =\, &
-0.051\pm0.011(\text{param.})^{+0.026}_{-0.005}(\text{scale})
\pm0.012(\text{tw3})\\
&+[-0.031\pm0.008(\text{param.})^{+0.024}_{-0.031}(\text{scale})
\pm0.012(\text{tw3})]i\\
a_{2,\text{II}} =\, &
0.15\pm0.03(\text{param.})^{+0.01}_{-0.04}(\text{scale})
\pm0.04(\text{tw3})\\
&+[0.045\pm0.012(\text{param.})^{+0.040}_{-0.033}(\text{scale})
\pm0.038(\text{tw3})]i.
\end{split}
\label{num10}
\end{equation}
The first error comes from the error of the input parameters in table
\ref{params}. The scale uncertainty is obtained by varying
$\mu_\text{hc}$ between 1GeV and 6GeV. The error labelled by
$\text{tw3}$ gives the error of the twist-3 contribution, which is
obtained by varying $\rho_H$ between 0 and 1 and $\phi_H$ between 0
and $2\pi$. Within the scale uncertainty (\ref{num10}) is compatible
with \cite{Beneke:2005vv}. 

It is important to remark that the result I obtained in QCD comes 
with formally large logarithms $\ln\lqcd/m_b$. 
In contrast to the SCET calculation of
\cite{Beneke:2005vv,Kivel:2006xc} it is not possible to resum
these logarithms by a pure QCD calculation. Without resummation,
however, these logarithms might spoil perturbation theory. Regarding 
this fact it is the more important that the error arising from the scale 
uncertainty in (\ref{num10}) is small enough for perturbation theory 
to be valid. 

\subsection{Branching ratios}
\begin{figure}
\begin{center}
\input{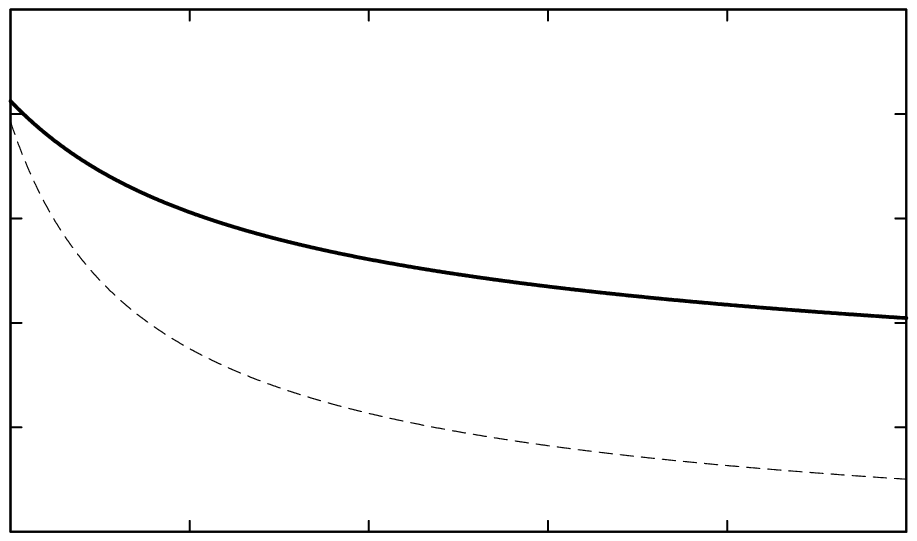}
\input{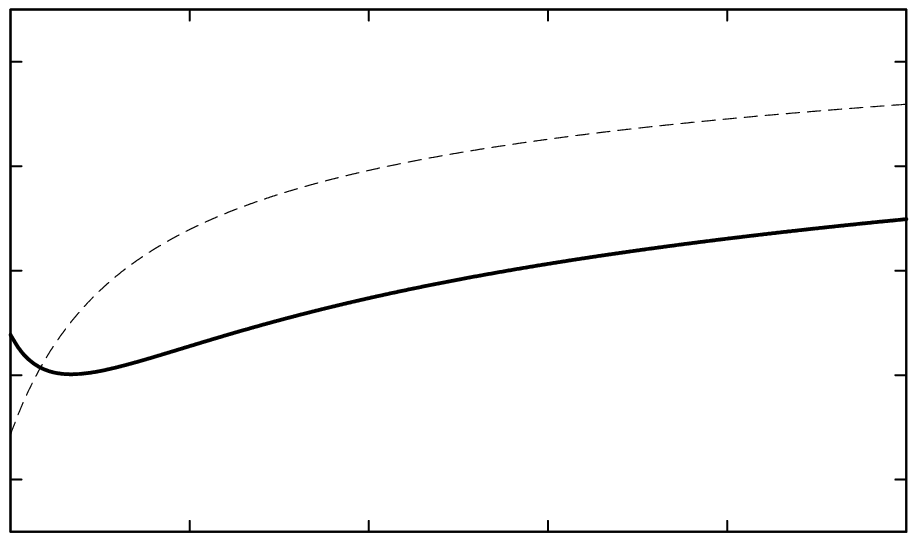}
\input{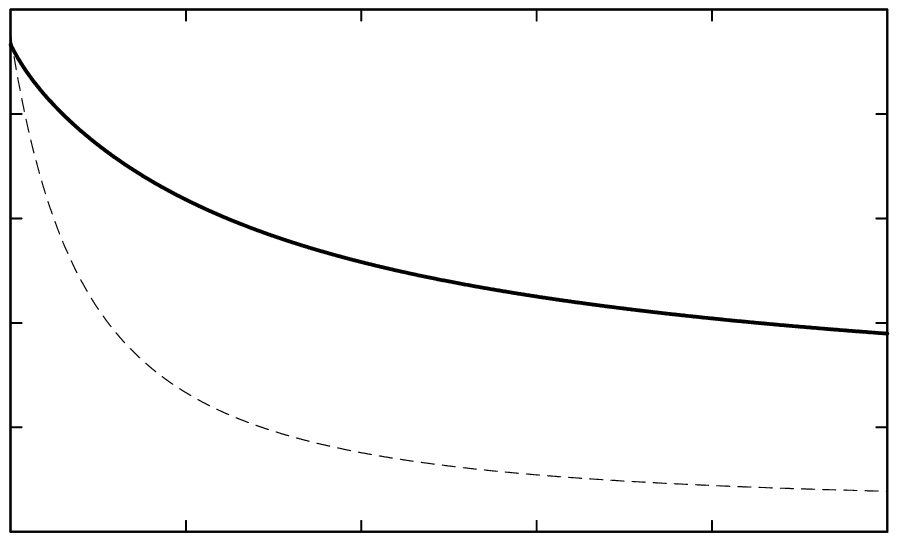}
\end{center}
\caption{CP-averaged branching ratios as functions of the
  hard collinear scale $\mu_\text{hc}$ in units of $10^{-6}$. 
  In the graph with the dashed line only the leading order of the 
  hard spectator scattering is contained, while in the solid line 
  hard spectator scattering is taken into account up to NLO.
}
\label{numpic2}
\end{figure}
The dependence of the CP-averaged branching ratios on the hard
collinear scale is shown in fig.~\ref{numpic2}. It is obvious that the
NLO corrections reduce this dependence significantly.

\begin{figure}
\begin{center}
\input{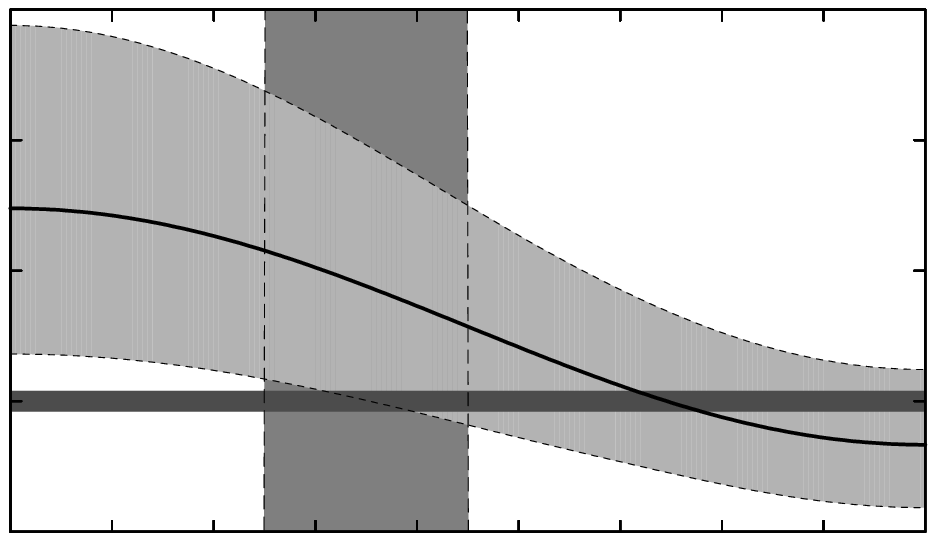}
\input{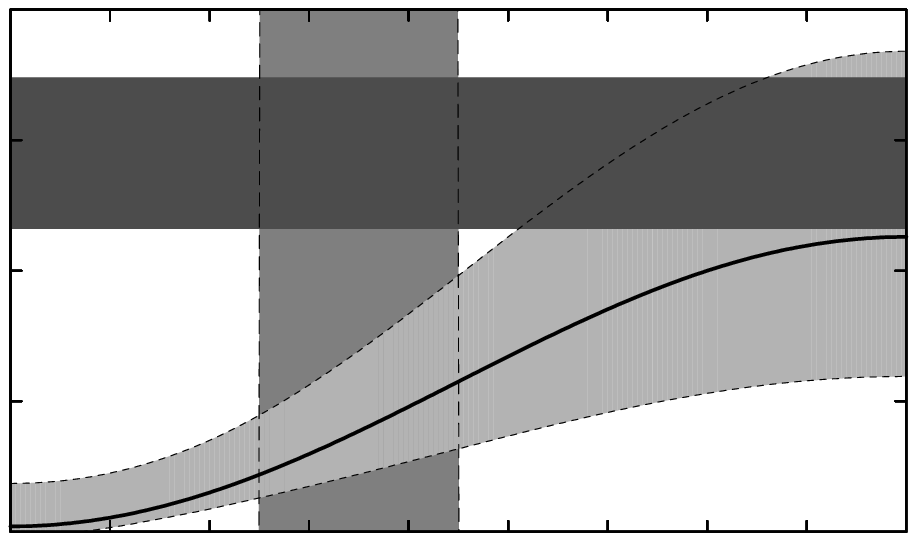}
\end{center}
\caption{CP-averaged branching ratios as functions of the
  CKM-angle $\gamma$ in units of $10^{-6}$. The light-grey band gives
  the uncertainty from the errors of table \ref{params} and from the
  twist-3 and the annihilation contributions. The solid inner line
  gives the central value. The horizontal dark-grey band gives the
  experimental value according to \cite{Barberio:2006bi} and the
  vertical grey band gives the value of $\gamma$ from table \ref{params}
  within the error ranges. 
}
\label{numpic3}
\end{figure}

From the parameter set in table \ref{params} we obtain the following
CP-averaged branching ratios
\begin{equation}
\begin{split}
10^6\text{BR}(B^+\to\pi^+\pi^0)\;=\;&
6.05^{+2.36}_{-1.98}(\text{had.})^{+2.90}_{-2.33}
(\text{CKM})^{+0.18}_{-0.31}(\text{scale})
\pm0.27(\text{sublead.})\\
10^6\text{BR}(B^0\to\pi^+\pi^-)\;=\;&
9.41^{+3.56}_{-2.99}(\text{had.})^{+4.00}_{-3.46}
(\text{CKM})^{+1.07}_{-3.93}(\text{scale})
^{+1.13}_{-0.70}(\text{sublead.})\\
10^6\text{BR}(B^0\to\pi^0\pi^0)\;=\;&
0.39^{+0.14}_{-0.12}(\text{had.})^{+0.20}_{-0.17}
(\text{CKM})^{+0.17}_{-0.06}(\text{scale})
^{+0.20}_{-0.08}(\text{sublead.}).
\end{split}
\label{num11}
\end{equation}
The origin of the errors are the uncertainties of the hadronic
parameters and the CKM parameters, the scale dependence and the
subleading power contributions, i.e. twist-3 and annihilation
contributions. The error arising from the scale dependence was
estimated by varying $\mu_b$ between 2GeV and 8GeV and $\mu_\text{hc}$
between 1GeV and 6GeV. If we compare (\ref{num11}) to the experimental
values \cite{Barberio:2006bi}:
\begin{eqnarray}
10^6\text{BR}(B^+\to\pi^+\pi^0) &=& 5.5\pm0.6
\nonumber\\
10^6\text{BR}(B^0\to\pi^+\pi^-) &=& 5.0\pm0.4
\nonumber\\
10^6\text{BR}(B^0\to\pi^0\pi^0) &=& 1.45\pm0.29
\label{num12}
\end{eqnarray}
we note that $\text{BR}(B^+\to\pi^+\pi^0)$ is in good agreement with
the data. This quantity is almost independent of $\gamma$.
The other branching ratios, which come with large errors, 
depend strongly on $\gamma$. This dependence is shown in
fig.~\ref{numpic3}. The light-grey band gives the uncertainty that is
defined in the same way as the errors in (\ref{num11}), where
different errors are added in quadrature. The solid inner line gives
the central value. The experimental values are
represented by the horizontal band, whereas the vertical band gives
the value of $\gamma$. It is obvious that the errors of the branching
fractions are too large for a reasonable determination of $\gamma$.

%
%
%
For $B^+\to\pi^+\pi^0$ and $B^0\to\pi^+\pi^-$ QCD-factorization is
expected to work well, because at tree level Wilson coefficients occur
in the so called colour allowed combination $C_1+C_2/N_c\sim 1$, while
$B^0\to\pi^0\pi^0$ comes at tree level with $C_2+C_1/N_c\sim 0.2$ such
that subleading power corrections are expected to be more
important. On the other hand there are big uncertainties in the
parameters occurring in the combinations $|V_{ub}|f^{B\pi}_+$,
$\frac{f_B}{f^{B\pi}_+\lambda_B}$ and $a_2^\pi$. In \cite{Beneke:2003zv} 
and \cite{Beneke:2005vv} these parameters were fitted by the 
experimental values (\ref{num12}) of $\text{BR}(B^+\to\pi^+\pi^0)$ 
and $\text{BR}(B^0\to\pi^+\pi^-)$. Setting 
\begin{equation}
a_2^\pi(1\text{GeV})=0.39
\label{num13}
\end{equation} 
leads to 
\begin{eqnarray}
|V_{ub}|f^{B\pi}_+ &\to&  0.80
\left(|V_{ub}|f^{B\pi}_+ \right)_\text{default}
\nonumber\\
\frac{f_B}{f^{B\pi}_+\lambda_B} &\to&
2.89
\left(\frac{f_B}{f^{B\pi}_+\lambda_B}\right)_\text{default}.
\label{num14}
\end{eqnarray}
This leads to the following branching ratios:
\begin{eqnarray}
10^6\text{BR}(B^+\to\pi^+\pi^0) &=&
5.5\pm0.2(\text{param.})^{+0.5}_{-0.3}(\text{scale})\pm0.6(\text{sublead.})
\nonumber\\
10^6\text{BR}(B^0\to\pi^+\pi^-) &=&
5.0^{+0.8}_{-0.9}(\text{param.})^{+0.9}_{-0.2}(\text{scale})
^{+0.9}_{-0.6}(\text{sublead.})
\nonumber\\
10^6\text{BR}(B^0\to\pi^0\pi^0) &=&
0.77\pm 0.3(\text{param.})^{+0.2}_{-0.3}(\text{scale})
^{+0.3}_{-0.2}(\text{sublead.}).
\label{num15}
\end{eqnarray}
The uncertainties of the quantities that occurred in (\ref{num13}) 
and (\ref{num14}) have not been considered in the estimation of the
errors in (\ref{num15}). The $B^0\to\pi^0\pi^0$ branching ratio
obtained in (\ref{num15}) is compatible with the value obtained in 
\cite{Beneke:2005vv}. Though it is too low, due to the
theoretical and experimental errors it is compatible with
(\ref{num12}).

\begin{figure}
\begin{center}
\input{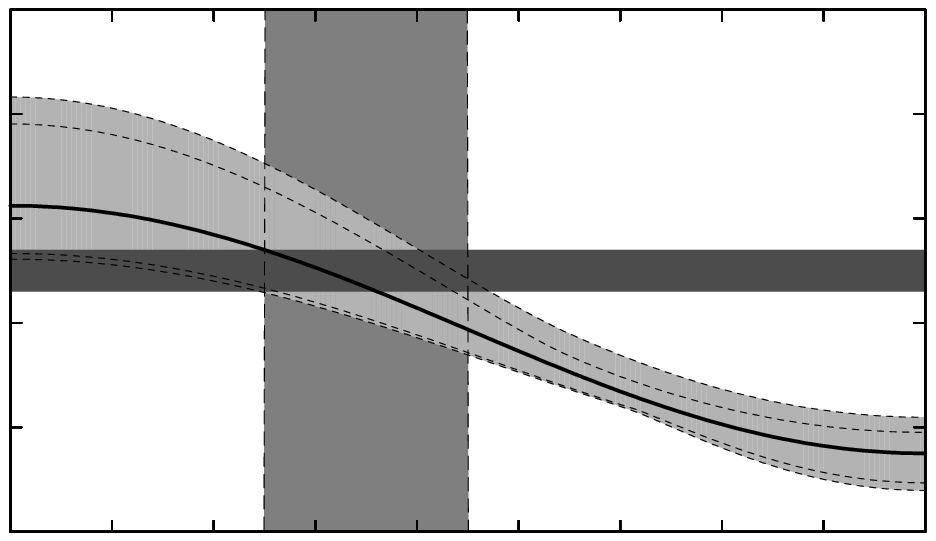}
\input{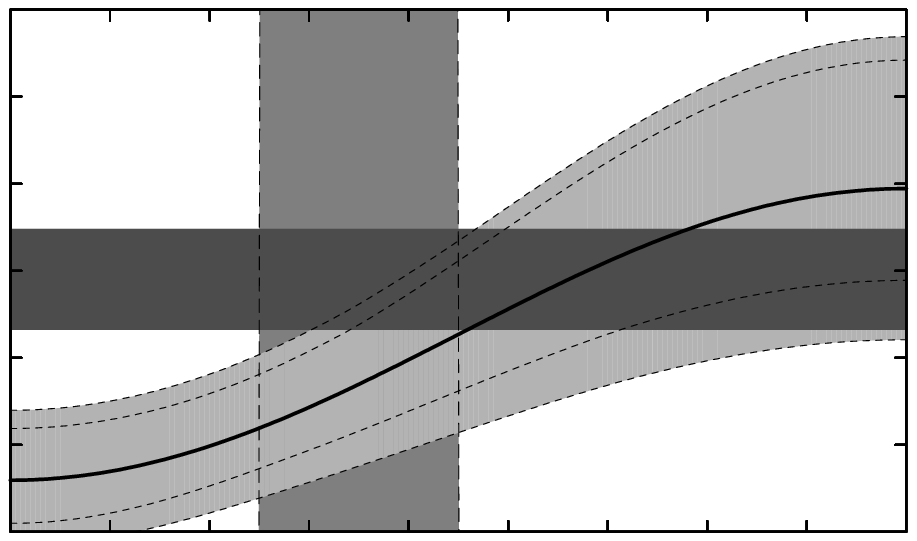}
\end{center}
\caption{CP-averaged branching ratios as functions of the
  CKM-angle $\gamma$ in units of $10^{-6}$ with the input parameters
  (\ref{num13}) and (\ref{num14}). The dashed lines inside of the
  light-grey band give the error coming from subleading power
  contributions, while the dashed lines at the border of the grey
  bands are included to lead the eye. The meaning of the other curves 
  and bands is the same as in fig.~\ref{numpic3} besides the fact 
  that the parameters occurring in (\ref{num13}) and (\ref{num14})
  were not included in the error estimation. 
}
\label{numpic4}
\end{figure}

There are two different sources of errors. On the one hand 
for errors that are due to uncertainties of input parameters and
the renormalisation scale there is at least in principle no lower limit. 
On the other hand errors arising from
subleading power corrections, i.e.\ twist-3 and annihilation
contributions, cannot be reduced in the framework of
QCD-factorization. Fig.~\ref{numpic4} shows the branching fractions of
$B^0\to\pi^+\pi^-$ and $B^0\to\pi^0\pi^0$ as functions of
$\gamma$. The errors arising from subleading power contributions are
represented by the dashed lines inside of the light-grey error
band. While in the case of $B^0\to\pi^+\pi^-$ this remaining error
might be small enough for non-trivial phenomenological statements
about $\gamma$, in the case of $B^0\to\pi^0\pi^0$ there remains an
error of about 30\%.
%
%
%

\chapter{Conclusions}
In the last decades $B$ physics has proven a promising field to
determine parameters of the flavour sector with high precision. 
It is expected that in the next few years the angles $\alpha$
and $\gamma$, which are directly connected to the complex phase of 
the CKM matrix, 
will be measured with an accuracy at the percent level. Furthermore
the discovery of physics beyond the standard model will be possible.

On the theoretical
side QCD factorization has turned out to be an appropriate tool to
calculate $B$ decay modes from first principles, because it allows for
systematic disentanglement of the perturbative physics and the
non-perturbative physics. The present calculation showed that the hard
spectator scattering amplitude factorizes up to and including 
$\mathcal{O}(\alpha_s^2)$, i.e.\ all infrared divergences cancel and
there are no remaining endpoint singularities. The former point is obvious
after the explicit calculation of $T^\text{II}$ and the latter point
was shown by evaluating the convolution integral (\ref{factform})
analytically. The explicit expressions for the hard spectator scattering kernel
(\ref{T1re})-(\ref{T2im}) confirmed the result of
\cite{Beneke:2005vv,Kivel:2006xc}. So they are also a confirmation that 
the leading power of the amplitudes can be obtained by performing the
power expansion at the level of Feynman integrals rather than at the
level of the QCD Lagrangian using an effective theory like SCET, which was
done in \cite{Beneke:2005vv,Kivel:2006xc}. 

The main challenges in the evaluation of Feynman integrals, which were
made possible with the help of tools like integration by parts
identities and differential equation techniques, were due to the fact
that the Feynman integrals came with up to five external legs and
three independent rations of scales. Many steps in the calculations 
of section \ref{clodi} might look like cookery. However I dare say 
calculating Feynman integrals is cookery.

One motivation to calculate the $\mathcal{O}(\alpha_s^2)$ corrections
of the hard spectator interactions separately is the fact, that
the LO of this class of diagrams starts at $\mathcal{O}(\alpha_s)$
such that in order to fix the scale we need the NLO correction. The
numerical results of section \ref{qcdampl} show that the NLO reduces
the scale dependence significantly. This is even more important with
respect to large logarithms that arise because of the fact that next to
the $m_b$-scale also the hard-collinear scale $\sqrt{\lqcd m_b}$
enters the hard spectator scattering amplitude. In contrast to the
effective theory ansatz the QCD calculation of this work does not
allow the resummation of these logarithms. 
This is why it is a crucial point, that the NLO is numerically
important but small enough for perturbation theory to be valid.

Next to the scale dependence a main source of uncertainty is due to
the fact that we do not know hadronic quantities well enough. This
might be improved in the next few years by lattice calculations and
even determination of the hadronic input parameters in experiment. 
Also a better control of power corrections would allow to obtain
much more precise predictions from QCD factorization.

Finally it is important to note that the present calculation is not
the complete order $\alpha_s^2$ result as the contributions of penguin 
contractions and the 
effective penguin operators where not considered in this thesis. 
Actually they play a dominant role in the branching ratios of 
$B\to K\pi$ and CP asymmetries of $B\to\pi\pi$ and should be taken 
into account in phenomenological applications. While writing down this 
thesis the order $\alpha_s^2$ of these contributions has been recently 
published in \cite{Beneke:2006mk}.
Also the $\mathcal{O}(\alpha_s^2)$ corrections of $T^\text{I}$ were not part of
this thesis. These contributions have been calculated in
\cite{Bell:2007tz,Bell:2007tv}. 

So the calculation of the present thesis is a small but very important
tessera in the mosaic of theoretical $B$-physics.

\begin{appendix}
\chapter{CAS implementation of IBP identities\label{ibpalgorithm}}
\section{User manual}
This section will give an introduction how to use my Mathematica packages 
{\ttfamily lorentz.m} and {\ttfamily ibp.m}. These packages use the
rules of section \ref{ibpmethod} and the algorithm of
\cite{Laporta:2001dd}. You can download these files from
\begin{verbatim}
    http://www.theorie.physik.uni-muenchen.de/~pilipp
\end{verbatim}
I assume that these files are located on your hard disk in the
directory {\ttfamily path}.
After you have started your Mathematica notebook with the two lines
\begin{verbatim}
   <<path/lorentz.m;
   <<path/ibp.m;
\end{verbatim}
you have to set some variables. Because my program distinguishes
between Lorentz vectors and scalars we have to define which variables
are of the type vector. This is done with the function 
\begin{verbatim}
   AddMomenta[p1,...,pn]
\end{verbatim}
which defines the variables {\ttfamily p1,...,pn} to be of the type
vector. The function
\begin{verbatim}
   RemMomenta[p1,...,pn]
\end{verbatim}
removes the attribute vector from {\ttfamily p1,...,pn} and 
\begin{verbatim}
   ShowMomenta[]
\end{verbatim}
gives list of all vector variables. Per default the variables
{\ttfamily p}, {\ttfamily q} and {\ttfamily l} are defined to be
vector variables.

The syntax of defining scalar products is the same as in Tracer
\cite{Jamin:1991dp}. The {\ttfamily OnShell}-command
\begin{verbatim}
   OnShell[on,{p1,0},{p2,p3,m},...]
\end{verbatim}
defines the scalar products $\mathtt{p1}\cdot\mathtt{p1} =0$ and 
$\mathtt{p2}\cdot \mathtt{p3} = \mathtt{m}$. By default there are the
following definitions:
\begin{verbatim}
   OnShell[on,{p,0},{q,0},{l,0},{p,q,1/2},{p,l,xi/2},{q,l,theta/2}]
\end{verbatim}
To undo the onshell definition use the flag
{\ttfamily off} instead of {\ttfamily on}.

An integral of the form (\ref{ibp1}) contains the set of momenta
$\{p_1,\ldots,p_n\}$, which are in general linear combinations of
basis momenta e.g.\ $\{0,p^\mu,p^\mu+yq^\mu\}$ where the basis momenta
are $\{p^\mu,q^\mu\}$. To tell Mathematica which variables are the
basis momenta we have to define the variable {\ttfamily MomBasis}. In
our example we set:
$$
\mathtt{  MomBasis=\{p,q\};}
$$
After this definition the function 
$$
\mathtt{ExternalMomenta[}p_1,\ldots,p_n\mathtt{];}
$$
has to be called to tell Mathematica that $p_1,\ldots,p_n$ are
the momenta which appear in the Feynman integrals. In the above
example:
$$
\mathtt{ExternalMomenta[0,p,p+y*q];}
$$

Feynman integrals are represented by the function {\ttfamily FInt}.
This function will be simplified applying rule \ref{ru1}, \ref{ru1.2}
and rule \ref{ru2} of section \ref{ibpmethod}. After the call of the
function
$\mathtt{ExternalMomenta[}p_1,\ldots,p_n\mathtt{]}$
the momentum $p_i$ is represented by the position $i$
at which it appears in the argument list.
So the integral (\ref{ibp1}) is represented by 
\begin{equation}
\begin{split}
\mathtt{FInt[}&
\mathtt{\{\{}
i_1,M_1^2,m_1
\mathtt{\}},
\ldots,
\mathtt{\{}
i_t,M_t^2,m_t
\mathtt{\}\},\{\{}
\tilde{i}_1,\tilde{M}^2_1,\tilde{m}_1
\mathtt{\}},
\ldots,
\mathtt{\{}
\tilde{i}_u,\tilde{M}^2_u,\tilde{m}_u
\mathtt{\}\},}
\\
&
\mathtt{\{\{}
j_1,s_1
\mathtt{\}},
\ldots,
\mathtt{\{}
j_l,s_l
\mathtt{\}\}]}
\end{split}
\nonumber
\end{equation}
Because most integrals do not have propagators of the form $k\cdot
p+M^2$, the second argument of {\ttfamily FInt} can be dropped such that 
$$
\mathtt{FInt[\{\{}i_1,M_1^2,m_1
\mathtt{\}},\ldots,\mathtt{\{}i_t,M_t^2,m_t\mathtt{\}\},\{\{}
j_1,s_1\mathtt{\}},\ldots,\mathtt{\{}j_l,s_l\mathtt{\}\}]}
$$
represents the integral
$$
\intd \frac{s_1^{n_1}\ldots s_l^{n_l}}{D_1^{m_1}\ldots D_t^{m_t}}.
$$
For example: We want to represent the integral
\begin{equation}
\intd \frac{k\cdot(p+yq)}{k^2(k+p)^2(k+p+yq)^2}.
\label{man1}
\end{equation}
After the above call of {\ttfamily ExternalMomenta[0,p,p+y*q]}
the momenta {\ttfamily 0, p, p+y*q} are represented by the numbers 1,
2 and 3 respectively. So (\ref{man1}) is represented by 
$$
\mathtt{
FInt[\{\{1,0,1\},\{2,0,1\},\{3,0,1\}\},\{\{3,1\}\}]
}
$$
which is transformed into 
$$
\mathtt{
-1/2*y*FInt[\{\{1,0,1\},\{2,0,1\},\{3,0,1\}\},\{\}]
}
$$
because rule \ref{ru1} and rule \ref{ru2} are applied and scaleless
integrals vanish in dimensional regularisation.

The identities from rule \ref{ru3} to rule \ref{ru5} are created by
the function {\ttfamily IBP}. This function takes three or five
arguments and is called by 
$$\mathtt{
IBP[Denom1, Denom2, la1, la2, lb]
}
$$
or 
$$
\mathtt{
IBP[Denom, la, lb]
}
$$  
which is a shortcut for
$$
\mathtt{
IBP[Denom, \{\}, la, \{\}, lb]
}.
$$
The first two arguments of {\ttfamily IBP} 
{\ttfamily Denom1} and {\ttfamily Denom2} are
lists that take the form $\{\{i_1,M_1^2\},\ldots,\{i_t,M_t^2\}\}$ and  
$\{\{\tilde{i},\tilde{M}^2_1\},\ldots,\{\tilde{i}_u,\tilde{M}^2_u\}\}$
respectively and describe the topology i.e.\ they tell Mathematica to create
identities of integrals which contain the propagators $D_1,\ldots,D_t$
and $\tilde{D}_1,\ldots,\tilde{D}_u$ respectively. The powers $m_i$ and
$\tilde{m}_i$ to which the propagators $D_i$ and $\tilde{D}_i$ have to
appear, are given by the next two arguments {\ttfamily la1} and
{\ttfamily la2}. These are lists whose elements are of the form
$\mathtt{\{}n,m\mathtt{\}}$: For all integrals with $n$ different propagators
of the form $D$ and $\tilde{D}$ respectively 
identities of the form (\ref{ibp5}), (\ref{ibp6}) and (\ref{ibp8}) are 
created for all integrals, where $\sum_{k = 1}^n (m_k-1) \le m$
and $\sum_{k = 1}^n (\tilde{m}_k-1) \le m$ respectively. The third argument
{\ttfamily lb} has the same form as {\ttfamily la1} and {\ttfamily
la2} and tells the program how many scalar products of the form
$k\cdot p$ should be in the numerator. Here $n$ stands for the number
of different propagators of the form $D$ \emph{and} $\tilde{D}$ and
the integrands have to fulfil the condition
$\sum_{k=1}^l n_{j_k} \le m$.
For all of the lists {\ttfamily la1}, {\ttfamily la2} and
{\ttfamily lb} the default value for $m$ is 0 e.g.\ {\ttfamily
\{\{2,1\},\{3,0\}\}} and {\ttfamily \{\{2,1\}\}} lead to the same
result.
  
In our example a
convenient call of {\ttfamily IBP} would be 
$$\mathtt{
subslist = IBP[\{\{1,0\},\{2,0\},\{3,0\}\},\{\{2,1\},\{3,0\}\},\{\{2,1\},\{3,0\}\}];
}
$$
which is equivalent
$$\mathtt{
subslist = IBP[\{\{1,0\},\{2,0\},\{3,0\}\},\{\{2,1\}\},\{\{2,1\}\}];
}
$$
because the default value of the powers of propagators is 0.
The command 
$$\mathtt{
FInt[\{\{1,0,1\},\{2,0,1\},\{3,0,1\}\},\{\{3,1\}\}]//.subslist
}
$$
reduces our integral to 
$$\mathtt{
-(1-2*e)*FInt[\{\{1,0,1\},\{3,0,1\}\},\{\}]/2/e
}
$$
where $\mathtt{e}=(4-d)/2$.  
 
\section{Implementation}
In this section I will describe in detail how to implement the rules
of section \ref{ibpmethod} in Mathematica . I will follow the
algorithm  described in \cite{Laporta:2001dd}, the reader is expected to be
familiar with this algorithm.  

First of all we have to tell our 
computer algebra system how to handle Lorentz vectors. We will write
all the definitions into the file {\ttfamily lorentz.m}. It proves to 
be useful to distinguish between Lorentz vectors and scalar
variables. In a CAS which does not know about type declarations of
variables  this is done by putting all the vector variables in a 
list we call {\ttfamily MomList}. So the first part of the file 
{\ttfamily lorentz.m} looks like this:

\begin{verbatim}
BeginPackage[ "LORENTZ`" ]; 

(*Pattern variables*)
Unprotect[a,b,c,mom,mom1,mom2,ip,rest];
Clear[a,b,c,mom,mom1,mom2,ip,rest];
Protect[a,b,c,mom,mom1,mom2,ip,rest];


Unprotect[d,e];
Clear[d,e];

d = 4 - 2*e;(*Dimension*)

Protect[d,e];


Unprotect[MomList]; MomList := {}; 
(*List of variables which are defined to be momenta; 
all other variables are handled as scalars*)
Protect[MomList];

Unprotect[AddMomenta];
Clear[AddMomenta];
AddMomenta[qlist___] :=
  (Unprotect[MomList];
    MomList = Union[MomList, {qlist}];
    Protect[MomList];)
Protect[AddMomenta];

Unprotect[RemMomenta];
Clear[RemMomenta];
RemMomenta[qlist___] :=
  (Unprotect[MomList];
    MomList = Complement[MomList, {qlist}];
    Protect[MomList];)
Protect[RemMomenta];

Unprotect[ClearMomenta];
Clear[ClearMomenta];
ClearMomenta[] :=
  (Unprotect[MomList]; MomList = {}; Protect[MomList];)
Protect[ClearMomenta];

Unprotect[ShowMomenta];
Clear[ShowMomenta];
ShowMomenta[] := Return[MomList];
Protect[ShowMomenta];
\end{verbatim}

After we have defined which variables are used as pattern variables
and we have set the dimension variable {\ttfamily d = 4-2*e}, we
introduce the protected list variable {\ttfamily MomList} which is
manipulated by the functions {\ttfamily AddMomenta}, 
{\ttfamily RemMomenta}, {\ttfamily ClearMomenta} and {\ttfamily
  ShowMomenta}. 

In the next step we define the function {\ttfamily IsVector} which
tells us if a variable is of the type vector (i.e. it is contained in
{\ttfamily MomList}) or scalar:

\begin{verbatim}
Unprotect[IsVector];
Clear[IsVector];
IsVector[a_ + b_] := IsVector[a] || IsVector[b];
IsVector[a_ b_] := IsVector[a] || IsVector[b];
IsVector[a_ /; MemberQ[MomList, a]] := True; (*that's the point*)
IsVector[a_] := False;
Protect[IsVector];
\end{verbatim}

Following the conventions of ``Tracer'' \cite{Jamin:1991dp} we define a
scalar product between Lorentz vectors which gets the attribute
{\ttfamily Orderless}. The scalar product is defined to be linear,
which makes the distinction between scalar and vector variables necessary.

\begin{verbatim}
Unprotect[SP]; 
Clear[SP];
SetAttributes[SP, Orderless]; (*scalarproduct of two Lorentzvectors*);
SP[(a_ /; ! IsVector[a])*b_, c_] := a*SP[b, c]; 
SP[a_ + b_, c_] := SP[a, c] + SP[b, c]; 
SP[0, _] := 0;
Protect[SP];
\end{verbatim}

The function {\ttfamily OnShell} is defined as in \cite{Jamin:1991dp}
i.e.\ we define the scalar products $p\cdot q = \frac{1}{2}$ and
$p^2=0$ by the command {\ttfamily OnShell[on,\{p,0\},\{p,q,1/2\}]}.

\begin{verbatim}
Off[General::spell1];
Unprotect[OnShell];
Clear[OnShell]; 
OnShell[flag_, list___] :=
  (*Defined as in Tracer*)
  Module[{l, i},
    l = {list};
    Unprotect[SP];
    Switch[flag,
      on, For[i = 1, i <= Length[l], i++,
        Switch[Length[l[[i]]],
          2, SP[l[[i]][[1]], l[[i]][[1]]] = l[[i]][[2]],
          3, SP[l[[i]][[1]], l[[i]][[2]]] = l[[i]][[3]];
          ]
        ],
      off, For[i = 1, i <= Length[l], i++,
        Switch[Length[l[[i]]],
          2, SP[l[[i]][[1]], l[[i]][[1]]] =.,
          3, SP[l[[i]][[1]], l[[i]][[2]]] =.;
          ]
        ]
      ];
    Protect[SP];
    ];
Protect[OnShell];
\end{verbatim} 

The last function we introduce is {\ttfamily Project}. This function,
applied to a linear combination of vector variables and a vector
variable {\ttfamily p}, gives the coefficient of {\ttfamily p} in that
linear combination. E.g.\ assume {\ttfamily p} and {\ttfamily q} are
vector variables and {\ttfamily x} and {\ttfamily y} are scalars then 
{\ttfamily Project[x*p+y*q,p]} is simplified to {\ttfamily x}. This is
the definition of {\ttfamily Project}:

\begin{verbatim}
Unprotect[Project];
Clear[Project];
Project[a_+b_,mom_]/;
MemberQ[ShowMomenta[],mom]:=Project[a,mom]+Project[b,mom];
Project[a_*mom1_,mom2_]/;
MemberQ[ShowMomenta[],mom2]&&IsVector[mom1]&&!IsVector[a]:=
a*Project[mom1,mom2];
Project[0,mom_]:=0;
Project[mom_,mom_]/;MemberQ[ShowMomenta[],mom]:=1;
Project[mom1_,mom2_]/;
MemberQ[ShowMomenta[],mom2]&&MemberQ[ShowMomenta[],mom1]:=0;
Protect[Project];
\end{verbatim}

The end of the file {\ttfamily lorentz.m} is special for the 
calculation in this thesis 
i.e.\ we introduce the Lorentz vectors {\ttfamily p}, {\ttfamily q} 
and {\ttfamily l} whose scalar products fulfil our kinematical
conditions:

\begin{verbatim}
(*-------------------other definitions--------------------*)

AddMomenta[p, q, l];

OnShell[on, {p,p,0}, {q,q,0}, {p,q,1/2}, {l,0}, {p,l,xi/2},
{q,l,theta/2}];



(*-------------------end other definitions-----------------*)

EndPackage[]
\end{verbatim}

We will write the definitions which handle the reduction of Feynman
integrals into the file {\ttfamily ibp.m}. The first part of this file
covers the global variables which will be described in more detail
when they will be used. The variable {\ttfamily MomBasis} which has to
be defined by the user contains all the basis momenta i.e.\ all
external momenta which appear in the integrals have to consist of a
linear combination of the components of {\ttfamily MomBasis}.

\begin{verbatim}
(*Patterns*)
Unprotect[Denom,Denom1,Denom2,Num,arg,
      expr1,expr2,ip,i1p,i2p,i3p,i4p,intp,n1p,n2p,Mp,M1p,M2p,
      a,b,c,dp,a1,b1,c1,d1,e1,a2,b2,c2,d2,M1,M2,m1,m2,mp,p1,p2,L,sp];

Clear[Denom,Denom1,Denom2,Num,arg,
      expr1,expr2,ip,i1p,i2p,i3p,i4p,intp,n1p,n2p,Mp,M1p,M2p,
      a,b,c,dp,a1,b1,c1,d1,e1,a2,b2,c2,d2,M1,M2,m1,m2,mp,p1,p2,L,sp];

Protect[Denom,Denom1,Denom2,Num,arg,
      expr1,expr2,ip,i1p,i2p,i3p,i4p,n1p,n2p,intp,Mp,M1p,M2p,
      a,b,c,dp,a1,b1,c1,d1,e1,a2,b2,c2,d2,M1,M2,mp,m1,m2,p1,p2,L,sp];


(*all global variables*)
MomBasis = {p,q,l};

Unprotect[eqlist];
eqlist = {};(*IBP1 writes into eqlist*)
Protect[eqlist];

Unprotect[Mom];
Mom = {};
Protect[Mom];
(*global list of external momenta. 
  This list is used by all other functions.
  This list is set by the function ExternalMomenta (see below),
  do not edit! *)

Unprotect[LinIndepMom];
LinIndepMom = {};
Protect[LinIndepMom];
(*Gives the position of the linearly independent momenta in Mom.
  This list is set by the function ExternalMomenta (see below), 
  do not edit!*)
\end{verbatim}

If we define a set of external momenta, which appear in our integrals,
not all of those will be linearly independent. To find the linearly
independent ones i.e.\ a basis of our external momenta we define the
function {\ttfamily FindBasis}. This function applied on a list of
momenta gives a list of the position of the linearly independent ones.
The algorithm is as follows: Consider a set of vectors
$\{v_1,\ldots,v_n\}$ from which we want to choose a minimal subset of
linearly independent vectors which form a basis of
$\text{span}\{v_1,\ldots,v_n\}$. We look for a most general solution
of the equation system
\begin{equation}
v_i^j c^i = 0.
\label{lineq}
\end{equation}
If one of the $c^i$ is not necessarily $0$ i.e.\ $v_i$ can be expressed
by a linear combination of the other vectors we remove $v_i$ from the
set $\{v_1,\ldots,v_n\}$ and repeat the procedure until (\ref{lineq})
gets the unique solution $c^i=0$ for all $i$. Usually the basis of
$\text{span}\{v_1,\ldots,v_n\}$ is not unique. In this case we have an
ambiguity, which of the vectors we can remove. The function below
removes this vector which comes first in the list $\{v_1,\ldots,v_n\}$
i.e. the vectors which are more behind have a higher precendence to be
a member of the basis which is returned.
This is the implementation in Mathematica:

\begin{verbatim}
Unprotect[FindBasis];
Clear[FindBasis];

FindBasis[{ip_}] := {}/;(Mom[[ip]] == 0);
FindBasis[p_List] :=
    (*gives the position of the linearly independend momenta in p
      reads the variable MomBasis. If this is not unique, a vector
      p[[i1]] has a higher precendence than p[[i2]] iff i1>i2. The
      vector p[[i]] is an integer s.t. Mom[[ p[[i]] ]] is the
      corresponding vector*)
    Catch[
      Module[{pc, Ilist, V, C, c, i, j},
        
        pc = Array[Function[{i}, {Mom[[ p[[i]] ]], i}], Length[p]];
        (*maps every momentum onto a position in the list p*)
        
        Ilist = Range[Length[p]];
        (*result list*)
        
        While[True,
          V = Array[
              Function[{i, j}, Project[pc[[i, 1]], MomBasis[[j]]] ],
              {Length[pc], Length[MomBasis]}];
          C = Array[c, {Length[pc]}];
          
          Off[Solve::svars];
          C = C /. (Solve[C.V == 0, C][[1]]);
          On[Solve::svars];
          
          For[i = 1, i <= Length[C], i++,
            (*Because we start with lowest value i=1 and increase i,
            we make sure that the vectors in the begining of p are
            dropped out first, i.e. p[[i1]] has a higher precendence
            than p[[i2]] iff i1 > i2.*)
            If[! MatchQ[C[[i]], 0],
              Ilist = Drop[Ilist, Position[Ilist, pc[[i, 2]]][[1]]];
              pc = Drop[pc, {i}];
              Break[];
              ];
            If[i >= Length[C], Throw[Ilist]];
            ];
          
          ];
        
        ]
      ];

Protect[FindBasis];
\end{verbatim}
In the next part we define some global functions.
The function {\ttfamily MakeComb} takes the integer
arguments {\ttfamily n} and {\ttfamily k} and gives a list of all 
subsets of {\ttfamily \{1,...,n\}} with length {\ttfamily k}. 
The functions {\ttfamily GetBasis}, {\ttfamily RepMom} and {\ttfamily
ruleI} are well commented such that we do not explain them. These functions 
are defined such that they remember the value they have already
calculated. So there are the commands {\ttfamily defGetBasis},
{\ttfamily defRepMom} and {\ttfamily defruleI} which redefine those
functions and which are evaluated in the function {\ttfamily
ExternalMomenta}. The argument of {\ttfamily ExternalMomenta} is a
list $\{p_1,\ldots,p_n\}$ of the external momenta which appear in the
integrals. By calling the function 
$\mathtt{ExternalMomenta[}p_1,\ldots,p_n\mathtt{]}$ the variable 
$\mathtt{Mom}$ is set to $\{p1,\ldots,p_n\}$ and to the variable
{\ttfamily LinIndepMom} a list of the position of the linearly
independent momenta in $\{p1,\ldots,p_n\}$ is assigned.
 
\begin{verbatim}
Unprotect[MakeComb];
Clear[MakeComb];
MakeComb[n_, k_] :=
    (*generates all subsets of {1, ..., n} with k elements *)
   
     Catch[
      Module[{L0, L1, i, j, l},
          L0 = Array[{#} &, n];(*L0 = {{1}, ..., {n}}*)
          L1 = {};
          For[l = 1, l <= k - 1, l++,
            For[j = 1, j <= Length[L0], j++,
              For[i = Last[L0[[j]]] + 1, i <= n, i++,
                  L1 = Append[L1, Append[L0[[j]], i]];
                  ];
              ];
            L0 = L1;
            L1 = {};
            ];
          Throw[L0];
          ];
      ];
Protect[MakeComb];


Unprotect[partfrac,defpartfrac];
(*This function gives a list {True,{c1,...,cn}} s.t. 
c1+...+cn=0,
c1*Mom[[mom[[1]]]]+...+cn*Mom[[mom[[n]]]]=0,
c1*(mom[1]-mass[1])+...+cn*(mom[n]-mass[n])=1
If this not possible it gives back {False,{}}
*)
Clear[partfrac,defpartfrac];

defpartfrac:=(
partfrac[mom_,mass_]:=
Catch[
Unprotect[partfrac];
Module[{res},
res=(
partfrac[mom,mass]=
Catch[
Module[{coeff,coefflist,solvelist,eql,i1,j1,i},
   Off[Solve::svars];
       i = Length[mom];
       coefflist = Array[coeff,i];
        eql = 
         Array[Function[{j1}, 
            Sum[coeff[i1]*Project[Mom[[ mom[[i1]] ]], MomBasis[[j1]]], 
                {i1, 1, i}] == 0], Length[MomBasis]];
        eql = Append[eql, Sum[coeff[i1], {i1, 1, i}] == 0];
        eql = Append[eql, Sum[coeff[i1]*(
              SP[Mom[[mom[[i1]]]],Mom[[mom[[i1]]]]]-mass[[i1]]
              ),{i1,1,i}] == 1];
        solvelist = Solve[eql, Array[coeff, i]]; 
        If[MatchQ[solvelist,{}],Throw[{False,{}}]];
        solvelist=solvelist[[1]];
        coefflist = coefflist /. solvelist;
        coefflist = coefflist /. coeff[_] -> 0;
   On[Solve::svars];
   Throw[{True,coefflist}];
](*endMod*);
](*endCatch*)
);
Protect[partfrac];
Throw[res];
];(*endMod*)
];(*endCatch*)
);
Protect[partfrac,defpartfrac];

Unprotect[defGetBasis,GetBasis];
Clear[defGetBasis,GetBasis];

defGetBasis := 
( GetBasis[Denom1_List,Denom2_List]:=
  (*Gets Denom1 and Denom2 of the form 
    {{p1,M1^2,m1},...,{pt,Mt^2,mt}} and 
    {{q1,Mq1^2,mq1},....,{qr,Mqr^2,mqr}} resp. and finds the 
    linear independent momenta {q_k1,...,q_kl} of {q1,...,qr} 
    and {p_i1,...,p_il} of {p1,....,pt}, which are completed 
    to a basis s.t. as much as possible vectors of 
    {q_k1,....,q_kl} are in the basis list*)
  
   Catch[
    Unprotect[GetBasis];
    Module[{res, p1,p2,p, bl1,bl2, i},
     p1=Array[ Denom1[[#,1]]&, Length[Denom1] ];
     p2=Array[ Denom2[[#,1]]&, Length[Denom2] ];
     p = Join[p1,p2];
     For[i=1,i<=Length[Mom],i++,
      If[!MemberQ[p,i],p=Prepend[p,i]];
     ];
    
     bl1 = FindBasis[p]; 
     bl2 = Array[ p[[ bl1[[#]] ]]&, Length[bl1] ];
     res = (GetBasis[Denom1,Denom2] = bl2);  
     Protect[GetBasis];
     Throw[res];
    ](*endMod*);
   ](*endCatch*);
);
Protect[defGetBasis,GetBasis];
 
Unprotect[RepMom,defRepMom];
Clear[RepMom,defRepMom];
defRepMom := (
 RepMom[mom_,basis_List] :=
 (*RepMom takes as arguments a momentum p and a basis {p1,...,pk}. 
   It gives a list {c1,...,ck} s.t. p = c1*p1+...+ck*pk *)
  Catch[
   Module[{coeff,coefflist,solvelist,eql, i,i1,i2},
    Unprotect[RepMom];

    Off[Solve::svars];
    eql = Array[ 
           Sum[ 
            coeff[i1]*Project[Mom[[ basis[[i1]] ]], MomBasis[[#]] ],
            {i1,1,Length[basis]} ] ==
           Project[Mom[[mom]],MomBasis[[#]] ]&,
           Length[MomBasis] 
          ];
    eql = eql //. 
            {(expr1_ == expr2_) /; FreeQ[{expr1, expr2}, coeff[_]] :> 
            MatchQ[expr1, expr2]};
	    (*An expression of the form "x==x" in a list of equations 
              is not automatically 
	      transformed into "True": This has to be done by hand*)
    coefflist = Array[  coeff, Length[basis]  ];
    solvelist = Solve[eql, coefflist];
    coefflist = coefflist /. solvelist;
    coefflist = coefflist /. coeff[_] -> 0; 
    On[Solve::svars];

    RepMom[p,basis] = coefflist;
    Throw[ coefflist[[1]] ];

    Protect[RepMom];
    Throw[res];
   ](*endMod*);
  ](*endCatch*);
); 


Unprotect[defruleI,ruleI];
Clear[defruleI,ruleI];
defruleI := (
 ruleI[mom_List]:=
 (* ruleI gives for the topology {i1,...,ik} a list of 
 coefficients (c1,...,ck) such that c1+...+ck = 1 and
 c1*Mom[[i1]]+...+ck*Mom[[ik]] = 0. The output is of the form
 {True,{c1,...,ck}} or {False} if this is not possible *)
  Catch[
   Module[{res, coefflist,coeff,eql, i,j,i1,j1},
    Off[Solve::svars];
    coefflist = Array[coeff, Length[mom]];
    eql = 
     Array[Function[{j1}, 
      Sum[coeff[i1]*Project[Mom[[ mom[[i1]] ]], MomBasis[[j1]]], 
                {i1, 1, Length[mom]}] == 0], Length[MomBasis]];
    eql = Append[eql, Sum[coeff[i1], {i1, 1, Length[mom]}] == 1];
    coefflist = coefflist /. Solve[eql, Array[coeff, Length[mom]]];
    coefflist = coefflist /. coeff[_] -> 0;
    (*Set all the coefficients which remain after solving the 
      system of equations to zero, if there is no solution of 
      this system of equations it follows
      coefflist = {coeff[1], coeff[2], coeff[3]} -> {0, 0, 0}.
      As the linear equation system is inhomogeneous, coefflist 
      can only be substituted to {0, 0, 0} if there is no 
      solution: 
      The momenta are linearly independend or the condition 
      sum coeff[i] = 1 cannot be satisfied*)
    On[Solve::svars];
      
    If[(coefflist //. {0, r___} -> {r}) == {},
     (*first case : k^2 not reducible*)
     res={False};, 
     (*else - part : k^2 reducible*)
     coefflist = coefflist[[1]];
     res={True,coefflist};
    ];(*endif*)
    
    Unprotect[ruleI];
    ruleI[mom]=res;
    (*remember previously calculated values*)
    Protect[ruleI];

    Throw[res];

   ](*endMod*);
  ](*endCatch*);
);
Protect[defruleI,ruleI];

Clear[ExternalMomenta];
ExternalMomenta[L___] :=
    (*sets the variables Mom and LinIndepMom*)
    Module[{i,i1,j,j1,j2,k,eql,coeff,coefflist,
            L1,L2,L3,L4,cmom,base,solvelist},

     Unprotect[Mom, LinIndepMom];
     Clear[Mom,LinIndepMom];
     Mom = {L};
     (*The variable Mom has to be set. Otherwise you are not allowed
     to use the function FindBasis *)
     LinIndepMom = FindBasis[Range[Length[Mom]]];
     Protect[Mom,LinIndepMom];

     Unprotect[partfrac];
     Clear[partfrac];
     defpartfrac;
     (*define the function partfrac*)
     Protect[partfrac];

     Unprotect[GetBasis];
     Clear[GetBasis];
     defGetBasis;
     Protect[GetBasis];

     Unprotect[RepMom];
     Clear[RepMom];
     defRepMom;
     Protect[RepMom];

     Unprotect[ruleI];
     Clear[ruleI];
     defruleI;
     Protect[ruleI];

  ];
Protect[ExternalMomenta]; 
\end{verbatim}

Feynman integrals are represented by the function {\ttfamily
FInt}. This function accesses the function {\ttfamily scaleless} which 
gives true if the topology is
scaleless. In this case {\ttfamily FInt} gives 0. {\ttfamily FInt}
uses the functions {\ttfamily GetBasis}, {\ttfamily RepMom} and
{\ttfamily ruleI} to get
simplified according to rule \ref{ru1}-\ref{ru2}.

\begin{verbatim}
(**********I Passarino Veltman reduction **************)
Unprotect[Scaleless];
Clear[Scaleless];

Scaleless[Denom_List]:=
Catch[
  Module[{pi,pj,i,j},
   For[i=1,i<=Length[Denom],i++,
    If[Denom[[i,2]]!=0,Throw[False]];
    pi=Mom[[ Denom[[i,1]] ]] - Mom[[ Denom[[1,1]] ]];
    For[j=2,j<=Length[Denom],j++,
     pj=Mom[[ Denom[[j]][[1]] ]] - Mom[[ Denom[[1]][[1]] ]];
     If[!MatchQ[Simplify[SP[pi,pj]],0],Throw[False];];
    ];
   ];
   Throw[True];
  ];
];

Scaleless[Denom1_List, Denom2_List]:=
Catch[
  Module[{p1,pi,pj,Mi,i,j},
   If[Length[Denom1]>0,
    p1 = Mom[[ Denom1[[1,1]] ]];
    For[i=1,i<=Length[Denom1],i++,
     If[Denom1[[i,2]]!=0,Throw[False]];
     pi=Mom[[ Denom1[[i,1]] ]] - p1;
     For[j=2,j<=Length[Denom1],j++,
      pj=Mom[[ Denom1[[j,1]] ]] - p1;
      If[!MatchQ[Simplify[SP[pi,pj]],0],Throw[False];];
     ];
     For[j=1,j<=Length[Denom2],j++,
      pj=Mom[[ Denom2[[j,1]] ]];
      If[!MatchQ[Simplify[SP[pi,pj]],0],Throw[False];];
     ];
    ];
   ,(*else part *)p1=0;](*endIf*);
   For[i=1,i<=Length[Denom2],i++,
    pi=Mom[[ Denom2[[i,1]] ]];
 
    If[!MatchQ[tres=Simplify[Denom2[[i,2]]-SP[pi,p1]],0],Throw[False]];
  
    For[j=1,j<=Length[Denom2],j++,
     pj=Mom[[ Denom2[[j,1]] ]];
     If[!MatchQ[Simplify[SP[pi,pj]],0],Throw[False];]; 
    ](*endFor[j]*);
   ](*endFor[i]*);
   

   Throw[True];

  ](*endMod*);
](*endCatch*); 

Protect[Scaleless];


Unprotect[FInt];
Clear[FInt];
FInt[Denom_List,Num_List]:=0/;Scaleless[Denom];
FInt[Denom1_List,Denom2_List,Num_List]:=0/;Scaleless[Denom1,Denom2];
(*scaleless integrals vanish*)

FInt[Denom_List,{},Num_List] := FInt[Denom,Num];

FInt[Denom_List, {a___, {ip_Integer, 0}, b___}] := FInt[Denom, {a, b}];
FInt[Denom1_List, Denom2_List, {a___, {ip_Integer, 0}, b___}] := 
FInt[Denom1, Denom2, {a, b}];
FInt[{a___, {ip_Integer, Mp_, 0}, b___}, Num_List] := FInt[{a, b}, Num];
FInt[Denom_List, {a___, {ip_Integer, Mp_, 0}, b___}, Num_List] := 
FInt[Denom, {a, b}, Num];
FInt[{a___, {ip_Integer, Mp_, 0}, b___}, Denom_List, Num_List] := 
FInt[{a, b}, Denom, Num];

FInt[{a1___,{i1p_Integer,Mp_,i2p_Integer},
     {i1p_Integer,Mp_,i3p_Integer},b1___},
     Num_List]:=
FInt[{a1,{i1p,Mp,i2p+i3p},b1},Num];

FInt[{a1___,{i1p_Integer,Mp_,i2p_Integer},
     {i1p_Integer,Mp_,i3p_Integer},b1___},
     Denom_List, Num_List]:=
FInt[{a1,{i1p,Mp,i2p+i3p},b1}, Denom, Num];

FInt[Denom_List,{a1___,{i1p_Integer,Mp_,i2p_Integer},
     {i1p_Integer,Mp_,i3p_Integer},b1___},
     Num_List]:=
FInt[Denom, {a1,{i1p,Mp,i2p+i3p},b1}, Num];

FInt[Denom_List,{a1___,{i1p_Integer,i2p_Integer},
                {i1p_Integer,i3p_Integer},b1___}]:=
FInt[Denom,{a1,{i1p,i2p+i3p},b1}];

FInt[Denom1_List, Denom2_List, {a1___,{i1p_Integer,i2p_Integer},
                {i1p_Integer,i3p_Integer},b1___}]:=
FInt[Denom1, Denom2, {a1,{i1p,i2p+i3p},b1}];

FInt[{a1___,{i1p_Integer,M1p_,i2p_Integer},
     {i3p_Integer,M2p_,i4p_Integer},b1___},
     Num_List]/; i1p > i3p :=
FInt[{a1,{i3p,M2p,i4p},{i1p,M1p,i2p},b1},Num];

FInt[Denom_List, {a1___,{i1p_Integer,M1p_,i2p_Integer},
     {i3p_Integer,M2p_,i4p_Integer},b1___},
     Num_List]/; i1p > i3p :=
FInt[Denom, {a1,{i3p,M2p,i4p},{i1p,M1p,i2p},b1},Num];

FInt[{a1___,{i1p_Integer,M1p_,i2p_Integer},
     {i3p_Integer,M2p_,i4p_Integer},b1___},
     Denom_List, Num_List]/; i1p > i3p :=
FInt[{a1,{i3p,M2p,i4p},{i1p,M1p,i2p},b1}, Denom, Num];

FInt[Denom_List, 
     {a___, {i1p_Integer, n1p_Integer},
     {i2p_Integer, n2p_Integer}, b___}] /; i1p > i2p :=
FInt[Denom, {a, {i2p, n2p}, {i1p, n1p}, b}];

FInt[Denom1_List, Denom2_List, 
     {a___, {i1p_Integer, n1p_Integer},
     {i2p_Integer, n2p_Integer}, b___}] /; i1p > i2p :=
FInt[Denom1, Denom2, {a, {i2p, n2p}, {i1p, n1p}, b}];







FInt[{{p1_, M1_, m1_}, Denom___}, {a___,{0, n_Integer},b___}] :=
    (*The entry {0, n} in the numerator denotes (k^2)^n in the 
      integrand where k is the integration variable*)
    
    FInt[{Denom}, {a, {0, n - 1}, b}] +
      (M1 - SP[ Mom[[p1]], Mom[[p1]] ])*
        FInt[{{p1, M1, m1}, Denom}, {a, {0, n - 1}, b}] -
      2*FInt[{{p1, M1, m1}, Denom}, {{p1, 1}, a, {0, n - 1}, b}];

FInt[{{p1_, M1_, m1_}, Denom1___}, Denom2_, 
      {a___,{0, n_Integer},b___}] :=
    (*The entry {0, n} in the numerator denotes (k^2)^n in the 
      integrand where k is the integration variable*)
    
    FInt[{Denom1}, Denom2, {a, {0, n - 1}, b}] +
      (M1 - SP[ Mom[[p1]], Mom[[p1]] ])*
        FInt[{{p1, M1, m1}, Denom1}, Denom2, {a, {0, n - 1}, b}] -
      2*FInt[{{p1, M1, m1}, Denom1}, Denom2, {{p1, 1}, a, {0, n - 1}, b}];


FInt[Denom_List, Num_List]:= 
(*Try to expand the denomintor into partial fractions*)
Module[{pf,lth},
Sum[pf[[2,i]]*
FInt[ReplacePart[Denom,Denom[[i,3]]-1,{i,3}],Num],{i,1,lth}]/;
(lth=Length[Denom];
pf=partfrac[Array[Denom[[#,1]]&,lth],Array[Denom[[#,2]]&,lth]];
pf[[1]])
];


(*reduction of the HQET-Propagators*)
FInt[Denom_List,{a___,{ip_Integer,Mp_,mp_Integer},b___},
                {a1___,{ip_Integer,sp_Integer},b1___}]:=
FInt[Denom,{a,{ip,Mp,mp-1},b},{a1,{ip,sp-1},b1}]-
Mp*FInt[Denom,{a,{ip,Mp,mp},b},{a1,{ip,sp-1},b1}];



(*reduction corresponding ruleI*)
FInt[{a___, {pl_Integer, Ml_, ml_Integer}, b___}, 
     {c___, {pl_Integer, nl_Integer}, dp___}] :=
Module[{res,unchanged,coefflist, coeff, denom, num, p, M, i, j, l, rI},
  (*Rule I*)
  res/;
  Catch[
   res=
   Catch[
     l = Length[{a}] + 1; (*Position of pl, Ml, ml; 
         i.e. denom[[l, 1]] = pl etc. with denom see below*)
      
     denom = {a, {pl, Ml, ml}, b};
     num = {c, {pl, nl}, dp};
     p = Array[denom[[#, 1]] &, Length[denom]];
     M = Array[denom[[#, 2]] &, Length[denom]];
      
      
     rI=ruleI[p];
     If[ !rI[[1]] , 
       (*first case : k^2 not reducible*)
       Throw[unchanged];,
       (*else - part : k^2 reducible*)
       Throw[1/2*Sum[(KroneckerDelta[l, j1 ] - rI[[2,j1]])*
             (FInt[
                   ReplacePart[denom, denom[[j1, 3]] - 1, {j1, 3}], 
                   {c, {pl, nl - 1}, dp}]
                 + (M[[j1]] - SP[  Mom[[ p[[j1]] ]], 
                                   Mom[[ p[[j1]] ]]  ])*
                   FInt[denom, {c, {pl, nl - 1}, dp}]),
           {j1, 1, Length[p]}]];
      ];(*endif*)
   ](*endCatch*);
   If[MatchQ[res,unchanged],Throw[False]];
   Throw[True]; 
  ](*endCatch*)
](*endMod*);

FInt[{a___, {pl_Integer, Ml_, ml_Integer}, b___},Denom_List, 
     {c___, {pl_Integer, nl_Integer}, dp___}] :=
Module[{res,unchanged,coefflist, coeff, denom, num, 
        p, M, i, j, l, rI},
  (*Rule I*)
  res/;
  Catch[
   res=
   Catch[
      
     l = Length[{a}] + 1; (*Position of pl, Ml, ml; 
         i.e. denom[[l, 1]] = pl etc. with denom see below*)
      
     denom = {a, {pl, Ml, ml}, b};
     num = {c, {pl, nl}, dp};
     p = Array[denom[[#, 1]] &, Length[denom]];
     M = Array[denom[[#, 2]] &, Length[denom]];
      
      
     rI=ruleI[p];
     If[ !rI[[1]], 
       (*first case : k^2 not reducible*)
       Throw[unchanged];,
        (*else - part : k^2 reducible*)
         Throw[1/2*Sum[(KroneckerDelta[l, j1 ] - rI[[2,j1]])*
               (FInt[
                     ReplacePart[denom, denom[[j1, 3]] - 1, {j1, 3}], 
                     Denom, {c, {pl, nl - 1}, dp}]
                   + (M[[j1]] - SP[  Mom[[ p[[j1]] ]], 
                                     Mom[[ p[[j1]] ]]  ])*
                     FInt[denom, Denom, {c, {pl, nl - 1}, dp}]),
             {j1, 1, Length[p]}]];
        ];(*endif*)
   ](*endCatch*);
   If[MatchQ[res,unchanged],Throw[False]];
   Throw[True]; 
  ](*endCatch*)
](*endMod*);



FInt[Denom1_List,Denom2_List,Num_List]:=
(*decompose the momenta of Num into a unique set of momenta 
  given by Denom2, Denom1 and some futher momenta which complete 
  the momenta of Denom2 and Denom1 to a basis*)
Module[{res,unchanged, base,pn,rep, i,j },
 res/;
 Catch[
  If[MatchQ[Num,{}],Throw[False]];
  res=Catch[
      base = GetBasis[Denom1,Denom2];
      pn = Array[Num[[#,1]]&,Length[Num]];

      For[i = 1, i <= Length[Num], i++,
         If[MemberQ[ base, pn[[i]] ], Continue[] ];
         (*The momentum in the numerator already appears in 
           the corresponding basis*)

         rep = RepMom[pn[[i]],base]; 
           
         Throw[Sum[
               rep[[j]]*
               FInt[ Denom1, Denom2, 
                     Prepend[
                       ReplacePart[Num, Num[[i, 2]] - 1, {i, 2}],
                       {base[[j]],1}]],{j, 1, Length[base]}] ]; 
      ](*endFor[i]*);
      Throw[unchanged];   
   ](*endCatch*);
  If[MatchQ[res,unchanged],Throw[False]];
  Throw[True]; 
 ](*endCatch*)
](*endMod*);


FInt[Denom_List,Num_List]:=
(*decompose the momenta of Num into a unique set of momenta 
  given by Denom and some futher momenta which complete the 
  momenta of Denom to a basis*)
Module[{res,unchanged, base,pn,rep, i,j },
 res/;
 Catch[
  If[MatchQ[Num,{}],Throw[False]];
  res=Catch[
      base = GetBasis[Denom,{}];
      pn = Array[Num[[#,1]]&,Length[Num]];

      For[i = 1, i <= Length[Num], i++,
         If[MemberQ[ base, pn[[i]] ], Continue[] ];
         (*The momentum in the numerator already appears in 
           the corresponding basis*)

         rep = RepMom[pn[[i]],base];    
         Throw[Sum[
               rep[[j]]*
               FInt[ Denom, 
                     Prepend[
                       ReplacePart[Num, Num[[i, 2]] - 1, {i, 2}],
                       {base[[j]],1}]],{j, 1, Length[base]}] ]; 
      ](*endFor[i]*);
      Throw[unchanged];   
   ](*endCatch*);
  If[MatchQ[res,unchanged],Throw[False]];
  Throw[True]; 
 ](*endCatch*)
](*endMod*);
\end{verbatim}

The implementation of (\ref{ibp4}) is given by these two rules:
\begin{verbatim}
(*these two rules should be applied last*)
FInt[{a1___,{p1_Integer, M1_, m1_Integer}, b1___, 
      {p2_Integer, M2_, m2_Integer}, c1___}, 
     {d1___, {p2_Integer, n2_Integer}, e1___}] :=
    Catch[
      Module[{coefflist, denom, i, j},
          denom = {a1,{p1, M1, m1}, b1, {p2, M2, m2}, c1};
                    
          Throw[1/2*(
                     FInt[{a1,{p1, M1, m1}, b1, {p2, M2, m2 - 1}, c1}, 
                          {d1, {p2, n2 - 1}, e1}] -        
                     FInt[{a1,{p1, M1, m1 - 1}, b1, {p2, M2, m2},c1}, 
                          {d1, {p2, n2 - 1}, e1}]
                     + (M2 - M1 + SP[  Mom[[p1 ]] , Mom[[ p1 ]]  ] - 
                        SP[  Mom[[p2]] , Mom[[ p2 ]]  ])
                        *FInt[denom, {d1, {p2, n2 - 1}, e1}]
                     + 2*FInt[denom, {{p1, 1}, d1, {p2, n2 - 1}, e1}])];
          ](*endMod*);
      ](*endCatch*)/;
      MemberQ[GetBasis[{a1,{p1,M1,m1},b1,{p2,M2,m2},c1},{}],p1];

FInt[{a1___,{p1_Integer, M1_, m1_Integer}, b1___, 
      {p2_Integer, M2_, m2_Integer}, c1___}, Denom_,
     {d1___, {p2_Integer, n2_Integer}, e1___}] :=
    Catch[
      Module[{coefflist, denom, i, j},
          denom = {a1,{p1, M1, m1}, b1, {p2, M2, m2}, c1};
                    
          Throw[1/2*(
                     FInt[{a1,{p1, M1, m1}, b1, {p2, M2, m2 - 1}, c1}, 
                          Denom, {d1, {p2, n2 - 1}, e1}] -        
                     FInt[{a1,{p1, M1, m1 - 1}, b1, {p2, M2, m2},c1}, 
                          Denom, {d1, {p2, n2 - 1}, e1}]
                     + (M2 - M1 + SP[  Mom[[p1 ]] , Mom[[ p1 ]]  ] - 
                        SP[  Mom[[p2]] , Mom[[ p2 ]]  ])
                        *FInt[denom, Denom, {d1, {p2, n2 - 1}, e1}]
                     + 2*FInt[denom, Denom, {{p1, 1}, d1, {p2, n2 - 1}, 
                              e1}]
                    )
               ];
          ](*endMod*);
      ](*endCatch*)/;
      MemberQ[GetBasis[{a1,{p1,M1,m1},b1,{p2,M2,m2},c1},Denom],p1];


\end{verbatim}

The next step is to implement rule \ref{ru3}-\ref{ru5}. We define the
function {\ttfamily IBP1} which takes as a starting point an integral
as in (\ref{ibp1}) and is given the two arguments {\ttfamily
  \{\{i1,M1\^{}2,m1\},...,\{it,Mt\^{}2,mt\}\}} and {\ttfamily
  \{\{j1,s1\},...,\{jl,sl\}\}} like {\ttfamily FInt}. For the integral
defined by these arguments the identities (\ref{ibp5}), (\ref{ibp6}) and
(\ref{ibp8}) are generated and written into the variable {\ttfamily
  eqlist}. 

\begin{verbatim}
Unprotect[IBP1];
Clear[IBP1];
IBP1[Denom_List, Num_List] := IBP1[Denom, {}, Num];

IBP1[Denom1_List, Denom2_List, Num_List] :=    
    Module[{expr, s, t, l, p1, p2, M1, M2, m1, m2, r, n, 
            i1, i2, i, j, idl},
      
      Unprotect[eqlist];
      
      p1 = Array[Denom1[[#, 1]] &, Length[Denom1]];
      M1 = Array[Denom1[[#, 2]] &, Length[Denom1]];
      m1 = Array[Denom1[[#, 3]] &, Length[Denom1]];
      p2 = Array[Denom2[[#, 1]] &, Length[Denom2]];
      M2 = Array[Denom2[[#, 2]] &, Length[Denom2]];
      m2 = Array[Denom2[[#, 3]] &, Length[Denom2]];
      n = Array[Num[[#, 2]] &, Length[Num]];
      s = Sum[Num[[i, 2]], {i, 1, Length[Num]}];
      r = Sum[Denom1[[i, 3]], {i, 1, Length[Denom1]}];
      t = Length[Denom1];
      l = Length[Num];
      
      
      
      (* Identity I *)
      expr = (Collect[(d + s - 2*r)*FInt[Denom1, Denom2, Num] -
                  Sum[2 *m1[[i]]*(
                    (M1[[i]] - SP[Mom[[p1[[i]]]], Mom[[p1[[i]]]]])*
                        FInt[ReplacePart[Denom1, 
                                Denom1[[i, 3]] + 1, {i, 3}], 
                             Denom2, Num]- 
                      FInt[ReplacePart[Denom1, 
                             Denom1[[i, 3]] + 1, {i, 3}], 
                           Denom2, Append[Num, {p1[[i]], 1}]]),
                   {i, 1, t}]-
                  Sum[m2[[i]]*
                      FInt[Denom1,ReplacePart[Denom2,
                                    Denom2[[i,3]]+1,{i,3}],
                           Append[Num,{p2[[i]],1}]],
                   {i,1,Length[Denom2]}],
              FInt[___]]
          );
      eqlist = {expr};
      
      (* Identity II *)
      For[i1 = 1, i1 <= Length[LinIndepMom], i1++,
        i = LinIndepMom[[i1]];
        expr = 
            (Collect[
              Sum[  
                  n[[j]]*SP[ Mom[[i]], Mom[[ Num[[j, 1]] ]] ]*
                    FInt[Denom1, Denom2,
                      ReplacePart[Num, Num[[j, 2]] - 1, {j, 2}]], 
               {j, 1, l}]-
              Sum[                 
                  2*m1[[j]]*
                  (SP[ Mom[[i]], Mom[[ p1[[j]] ]] ]*
                   FInt[ReplacePart[Denom1, Denom1[[j, 3]] + 1, {j, 3}], 
                        Denom2, Num] + 
                   FInt[ReplacePart[Denom1, Denom1[[j, 3]] + 1, {j, 3}], 
                        Denom2, Append[Num, {i, 1}]]), 
               {j, 1, t}]-
              Sum[m2[[j]]*SP[ Mom[[i]], Mom[[p2[[j]] ]] ]*
                  FInt[Denom1,
                       ReplacePart[Denom2, Denom2[[j, 3]] + 1, {j, 3}],
                       Num],
               {j, 1, Length[Denom2]}],
              FInt[___]]);
        eqlist = Append[eqlist, expr];
        ];(*endFor[i1]*)
      
      (* Identity III *);
      idl = MakeComb[Length[LinIndepMom], 2];
      For[i = 1, i <= Length[idl], i++,
        i1 = LinIndepMom[[ idl[[i, 1]] ]];
        i2 = LinIndepMom[[ idl[[i, 2]] ]];
        expr = (Collect[
              Sum[2*n[[j]]*(SP[Mom[[Num[[j, 1]]]], Mom[[i2]]]*   
                          FInt[Denom1, Denom2, 
                               Append[
                                ReplacePart[Num, 
                                  Num[[j, 2]] - 1, {j, 2}], 
                                {i1,  1}]] -
                          SP[Mom[[Num[[j, 1]]]], Mom[[i1]]]*    
                          FInt[Denom1, Denom2, 
                            Append[
                             ReplacePart[Num, 
                               Num[[j, 2]] - 1, {j, 2}], 
                             {i2, 1}]]),
               {j, 1, Length[Num]}]-
              
              Sum[4*m1[[j]]*(
                         SP[Mom[[Denom1[[j, 1]]]], Mom[[i2]]]*  
                         FInt[ReplacePart[Denom1, 
                                Denom1[[j, 3]] + 1, {j, 3}], 
                           Denom2, Append[Num, {i1, 1}]] -
              
                         SP[Mom[[Denom1[[j, 1]]]], Mom[[i1]]]*   
                         FInt[ReplacePart[Denom1, 
                                Denom1[[j, 3]] + 1, {j, 3}], 
                           Denom2, Append[Num, {i2, 1}]]),
               {j, 1, Length[Denom1]}] - 
              
              Sum[2*m2[[j]]*(
                         SP[Mom[[Denom2[[j, 1]]]], Mom[[i2]]]*
                         FInt[Denom1,
                           ReplacePart[Denom2, 
                             Denom2[[j, 3]] + 1, {j, 3}], 
                           Append[Num, {i1, 1}]] -
                         SP[Mom[[Denom2[[j, 1]]]], Mom[[i1]]]* 
                         FInt[Denom1,
                           ReplacePart[Denom2, 
                             Denom2[[j, 3]] + 1, {j, 3}], 
                           Append[Num, {i2, 1}]]),
               {j, 1, Length[Denom2]}],
            FInt[___]]
            );
        eqlist = Append[eqlist, expr];
        ](*endFor[i]*);
      
      
      Protect[eqlist];
      ](*endModule*);
Protect[IBP1];
\end{verbatim}

The function {\ttfamily IBP} follows the algorithm of \cite{Laporta:2001dd} to
generate a list of the IBP identities. After generation a new set of
identities by calling {\ttfamily IBP1} it replaces more complex
integrals by less complex ones. The complexity of integrals is defined
in \cite{Laporta:2001dd}.

\begin{verbatim}
Unprotect[Hf1];
Clear[Hf1];
Hf1[1, Mp_] := {{Mp}};
Hf1[l_, 0] := {Array[0 &, l]};
Hf1[l_, Mp_] :=
   (*generate all l - tuples {n1, ..., nl} s.t. n1 + ... + nl = Mp *)
  
      Catch[
      Module[{L1, L2, i, j},
          
          L2 = {};
          For[i = 0, i <= Mp, i++,
            L1 = Hf1[l - 1, Mp - i];
            For[j = 1, j <= Length[L1], j++,
              L2 = Append[L2, Join[{i}, L1[[j]]]];
              ];
            ];
          
          Throw[L2];
          ](*endMod*);
      ](*endCatch*);
Protect[Hf1];



Unprotect[BT];
Clear[BT];
BT[L1_List, L2_List] :=
    (*give an ordering to lists of integers*)
    
    Catch[
      Module[{i},
          For[i = 1, i <= Min[Length[L1], Length[L2]], i++,
            If[L1[[i]] > L2[[i]], Throw[True]];
            If[L1[[i]] < L2[[i]], Throw[False]];
            ];
          If[Length[L1] > Length[L2], Throw[True]];
          Throw[False];
          ];
      ];
Protect[BT];


Unprotect[Verbose];
Clear[Verbose];
Unprotect[verboseflag];
verboseflag=False;
Verbose[flag_]:=
(
Unprotect[verboseflag];
If[MatchQ[flag,on],verboseflag=True;];
If[MatchQ[flag,off],verboseflag=False;];
Protect[verboseflag]
);
Protect[verboseflag];


Protect[Verbose];

Unprotect[IBP];
Clear[IBP];

IBP[Denom_List, la_List, lb_List] :=
IBP[Denom, {}, la, {}, lb];

IBP[Denom1_List, Denom2_List, la1_List, la2_List, lb_List] :=
  Catch[
    Module[{dummy, i1, i2, j, k, l, m1, m2, n1, n2,
           Mp, Mp2, Md1, Md2, si, subslist, 
           cl1, cl2, hl, delist1, delist2, 
           momlist, num, denom1, denom2, maxf, 
           subsrule, a1, a2, b},
      
      For[i = 0, i <= Length[Denom1], i++,
        a1[i] = 0;
      ];(*Default value for a1*)

      For[i = 0, i <= Length[Denom1]+Length[Denom2], i++,
       b[i] = 0;
      ];(*Default value for b*)

      For[i = 0, i <= Length[Denom2], i++,
        a2[i] = -1;
      ];
      a2[0] = 0;
      (*Default value for a2*)
      
      
      For[i = 1, i <= Length[la1], i++,
        a1[la1[[i, 1]]] = la1[[i, 2]];
        ];
      
      For[i = 1, i <= Length[la2], i++,
        a2[la2[[i, 1]]] = la2[[i, 2]];
        ];
      
      For[i = 1, i <= Length[lb], i++,
        b[lb[[i, 1]]] = lb[[i, 2]];
        ];

      subslist = {}; (*FInt[ ...] -> ...*);
      For[n1 = 0, n1 <= Length[Denom1], n1++,
       cl1 = MakeComb[Length[Denom1], n1]; 

       For[n2 = 0, n2 <= Length[Denom2], n2++,
        cl2 = MakeComb[Length[Denom2], n2];
        
        For[i1 = 1, i1 <= Length[cl1], i1++,
         For[i2 = 1, i2 <= Length[cl2], i2++,
          
          For[Mp = 0, Mp <= b[n1+n2], Mp++,
           momlist = Hf1[Length[Mom], Mp];
          
           For[l = 1, l <= Length[momlist], l++,
            
            num = Array[{#, momlist[[l,#]]} &, Length[momlist[[l]] ] ];
            num = num//.{r1___,{i1_Integer,0},r2___} -> {r1,r2};
             
            For[Md1 = 0, Md1 <= a1[n1], Md1++,
             delist1 = Hf1[n1, Md1];
             For[Md2 = 0, Md2 <= a2[n2], Md2++,
              delist2 = Hf1[n2, Md2];
              For[m1 = 1, m1 <= Length[delist1], m1++,
               denom1 = 
                 Array[Join[Denom1[[ cl1[[i1,#]] ]], 
                            {1+delist1[[m1, #]]}] &,  n1];
               For[m2 = 1, m2 <= Length[delist2], m2++,
                denom2 = 
                 Array[Join[Denom2[[ cl2[[i2,#]] ]], 
                            {1+delist2[[m2, #]]}] &,  n2];

                (*Step 8*)
		If[(MatchQ[denom2,{}]&&
                    MatchQ[FInt[denom1,num]/.FInt->FIntin,
                           FIntin[denom1,num]])||
                   MatchQ[ FInt[denom1,denom2,num]/.FInt->FIntin,
                           FIntin[denom1,denom2,num] ],
		   If[verboseflag,Print[denom1,denom2,num]];
                   IBP1[denom1,denom2,num],
                   Continue[] ]; 
                (*Create IBP-identities from topologies, which cannot
		  be reduced by passarino veltman. *)

		Unprotect[eqlist];
		(*Step 9(a)*)
                For[j = 1, j <= Length[eqlist], j++,
                 eqlist[[j]] = eqlist[[j]]//.subslist;
                 eqlist[[j]] = Collect[eqlist[[j]], 
                                       HoldPattern[FInt[___]],
                                       Expand[ Together[#] ]&];
                 (*substitude all the known identities into the new 
                   IBP identities *) 

                 hl = {};
                 eqlist[[j]] //. {FInt[expr___] :> (
                  hl = Append[hl, FInt[expr]];
                  dummy)};
                 (*extract all the Feynmanintegrals from eqlist[[j]]
                   and write them into hl *)
                 If[MatchQ[hl,{}],Continue[]];

                 (*Step 9(b) *)
                 
                 hl = hl /. {FInt[De1_, De2_, Nu_] :> 
                   FIntin[De1, De2, Nu,
                        Join[{Length[De1]+Length[De2]},
                             {Sum[ De1[[k,3]], {k,1,Length[De1]}]+
                              Sum[ De2[[k,3]], {k,1,Length[De2]}]},
                             {Sum[ De2[[k,3]], {k,1,Length[De2]}]},
                             {Sum[ Nu[[k,2]], {k,1,Length[Nu]}]},
                             Array[De1[[#,1]]&,Length[De1]],
                             Array[De1[[#,3]]&,Length[De1]],
                             Array[De2[[#,1]]&,Length[De2]],
                             Array[De2[[#,3]]&,Length[De2]],
		 	     Array[Nu[[#,2]]&,Length[Nu]]] ]};
         
                 hl = hl /. {FInt[De_, Nu_] :> 
                   FIntin[De, Nu,
                        Join[{Length[De]},
                             {Sum[ De[[k,3]], {k,1,Length[De]}]},
                             {0},
                             {Sum[ Nu[[k,2]], {k,1,Length[Nu]}]},
                             Array[De[[#,1]]&,Length[De]],
                             Array[De[[#,3]]&,Length[De]],
		 	     Array[Nu[[#,2]]&,Length[Nu]]] ]};
                 (*Give to the Feynman integrals a specific weight
                   which measures the complexity of the integral*)
                 
                 maxf = 1;(*hl[[maxf]] will be the Feynmanintegral
                            with the highest complexity*)
                 For[ k=2, k<=Length[hl], k++,
                  If[BT[hl[[k]]/.FIntin[___,arg_]->arg,
                        hl[[maxf]]/.FIntin[___,arg_]->arg],
                     maxf=k](*endIf*);
                 ](*endFor[k]*);        
                 
                 hl[[maxf]] = hl[[maxf]]/.FIntin[De___,_]:>
                                          FInt[De];

                 (*Step 9(c) *)
                 If[Coefficient[eqlist[[j]],hl[[maxf]]]==0,
                  Throw[{eqlist[[j]],hl[[maxf]]}]](*endif*);
		 subsrule = {hl[[maxf]] -> 
                   Collect[
                           (-eqlist[[j]]/.hl[[maxf]]->0)/
			   Coefficient[eqlist[[j]],hl[[maxf]]],
			   HoldPattern[FInt[___]],Expand[Together] ]};
	         subslist = Join[subsrule,subslist];

                ](*endFor[j]*);  
 
               ](*endFor[m2]*);
              ](*endFor[m1]*);
             ](*endFor[Md2]*);
            ](*endFor[Md1]*); 
           
           ](*endFor[l]*);
          ](*endFor[Mp]*);

         ](*endFor[i2]*);
        ](*endFor[i1]*);
        
       ](*endFor[n2]*);
      ](*endFor[n1]*);
      Throw[subslist];
      ](*endMod*);
    ](*endCatch*);
Protect[IBP];
\end{verbatim}

\chapter{Master integrals}
\section{Integrals with up to three external lines}
\label{mi}
In this section I give explicit expression for the one-, two- and
three-point master integrals, which occur in my calculations. 
They are calculated in $d=4-2\epsilon$ dimensions.

There remains only one nonzero one-point integral:
\begin{equation}
A_0\equiv
\mu^{2\epsilon}\intd\frac{1}{k^2-m^2}=\ifp\Gamma(1+\epsilon)
\left(\frac{4\pi\mu^2}{m^2}\right)^\epsilon
m^2\left(\frac{1}{\epsilon}+1+\epsilon+\Op(\epsilon^2)\right)
\label{A0}
\end{equation}

The two-point integrals are:
\begin{eqnarray}
B_{s1}(x,y)&\equiv&
\mu^{2\epsilon}\intd\frac{1}{k^2(k+xp+yq)^2}
\nonumber\\
&=& \ifp\Gamma(1+\epsilon)
(4\pi\mu^2)^\epsilon
\bigg[\frac{1}{\epsilon}+2-\ln x-\ln y+i\pi+
\nonumber\\
&&\epsilon\big(4-\frac{2\pi^2}{3}-2\ln x-2\ln y
+\frac{1}{2}\ln^2x+\frac{1}{2}\ln^2y+\ln x\ln y
\nonumber\\
&&+i\pi(2-\ln x-\ln y)\big)+\Op(\epsilon^2)\bigg]
\label{Bs10}
\end{eqnarray}
\begin{eqnarray}
B_{s1}(x,y,\xi,\theta)&\equiv&
\mu^{2\epsilon}\intd\frac{1}{k^2(k+xp+yq-l)^2}
\nonumber\\
&=& \ifp\Gamma(1+\epsilon)
(4\pi\mu^2)^\epsilon
\bigg[\frac{1}{\epsilon}+2-\ln(xy-x\xi-y\theta)+i\pi+
\nonumber\\
&&\epsilon\big(4-\frac{2\pi^2}{3}
+\frac{1}{2}\ln^2(xy-x\xi-y\theta)-
2\ln(xy-x\xi-y\theta)
\nonumber\\
&&+i\pi(2-\ln(xy-x\xi-y\theta))\big)+\Op(\epsilon^2)\bigg]
\label{Bs1}
\end{eqnarray}
\begin{eqnarray}
B_{s2}(x,y)&\equiv&\mu^{2\epsilon}\intd\frac{1}{k^2(k+xp-yq)^2}
\nonumber\\
&=& \ifp\Gamma(1+\epsilon)(4\pi\mu^2)^\epsilon
\bigg[\frac{1}{\epsilon}+2-\ln x-\ln y+
\nonumber\\
&&\epsilon\big(4-\frac{\pi^2}{6}-2\ln x-2\ln y
+\frac{1}{2}\ln^2x+\frac{1}{2}\ln^2y+\ln x\ln y
\big)
\nonumber\\
&&+\Op(\epsilon^2)\bigg]
\label{Bs2}
\end{eqnarray}
\begin{eqnarray}
B_m(x,y) &\equiv&
\mu^{2\epsilon}\intd\frac{1}{(k^2+2k\cdot(p+q))(k+xp+yq)^2}
\nonumber\\
&=&\ifp\Gamma(1+\epsilon)(4\pi\mu^2)^\epsilon
\bigg[\frac{1}{\epsilon}+2+
\frac{(x+y-xy)\ln(x+y-xy)}{\bar{x}\bar{y}}+
\nonumber\\
&&\epsilon\big(4+2\frac{(x+y-xy)\ln(x+y-xy)}{\bar{x}\bar{y}}
-\frac{(x+y-xy)\ln^2(x+y-xy)}{\bar{x}\bar{y}}
\nonumber\\
&&-\frac{(x+y-xy)\li(\bar{x}\bar{y})}{\bar{x}\bar{y}}
\big)+\Op(\epsilon^2)\bigg]
\label{Bm0}
\end{eqnarray}
\begin{eqnarray}
B_m(x,y,\xi,\theta) &\equiv&
\mu^{2\epsilon}\intd\frac{1}{(k^2+2k\cdot(p+q-l))(k+xp+yq-l)^2}
\nonumber\\
&=&\ifp\Gamma(1+\epsilon)(4\pi\mu^2)^\epsilon
\bigg[\frac{1}{\epsilon}+2-
\frac{(1-\xi-\theta)\ln(1-\xi-\theta)}
{\bar{x}\bar{y}}
\nonumber\\
&&+\frac{(x+y-xy-\xi-\theta)\ln(x+y-xy-\xi-\theta)}
{\bar{x}\bar{y}}
\nonumber\\
&&+\epsilon\bigg(
4-2\frac{1-\xi-\theta}{\bar{x}\bar{y}}\ln(1-\xi-\theta)+
\frac{1-\xi-\theta}{2\bar{x}\bar{y}}\ln^2(1-\xi-\theta)
\nonumber\\
&&+2\frac{x+y-xy-\xi-\theta}{\bar{x}\bar{y}}\ln(x+y-xy-\xi-\theta)
\nonumber\\
&&-\frac{x+y-xy-\xi-\theta}{2\bar{x}\bar{y}}
\ln^2(x+y-xy-\xi-\theta)
\nonumber\\
&&+\frac{x+y-xy-\xi-\theta}{\bar{x}\bar{y}}
\li\frac{\bar{x}\bar{y}}{-x-y+xy-\xi-\theta}
\bigg)\nonumber\\
&&+\mathcal{O}(\epsilon^2)\bigg]
\label{Bm}
\end{eqnarray}
Note that (\ref{Bs10}) and (\ref{Bm0}) are the leading power of
(\ref{Bs1}) and (\ref{Bm}) resp.

The three-point integrals are:
\begin{eqnarray}
C_1(x,y)&\equiv&\intd\frac{1}{k^2(k+xp+yq)^2(k^2+2k\cdot(p+q))}
\nonumber\\
&=&-\ifp\frac{1}{x-y}\bigg[\li\frac{y(x-1)}{x}-\li\frac{x(y-1)}{y}
+\li(\bar{x})-\li(\bar{y})+
\nonumber\\
&&i\pi(\ln x-\ln y)+\Op(\epsilon)\bigg]
\label{C1}
\end{eqnarray}
\begin{eqnarray}
C_2(x,\xi)&\equiv&\intd\frac{1}{k^2(k+xp-\xi q)^2(k^2+2k\cdot(p+q))}
\nonumber\\
&=&-\ifp\frac{1}{x}\bigg[-\li(x)+\frac{2\pi^2}{3}+\frac{1}{2}\ln^2x
-\ln x\ln\bar{x}+\frac{1}{2}\ln^2\xi-\ln\xi\ln x+
\nonumber\\
&&\Op(\xi)+\Op(\epsilon)\bigg]
\label{C2}
\end{eqnarray}
\begin{eqnarray}
C_3(x,y,\theta)&\equiv&\intd\frac{1}{k^2(k+\theta p+yq)^2(k+xp+q)^2}
\nonumber\\
&=&\ifp\frac{1}{xy}\left(
\ln\theta\ln\bar{y}+\ln y\ln\bar{y}-\ln x\ln\bar{y}+2\li y\right)
\nonumber\\
&&+\Op(\theta)+\Op(\epsilon)
\label{C3}
\end{eqnarray}
\begin{eqnarray}
C_4(x,y,\xi) &\equiv& \intd\frac{1}{(k-xp)^2(k-p+\xi q)^2(k+yq)^2}
\nonumber\\
&=&\ifp\frac{1}{\bar{x}y}\left(
\ln x\ln\bar{x}+\ln\xi\ln x-\ln x\ln y+2\li(\bar{x})+i\pi\ln x\right)
\nonumber\\
&&+\Op(\xi)+\Op(\epsilon)
\label{C4}
\end{eqnarray}
\begin{eqnarray}
C_5(x,y,\theta) &\equiv& \intd\frac{1}{(k+xp)^2(k+(x-\theta)p+yq)^2
(k^2+2k\cdot((x-\theta)p+q))}
\nonumber\\
&=&\ifp\frac{1}{\bar{x}y}\bigg[
-\frac{\pi^2}{6}+\ln^2x+\ln y\ln\bar{y}-\frac{\ln^2 y}{2}
-\ln x\ln(x+y-xy)
\nonumber\\
&&+\ln y\ln(x+y-xy)-
\frac{\ln^2(x+y-xy)}{2}-\ln x\ln\theta
\nonumber\\
&&+\ln\theta\ln(x+y-xy)
-\li\frac{x(y-1)}{y}+
\li(y)+
\li\frac{y(x-1)}{x}
\bigg]
\nonumber\\
&&+\Op(\theta)+\Op(\epsilon)
\label{C5}
\end{eqnarray}
\section{Massive four-point integral}
\label{4pointmass}
We consider the following massive four-point integral in
$d=4-2\epsilon$ dimensions (fig.~\ref{4pointfig}):
\begin{figure}
\begin{center}
\resizebox{0.5\textwidth}{!}{\includegraphics{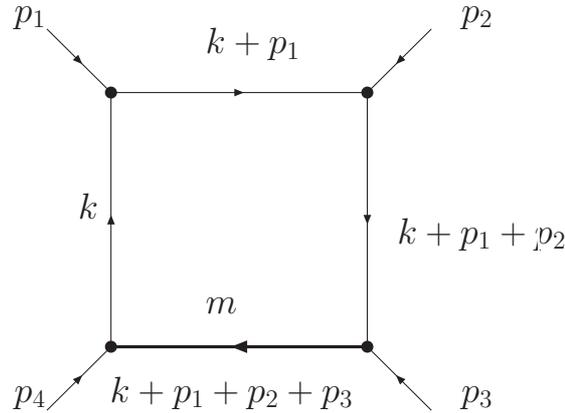}}
\end{center}
\caption{Basic one-loop four-point intergral. The massive line, which
carries the mass $m$, is indicated by the thick line.}
\label{4pointfig}
\end{figure}
\begin{equation}
I_4(p_1,p_2,p_3,p_4)=\mu^{2\epsilon}\intd
\frac{1}{D_1D_2D_3D_4}
\label{4p1}
\end{equation}
where
\begin{eqnarray}
D_1&=&k^2+i\eta
\nonumber\\
D_2&=&(k+p_1)^2+i\eta
\nonumber\\
D_3&=&(k+p_1+p_2)^2+i\eta
\nonumber\\
D_4&=&(k+p_1+p_2+p_3+p_4)^2-m^2+i\eta
\label{4p2}
\end{eqnarray}
Following \cite{Duplancic:2000sk} we introduce the external masses
\begin{equation}
p_i^2=m_i^2\quad (i=1,2,3,4)
\label{4p3}
\end{equation}
and the Mandelstam variables
\begin{equation}
s=(p_1+p_2)^2,\quad t=(p_2+p_3)^2.
\label{4p4}
\end{equation}
Furthermore we consider only the case, where 
\begin{equation}
m_2^2=0\quad \text{and}\quad m_4^2=m^2.
\label{4p5}
\end{equation}

The integral (\ref{4p1}) can be evaluated using the method of 
\cite{Duplancic:2000sk}. 
This paper gives explicit expressions for massless one-loop box
integrals. It is however possible to extend the single steps of this
paper to our case.

So finally we obtain: 
\begin{equation}
\begin{split}
&I_4(p_1,p_2,p_3,p_4)\equiv I_4(s,t,m_1^2,m_3^2,m^2)
=\frac{i}{(4\pi)^2}
\frac{\Gamma(1+\epsilon)(4\pi\mu^2)^\epsilon}{m^2(s-m_1^2)-st+m_1^2m_3^2}
\\
&\begin{split}
\quad\times\Bigg[&\frac{1}
{\epsilon}\left(\ln(-s-i\eta)+\ln(m^2-t-i\eta)
-\ln(m^2-m_3^2-i\eta)-\ln(-m_1^2-i\eta)\right)
\\
&+\ln^2(m^2-m_3^2-i\eta)+\ln^2(-m_1^2-i\eta)-\ln^2(-s-i\eta)
-\ln^2(m^2-t-i\eta)
\\
&+\ln(m^2-i\eta)\left(\ln(-s-i\eta)+\ln(m^2-t-i\eta)
-\ln(m^2-m_3^2-i\eta)-\ln(-m_1^2-i\eta)\right)
\\
&+2\li\left(1-\frac{m^2-t-i\eta}{-m_1^2-i\eta}\right)
-2\li\left(1-\frac{m^2-m_3^2-i\eta}{-s-i\eta}\right)
\\
&+2\li\left(1-(m_3^2-m^2+i\eta)f^m\right)
+2\li\left(1-(m_1^2+i\eta)f^m\right)
\\
&-2\li\left(1-(t-m^2+i\eta)f^m\right)
-2\li\left(1-(s+i\eta)f^m\right)
\Bigg],
\end{split}
\end{split}
\label{4p6}
\end{equation}
where
$f^m=\frac{s+t-m_1^2-m_3^2}{m^2(m_1^2-s)+st-m_1^2 m_3^2}$.

The case $m_1^2=0$ gives rise to further divergences and has to be
considered separately:
\begin{equation}
\begin{split}
&I_4(s,t,m_1^2=0,m_3^2,m^2)
=\frac{i}{(4\pi)^2}
\frac{\Gamma(1+\epsilon)(4\pi\mu^2)^\epsilon}{s(m^2-t)}
\\
&\begin{split}
\quad\times\Bigg[&
-\frac{3}{2\epsilon^2}+
\frac{1}
{\epsilon}\left(2\ln(m^2-t-i\eta)-\frac{1}{2}\ln(m^2-i\eta)
+\ln(-s-i\eta)-\ln(m^2-m_3^2-i\eta)\right)
\\
&+\frac{2\pi^2}{3}
+\frac{1}{4}\ln^2(m^2-i\eta)-\ln^2(m^2-t-i\eta)+\ln^2(m^2-m_3^2-i\eta)
-\ln^2(-s-i\eta)
\\
&+\ln(m^2-i\eta)\left(\ln(-s-i\eta)-\ln(m^2-m_3^2-i\eta)\right)
\\
&-2\li\left(1-\frac{m^2-m_3^2-i\eta}{-s-i\eta}\right)
+2\li\left(1-(m_3^2-m^2+i\eta)f^m\right)
\\
&-2\li\left(1-(t-m^2+i\eta)f^m\right)
-2\li\left(1-(s+i\eta)f^m\right)
\Bigg],
\end{split}
\end{split}
\label{4p7}
\end{equation}
where
$f^m=\frac{s+t-m_3^2}{s(t-m^2)}$.
\section{Massless five-point integral}
\label{5point}
We consider the following massless five-point integral:
\begin{equation}
E_0=\mu^{2\epsilon}\intd\frac{1}{
k^2(k+\bar{x}p)^2(k+p)^2(k+l)^2(k+l-yq)^2
}.
\label{5p1}
\end{equation}
Despite the fact that there occur only three linearly independent
momenta in $E_0$, (\ref{5p1}) cannot be decomposed into partial
fractions. However using a standard Feynman parametrisation we end up
with integrals which can be calculated by computer algebra
systems. Now the exact result is rather involved so I give just the
leading power. Higher powers can be obtained by using the methods of
section \ref{diffeq}.
\begin{equation}
\begin{split}
E_0\doteq
\ifp\frac{\Gamma(1+\epsilon)(4\pi\mu^2)^\epsilon}{\bar{x}y\theta\xi^2}
\bigg[&
-\frac{2}{\epsilon^2}+
\frac{-4+2\ln\xi+2\ln\theta+2i\pi}{\epsilon}
-\ln^2\xi-\ln^2\theta\\
&-2\ln\xi\ln\theta+4\ln\xi+4\ln\theta+\pi^2\\
&+2i\pi\left(2-2\ln\xi-2\ln\theta\right)
\bigg].
\label{5p2}
\end{split}
\end{equation} 
\chapter{Matching of $\lambda_B$}
\label{matchlb}
In this thesis we consider a pure QCD calculation. This
calculation is given in terms of $f_B/\lambda_B$. In order to compare our
calculation to HQET or SCET results we have to match the expression
$f_B/\lambda_B$ onto HQET. We use the definition of the $B$-meson
wave function of \cite{Lange:2003ff}, which is defined in the case of HQET by
the heavy quark field of $h_v$. So we define
\begin{equation}
if_Bm_B\phi_+^\text{QCD}(\omega)\equiv\frac{1}{2\pi}\int_0^\infty dt\,
e^{i\omega t}\langle 0|\bar{q}(z)[\ldots]\sh{n}_+\gamma_5 b(0)|
\bar{B}\rangle_{z^-,z_\perp=0}
\label{match1}
\end{equation}
in the case of QCD and 
\begin{equation}
i\hat{f}_Bm_B\phi_+^\text{HQET}(\omega)\equiv
\frac{1}{2\pi}\int_0^\infty dt\,
e^{i\omega t}\langle 0|\bar{q}(z)[\ldots]\sh{n}_+\gamma_5 h_v(0)|
\bar{B}\rangle_{z^-,z_\perp=0}
\label{match2}
\end{equation} 
in the case of HQET. Here the integration goes over $t=v\cdot z$ where
$v$ is the four-velocity of the $B$-meson.
We define the $B$-meson decay constant by
\begin{equation}
if_Bm_B=\langle0|\bar{q}(0)[\ldots]\sh{n}\gamma_5b(0)|\bar{B}\rangle
\label{match1.1}
\end{equation}
and analogously the HQET decay constant, which depends on the
renormalisation scale $\mu$, by
\begin{equation}
i\hat{f}_B(\mu)m_B=
\langle0|\bar{q}(0)[\ldots]\sh{n}\gamma_5h_v(0)|\bar{B}\rangle.
\label{match1.2}
\end{equation}
The matching coefficient is defined by
\begin{equation}
f_B\int_0^\infty d\omega\,\frac{\phi_+^\text{QCD}(\omega)}{\omega}
\equiv C_{\lambda_B}
\hat{f}_B\int_0^\infty
d\omega\,\frac{\phi_+^\text{HQET}(\omega)}{\omega}
\label{matcheq}
\end{equation}
such that we get writing the $\mu$-dependence explicitly
\begin{equation}
\frac{f_B}{\lambda_B^\text{QCD}(\mu)}=
C_{\lambda_B}(\mu)
\frac{\hat{f}_B(\mu)}{\lambda_B^\text{HQET}(\mu)}.
\label{matcheq2}
\end{equation}
\begin{figure}
\begin{center}
\resizebox{0.7\textwidth}{!}{\includegraphics{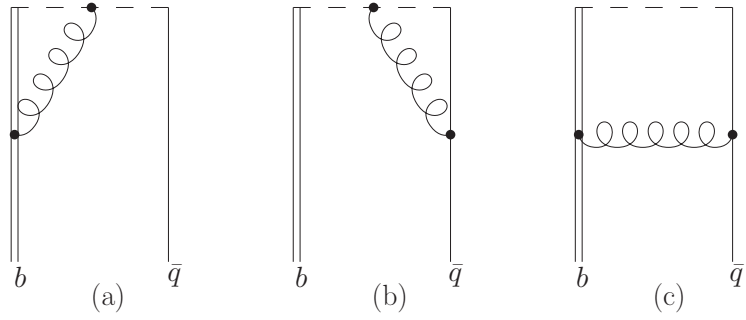}}
\end{center}
\caption{NLO contributions to $\lambda_B$. The double line stands for
  the $b$-quark field.}
\label{lbpic1}
\end{figure}
We get $C_{\lambda_B}$ to $\mathcal{O}(\alpha_s)$ by calculating the
convolution integrals occurring in (\ref{matcheq}) in both QCD and HQET
up to $\mathcal{O}(\alpha_s)$. The corresponding diagrams are shown in
fig.~\ref{lbpic1}. As in section \ref{wf} we can use the wave
functions (\ref{match1}), (\ref{match2}) defined by free quark states 
instead of hadronic states. We assign to the $b$-quark the
momentum $v(m_b-\tilde{\omega})$ ($-v\tilde{\omega}$ resp.) in the case of
pure QCD (HQET resp.) and $v\tilde{\omega}$ to the soft constituent quark,
where $v$ is the four velocity of the $B$-meson. At tree level we get
for both QCD and HQET the same wave function:
\begin{equation}
if_B\phi_+^{(0)}(\omega)=
N_c\delta(\omega-\tilde{\omega})\,\bar{q}\sh{n}_+\gamma_5\Psi
\label{match3}
\end{equation}
where the spinor $\Psi$ fulfils the condition
$\sh{v}\Psi=\Psi$, and our convolution integral is
\begin{equation}
if_B\int_0^\infty d\omega\,\frac{\phi_+^{(0)}(\omega)}{\omega}
=\frac{1}{N_c\tilde{\omega}}\,\bar{q}\sh{n}_+\gamma_5\Psi.
\label{match4}
\end{equation}
At NLO only the first diagram in fig.~\ref{lbpic1} needs to be
considered as the other two are in leading power identical for QCD and
HQET. The following expressions are given in the $\overline{\text{MS}}$
scheme, i.e.\ we redefine $\mu^2\to \mu^2 \frac{e^{\gamma_\text{E}}}{4\pi}$.
For the diagram in fig. \ref{lbpic1}(a) we get in QCD 
\begin{equation}
\frac{\alpha_s}{4\pi}C_FN_c\frac{1}{\tilde{\omega}}
\bar{q}\sh{n}_+\gamma_5\Psi\left(
\frac{2+2\ln\frac{\tilde{\omega}}{m_b}}{\epsilon}+
4\ln\frac{\mu}{m_b}+4-\frac{\pi^2}{6}-2\ln^2\frac{\tilde{\omega}}{m_b}
+4\ln\frac{\tilde{\omega}}{m_b}\ln\frac{\mu}{m_b}
\right)
\label{match5}
\end{equation}
and in HQET
\begin{equation}
\frac{\alpha_s}{4\pi}C_FN_c\frac{1}{\tilde{\omega}}
\bar{q}\sh{n}_+\gamma_5\Psi\left(
-\frac{1}{\epsilon^2}+\frac{2\ln\frac{\tilde{\omega}}{\mu}}{\epsilon}
-2\ln^2\frac{\tilde{\omega}}{\mu}-\frac{\pi^2}{4}
\right).
\label{match6}
\end{equation}
The wave function renormalisation constants of the heavy quark field
are given in the onshell scheme for the QCD $b$-field:
\begin{equation}
Z_{2b}^\frac{1}{2}=1+\frac{\alpha_s}{4\pi}C_F\left(
-\frac{1}{2\epsilon}-\frac{1}{\epsilon_\text{IR}}-3\ln\frac{\mu}{m_b}-2
\right)
\label{match7}
\end{equation}
and for the HQET field $h_v$:
\begin{equation}
Z_{2h_v}^\frac{1}{2}=1+\frac{\alpha_s}{4\pi}C_F\left(
\frac{1}{\epsilon}-\frac{1}{\epsilon_\text{IR}}\right).
\label{match8}
\end{equation}
The renormalisation of the $q$-field drops out in the
matching. Diagrammatically the matching equation (\ref{matcheq}) reads:
\begin{equation}
Z_{2b}^\frac{1}{2}\left(
\raisebox{-.5cm}{\resizebox{1cm}{!}{\includegraphics{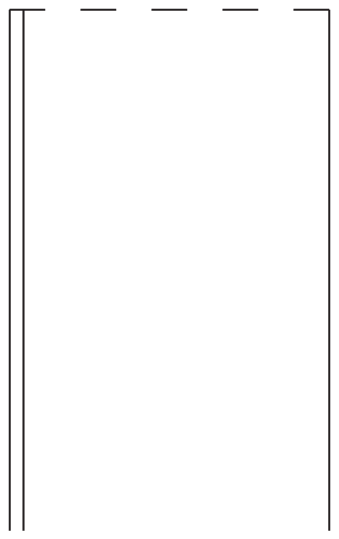}}}
+
\raisebox{-.5cm}{\resizebox{1cm}{!}{\includegraphics{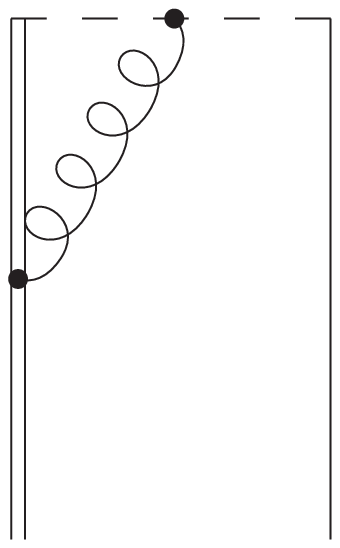}}}
\right)^\text{QCD}
=
C_{\lambda_B}
Z_{2h_v}^\frac{1}{2}\left(
\raisebox{-.5cm}{\resizebox{1cm}{!}{\includegraphics{lambdaB/match0}}}
+
\raisebox{-.5cm}{\resizebox{1cm}{!}{\includegraphics{lambdaB/match}}}
\right)^\text{HQET}.
\label{match9}
\end{equation}
Finally we obtain
\begin{equation}
C_{\lambda_B}(\mu)=1+\frac{\alpha_s}{4\pi}C_F\left(
2\ln^2\frac{\mu}{m_b}+\ln\frac{\mu}{m_b}+2+\frac{\pi^2}{12}
\right)
\label{match10}
\end{equation}
where we have renormalised the UV-divergences in the
$\overline{\text{MS}}$-scheme. 

\end{appendix}

  \backmatter

  \markboth{}{}

  \addcontentsline{toc}{chapter}{\protect Acknowledgements}

\chapter*{Acknowledgements}

There are many people who contributed to the present thesis.
First of all I would like to thank my advisor Prof. Gerhard Buchalla.
He made my PhD studies possible by supplying me with a scholarship of
the ``Graduiertenkolleg  Particle Physics at the Energy Frontier 
of New Phenomena''. I also want to thank him for many inspiring
discussions.

I am particularly indebted to Guido Bell for giving me useful hints
and for stimulating discussions, where I learned much about 
technical details in the calculations of Feynman integrals. I would 
also like to thank my room mate Matth\"aus Bartsch and Sebastian
J\"ager for many instructive and helpful discussions. 

Furthermore I want to thank Matth\"aus Bartsch, Guido Bell, Gerhard 
Buchalla and Sebastian J\"ager for proofreading the drafts 
and comments on the manuscript.

Finally I want to thank my parents, whose aid made my physics studies
possible.

\end{document}